\documentclass[a4paper,10pt]{article}
\usepackage[margin=0.75in]{geometry}
\usepackage{graphicx,tabularx,amsthm,amsmath,mathtools,amsfonts,bm,amssymb,setspace,color,float,epstopdf,caption,subcaption,appendix}
\usepackage[pdfpagemode=UseNone,pdfstartview={XYZ null null null}]{hyperref}
\renewcommand\theequation{\thesection.\arabic{equation}}

\definecolor{dark-red}{rgb}{0.4,0.15,0.15}
\definecolor{dark-blue}{rgb}{0.15,0.15,0.4}
\definecolor{medium-blue}{rgb}{0,0,0.5}
\hypersetup{colorlinks,linkcolor={dark-red},citecolor={dark-blue},urlcolor={medium-blue}}

\newtheorem{remark}{Remark}[section]
\newtheorem{prop}{Proposition}[section]

\newcommand{\bsub}{\begin{subequations}}
\newcommand{\esub}{\end{subequations}$\!$}

\newcommand{\R}{\mathbb{R}}

\newcommand{\Fb}{\mathbb{\mathcal F}}
\newcommand{\Nb}{\mathbb{\mathcal N}}

\newcommand{\eps}{\varepsilon}
\newcommand{\lam}{\lambda}

\newcommand{\hb}{\mbox{\boldmath$h$}}
\newcommand{\fb}{\mbox{\boldmath$\mu_p$}}
\newcommand{\xb}{\mbox{\boldmath$x$}}
\newcommand{\vb}{\mbox{\boldmath$v$}}
\newcommand{\eb}{\mbox{\boldmath$e$}}

\DeclareMathOperator{\csch}{csch}
\DeclareMathOperator{\sech}{sech}

\title{Stable Asymmetric Spike Equilibria for the Gierer-Meinhardt
  Model with a Precursor Field} \author{Theodore Kolokolnikov
  \thanks{Department of Mathematics and Statistics, Dalhousie
    University, Halifax, Canada. (corresponding author {\tt tkolokol@gmail.com})} \and Fr\'ed\'eric Paquin-Lefebvre
  \thanks{Department of Mathematics, UBC, Vancouver, Canada. ({\tt
      paquinl@math.ubc.ca}) } \and Michael J. Ward \thanks{Department
    of Mathematics, UBC, Vancouver, Canada. ({\tt ward@math.ubc.ca})} }

\begin{document}

\maketitle

\begin{abstract}
  Precursor gradients in a reaction-diffusion system are spatially
  varying coefficients in the reaction-kinetics. Such gradients have
  been used in various applications, such as the head formation in the
  Hydra, to model the effect of pre-patterns and to localize patterns
  in various spatial regions. For the 1-D Gierer-Meinhardt (GM) model
  we show that a simple precursor gradient in the decay rate of the
  activator can lead to the existence of stable, asymmetric, two-spike
  patterns, corresponding to localized peaks in the activator of
  different heights. This is a qualitatively new phenomena for the GM
  model, in that asymmetric spike patterns are all unstable in the
  absence of the precursor field. Through a determination of the
  global bifurcation diagram of two-spike steady-state patterns, we
  show that asymmetric patterns emerge from a supercritical
  symmetry-breaking bifurcation along the symmetric two-spike branch
  as a parameter in the precursor field is varied. Through a combined
  analytical-numerical approach we analyze the spectrum of the
  linearization of the GM model around the two-spike steady-state to
  establish that portions of the asymmetric solution branches are
  linearly stable. In this linear stability analysis a new class of
  vector-valued nonlocal eigenvalue problem (NLEP) is derived and analyzed.
\end{abstract}

\section{Introduction}\label{sec:intro}

We analyze the existence, linear stability, and bifurcation behavior
of localized steady-state spike patterns for the Gierer-Meinhardt
reaction-diffusion (RD) model in a 1-D domain where we have included a
spatially variable coefficient for the decay rate of the activator. We
will show that this spatial heterogeneity in the model, referred to as
a precursor gradient, can lead to the existence of stable {\em
  asymmetric} two-spike equilibria, corresponding to steady-state
spikes of different height (see the right panel of
Fig.~\ref{fig:pde_run1}). This is a qualitatively new phenomenon for
the GM model since, in the absence of a precursor field, asymmetric
steady-state spike patterns for the GM model are always unstable
\cite{ww_asy}. A combination of analytical and numerical methods is
used to determine parameter ranges where stable asymmetric
steady-state patterns for the GM model with a simple precursor field
can occur. We will show that these stable asymmetric equilibria emerge
from a symmetry-breaking supercritical pitchfork bifurcation of
symmetric spike equilibria as a parameter in the precursor field is
varied.

Precursor gradients have been used in various specific applications of
RD theory since the initial study by Gierer and Meinhardt
in \cite{gm} for modeling head development in the Hydra. For
other RD systems, precursor gradients have also been used in the
numerical simulations of \cite{holloway1} to model the formation and
localization of heart tissue in the Axolotl, which is a type of
salamander. Further applications of such gradients for the GM model
and other RD systems are discussed in \cite{holloway1},
\cite{holloway2}, \cite{m}, and \cite{hh}. With a precursor field, or
with spatially variable diffusivities, the RD system
does not generally admit a spatially uniform state. As a result, a
conventional Turing stability approach is not applicable and the
initial development of small amplitude patterns must be analyzed through
either a slowly-varying assumption or from full numerical simulations
(cf.~\cite{hund1}, \cite{page1}, \cite{page2}, \cite{krause2}).

In contrast to small amplitude patterns, in the singularly perturbed
limit of a large diffusivity ratio ${\mathcal O}(\eps^{-2})\gg 1$, many
two-component RD systems in 1-D admit spike-type solutions. In this
direction, there is a rather extensive analytical theory on the
existence, linear stability and slow dynamics of spike-type solutions
for many such RD systems in 1-D (see \cite{dk}, \cite{dkp},
\cite{iww}, \cite{iw}, \cite{swr} \cite{tzou_2}, \cite{tzou_3}, and
the references therein). To establish parameter regimes where
spike-layer steady-states are linearly stable, one must analyze the
spectrum of the operator associated with a linearization around the
spike-layer solution. In this spectral analysis one must consider both
the small eigenvalues of order ${\mathcal O}(\eps^2)$ associated with
near-translation invariance and the large ${\mathcal O}(1)$
eigenvalues that characterize any instabilities in the amplitudes of
the spikes. These latter eigenvalues are associated with nonlocal
eigenvalue problems (NLEPs), for which many rigorous results are
available (cf.~\cite{dgk3}, \cite{wei_surv}, \cite{ww}).

Despite these advances, the effect of spatially heterogeneous
coefficients in the reaction kinetics on spike existence, stability,
and dynamics is much less well understood. With a precursor gradient,
spike pinning can occur for the GM model (cf.~\cite{wmhgm},
\cite{winter}) and for the Fitzhugh-Nagumo model
(cf.~\cite{chen_nagumo}, \cite{heijster}), while a plant hormone
(auxin) gradient is predicted to control the spatial locations of root
formation in plant cells \cite{brena}. In other contexts, a spatial
heterogeneity can trigger a self-replication loop consisting of spike
formation, propagation, and annihilation against a domain boundary
\cite{krause}. More recently, clusters of spikes that are confined as
a result of a spatial heterogeneity have been analyzed in 1-D in
\cite{kx} and \cite{kw_siamrev} for the GM and Schnakenberg models,
respectively, and in \cite{kwei} for 2-D spot clusters of the GM
model. In these recent approaches the RD system with clustered spikes
is effectively approximated by a limiting equation for the spike
density.

In our study we will consider the dimensionless GM model in 1-D with
activator $a$ and inhibitor $h$, and with a smooth precursor
$\mu(x)>0$ in the decay rate of the activator, given for $\eps\ll 1$
by \bsub\label{gm:full}
\begin{align}
  a_t &=\eps^2 a_{xx} -\mu(x) a+ \frac{a^2}{h}\,,
   \qquad |x|<L \,, \quad  t>0\,; \qquad a_x(\pm L,t)=0 \,, \label{gm:fulla}\\
 \tau h_t &= h_{xx} - h +\eps^{-1}a^2 \,, \qquad |x|<L \,,
            \quad t > 0 \,; \qquad h_{x}(\pm L,t) = 0 \,. \label{gm:fullb}
\end{align}
\esub Although our analytical framework can be applied more
generally, we will exhibit stable asymmetric spike-layer steady-states
only for the specific precursor field
\begin{equation}
  \mu(x)=1+b x^2 \,, \label{intro:prec}
\end{equation}
where $b>0$ is a bifurcation parameter. In our formulation in
\eqref{gm:full}, we have for convenience fixed the inhibitor
diffusivity at unity and will use the domain length $L$ as the other
bifurcation parameter.

\begin{figure}[htbp]
\begin{center}
  \includegraphics[width=0.49\linewidth,height=4.5cm]{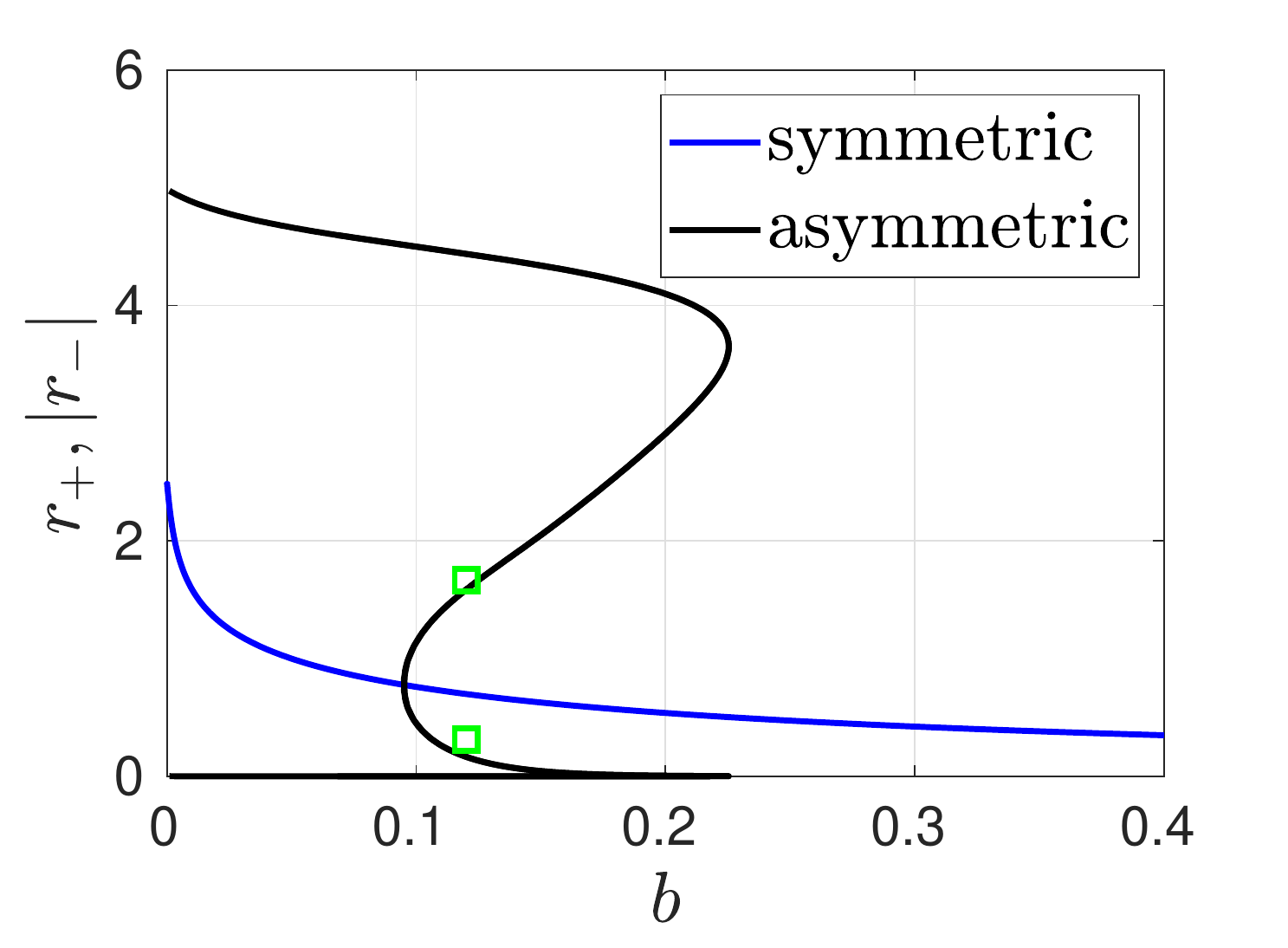}
  \includegraphics[width=0.49\linewidth,height=4.5cm]{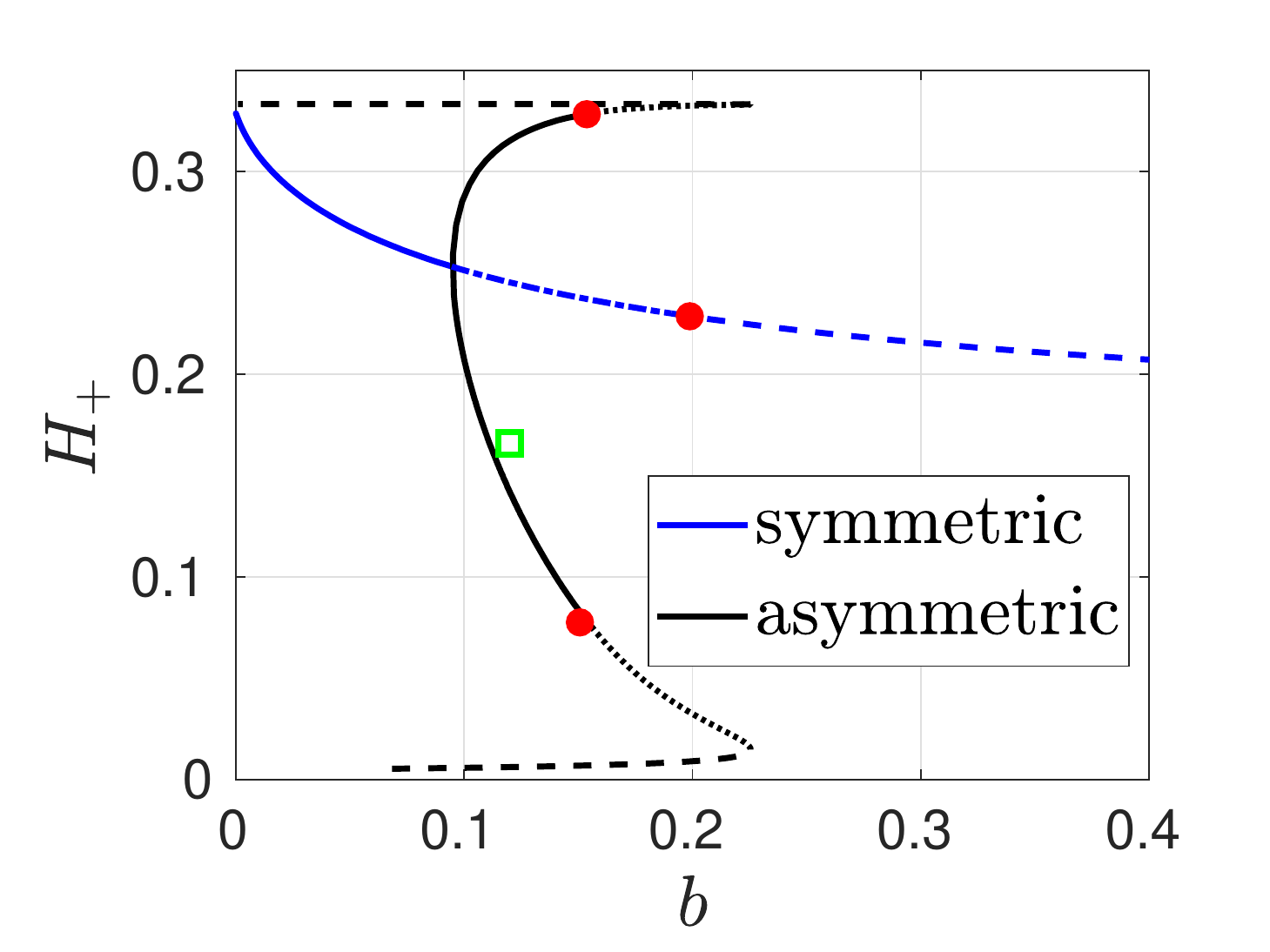}
  \caption{Left: steady-state spike locations $x_1=-r_{-}$ and
    $x_2=r_{+}$ for $L=5$ versus $b$ in \eqref{intro:prec}. Right:
    height $H_+$ of the rightmost spike versus $b$. Solid lines:
    linearly stable to both the small eigenvalues and the large (NLEP)
    eigenvalues when $\tau\ll 1$. Dash-dotted lines: unstable for the
    small eigenvalues but stable for the large eigenvalues when
    $\tau\ll 1$. Dashed line: stable to the small eigenvalues but
    unstable to the large eigenvalues when $\tau\ll 1$. Dotted line:
    unstable to both the small and large eigenvalues when $\tau\ll
    1$. Red dots: zero-eigenvalue crossings for the NLEP. Green
    squares: the stable steady-state observed in the full PDE
    simulation of \eqref{gm:full} shown in
    Fig.~\ref{fig:pde_run1}.}\label{fig:L_5}
\end{center}
\end{figure}

\begin{figure}[htbp]
\begin{center}
  \includegraphics[width=0.32\linewidth,height=4.5cm]{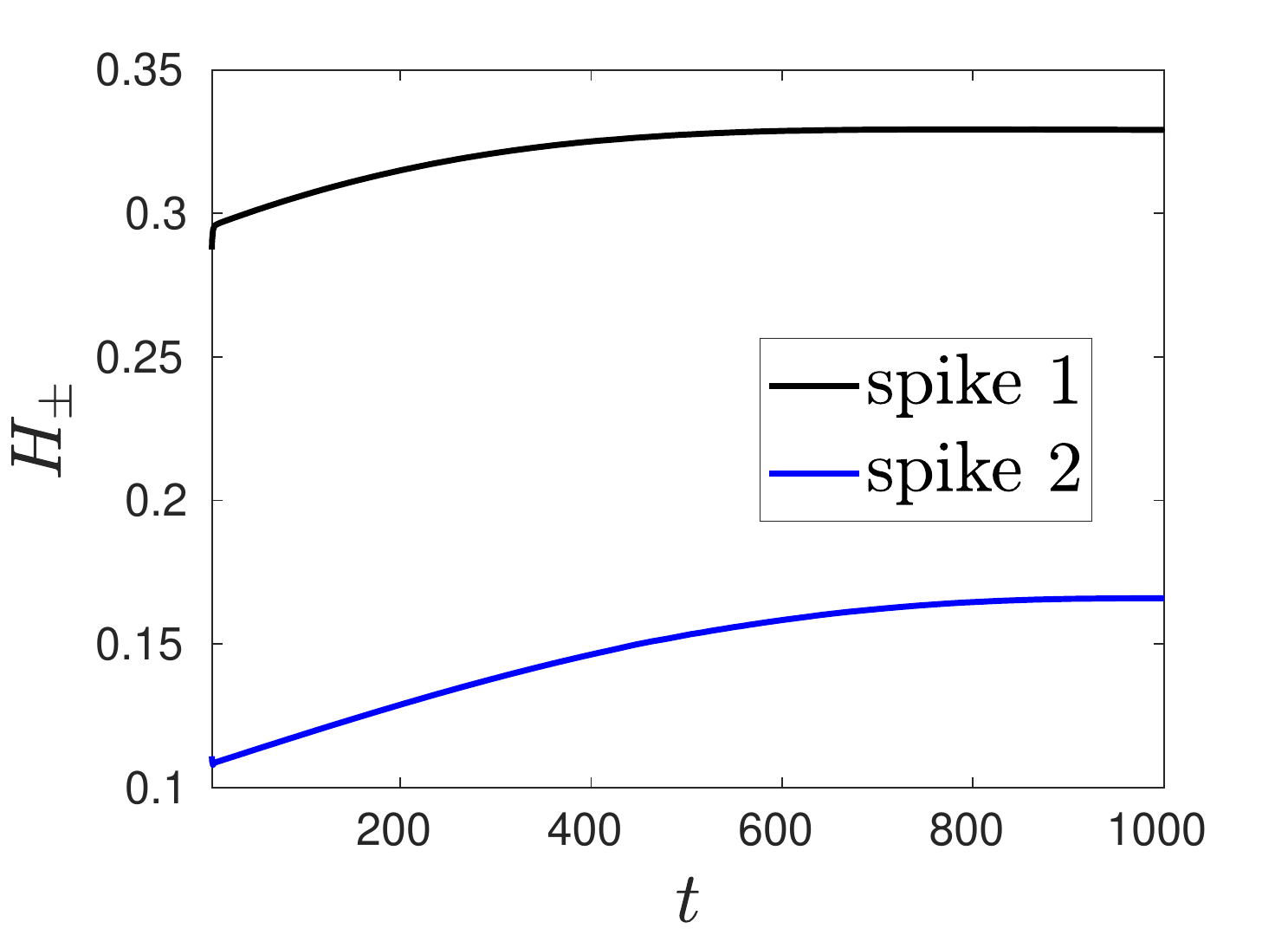}
  \includegraphics[width=0.32\linewidth,height=4.5cm]{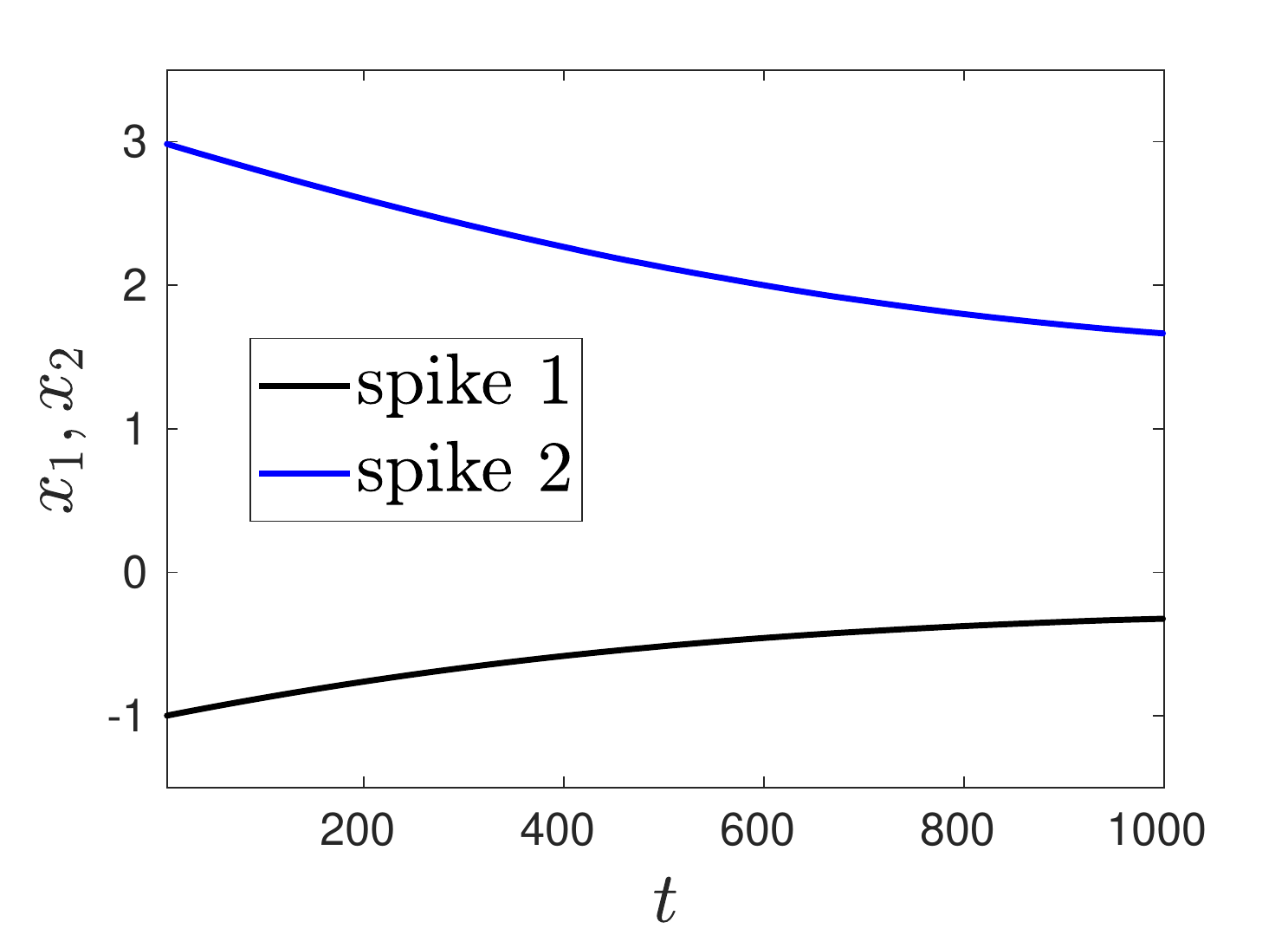}
  \includegraphics[width=0.32\linewidth,height=4.5cm]{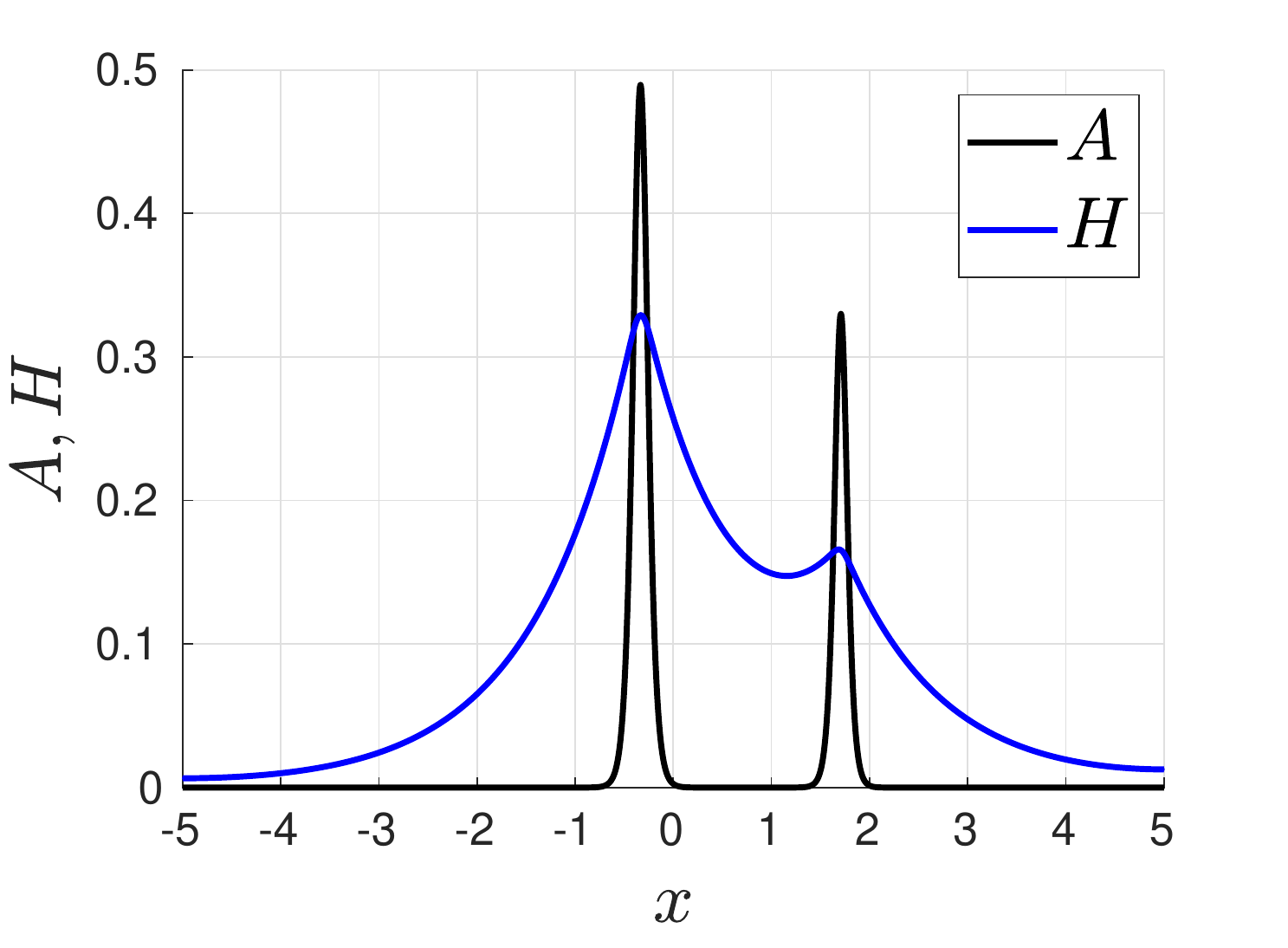}
  \caption{Time-dependent PDE simulations of \eqref{gm:full} with
    $L=5$, $\eps=0.05$, and $\tau=0.25$ for a precursor
    $\mu(x)=1+bx^2$ with $b=0.12$. Initial condition is a
    quasi-equilibrium two-spike solution with spike locations
    $x_1(0)=-1$ and $x_2(0)=3$. Spike heights (left panel), denoting
    maxima of the inhibitor field, and spike locations (middle panel)
    versus time. Right: the steady-state asymmetric two-spike
    equilibrium, stable to the small and large eigenvalues,
    corresponding to the green squares in
    Fig.~\ref{fig:L_5}.} \label{fig:pde_run1}
\end{center}
\end{figure}

In \S \ref{pre:dae} we use a matched asymptotic approach to derive a
differential algebraic system of ODEs (DAEs) for a collection of
spikes for \eqref{gm:full}, under the assumption that the
quasi-equilibrium spike pattern is stable on ${\mathcal O}(1)$
time-scales. The DAE system is written in terms of 1-D Green's
functions, or equivalently as a tridiagonal system. In \S
\ref{pre:bif} we provide two alternative approaches for computing
global branches of two-spike equilibria of the DAE system, for the
$\mu$ as given in \eqref{intro:prec}, and we formulate a generalized
matrix eigenvalue problem characterizing the linear stability of
branches of equilibria. Numerical results for steady-state spike
locations and spike heights, denoting maxima of the inhibitor field,
corresponding to global bifurcation branches of two-spike equilibria
are shown in \S \ref{pre:two_numer} in terms of the precursor
parameter $b$ and the domain half-length $L$. We show that the
asymmetric branches of two-spike equilibria emerge from a symmetry
breaking pitchfork bifurcation from the symmetric branch at a critical
value $b=b_p(L)$. For $b>0.076$, we show that this bifurcation is
supercritical, and that the bifurcating branches of asymmetric
equilibria are linearly stable as a steady-state solution of the DAE
dynamics.

In \S \ref{pre:comp} we derive a vector-valued NLEP characterizing
spike amplitude instabilities of steady-state spike patterns of
\eqref{gm:full}. For the case of symmetric two-spike equilibria, the
vector-valued NLEP can be diagonalized, and we obtain necessary and
sufficient conditions for the linear stability of these patterns when
$\tau$ in \eqref{gm:full} is sufficiently small. The resulting
stability thresholds are shown in the global bifurcation plots in \S
\ref{pre:two_numer}. However, for asymmetric two-spike equilibria, we
obtain a new vector-valued NLEP that cannot be diagonalized, and for
which the NLEP stability results in \cite{wei_surv} are not directly
applicable.  For this new NLEP we determine analytically parameter
values corresponding to zero-eigenvalue crossings, and for 
$\tau=0$ we numerically compute any unstable eigenvalues by using a
discretization of the vector-valued NLEP combined with a generalized
matrix eigenvalue solver.

In \ref{pre:numerics} we confirm our global bifurcation and linear
stability results through full PDE simulations of \eqref{gm:full}. As
an illustration of our results, in Fig.~\ref{fig:L_5} we plot the
spike locations and spike heights corresponding to steady-state
branches of symmetric and asymmetric two-spike equilibria in terms of
the precursor parameter $b$ for a domain half-length $L=5$. The two
branches of asymmetric two-spike equilibria result from an even
reflection of solutions through the origin $x=0$. In the right panel
of Fig.~\ref{fig:L_5}, where we plot the spike heights, we show the
linear stability properties for the small eigenvalues, as obtained
from the linearization of the DAE system, and for the large
eigenvalues, as determined from computations of the vector-valued
NLEP. The time-dependent PDE simulations shown in
Fig.~\ref{fig:pde_run1} confirm that a quasi-equilibrium two-spike
pattern tends to a stable asymmetric equilibrium on a long time
scale. The paper concludes with a brief discussion in \S
\ref{sec:discussion}.

\section{Derivation of the DAE System}\label{pre:dae}

We now derive a DAE system for the spike locations for an $N$-spike
quasi-equilibrium pattern, which is valid in the absence of any
${\mathcal O}(1)$ time-scale instability of the pattern. Since this
analysis is similar to that given in \cite{iw} with no precursor field
and in \cite{wmhgm} for a precursor field, but with only one spike, we
only briefly outline the analysis here.

The spike locations $x_j$, for $j=1,\ldots,N$, are assumed to satisfy
$|x_{j+1}-x_j|\gg {\mathcal O}(\eps)$, with
$|x_1+L|\gg {\mathcal O}(\eps)$ and $|L-x_N|\gg {\mathcal
  O}(\eps)$. As shown in \cite{iw} and \cite{wmhgm}, in the absence of
any ${\mathcal O}(1)$ time-scale instability of the spike amplitudes,
the spikes will evolve on the long time-scale $\sigma=\eps^2 t$, and
so we write $x_j=x_j(\sigma)$.

To derive a DAE system for $x_j(\sigma)$, for $j=1,\ldots,N$, we first
construct the solution in the inner region near the $j$-th spike.
We introduce the inner expansion
\begin{equation}\label{dae:inn_ex}
  a = A_0 + \eps A_1 + \ldots \,, \qquad h=H_0 + \eps H_1 + \cdots \,,
\end{equation}
where $A_i=A_i(y,\sigma)$ and $H_i=H_i(y,\sigma)$ for $i=0,1$ and
$y=\eps^{-1}(x-x_j)$. Upon substituting \eqref{dae:inn_ex} into
\eqref{gm:full}, and using
$a_t=-\eps x_j^{\prime}A_{0y} +{\mathcal O}(\eps^2)$ where
$x_j^{\prime}\equiv {dx_j/d\sigma}$, we collect powers of $\eps$ to
obtain the following leading-order problem on $-\infty<y<\infty$:
\begin{equation}\label{dae:a0}
   A_{0yy} - \mu_j A_0 + {A_0^2/H_0} = 0 \,, \qquad H_{0yy}=0 \,,
\end{equation}
where $\mu_j\equiv \mu(x_j)$. At next order, we conclude on $-\infty<y<\infty$
that
\bsub\label{dae:a1h1}
\begin{gather}
    {\mathcal L} A_1 \equiv  A_{1yy} - \mu_j A_1 + \frac{2A_0}{H_0} A_1 =
    \frac{A_0^2}{H_0^2} H_1 + y \mu^{\prime}(x_j) A_0 - x_j^{\prime} A_{0y}\,,
     \label{dae:a1}\\
    H_{1yy} =-A_0^2 \,. \label{dae:a2}
\end{gather}
\esub

From \eqref{dae:a0} we get that $H_0=H_{0j}(\sigma)$, where $H_{0j}$,
independent of $y$, is to be determined. In addition, the spike profile
is given by
\begin{equation}\label{dae:a0_sol}
  A_0 = \mu_j H_{0j} w\left(\sqrt{\mu_j} y\right) \, \qquad \mbox{where}
  \qquad  w(z)=\frac{3}{2}\sech^2({z/2}) \,,
\end{equation}
where $w(0)>0$ with $w^{\prime}(0)=0$, is the well-known homoclinic solution to
\begin{equation}
  w^{\prime\prime} -w + w^2 =0 \,, \quad -\infty<z<\infty \,, \quad
  w\to 0 \quad \mbox{as} \quad |z| \rightarrow\infty \,.
\end{equation}
Since ${\mathcal L} A_{0y}=0$, the solvability condition for
\eqref{dae:a1} is that
\begin{equation}\label{dae:solv_1}
  \begin{split}
    x_j^{\prime} \int A_{0y}^2\, dy & = \mu^{\prime}(x_j) \int y A_0 A_{0y} \, dy
    + \int \frac{A_0^2}{H_{0j}^2} H_1 A_{0y} \, dy \\
    & = \frac{\mu^{\prime}(x_j)}{2} \int y \left(A_0^2\right)_y \, dy +
    \frac{1}{3H_{0j}^2} \int \left(A_0^3\right)_y H_1 \, dy \\
    & = -\frac{\mu^{\prime}(x_j)}{2} \int A_0^2 \, dy -
    \frac{1}{3H_{0j}^2} \int A_0^3 H_{1y} \, dy \,,
  \end{split}
\end{equation}
where we have used integration by parts and the shorthand notation
$\int =\int_{-\infty}^{\infty}$. From a further integration by parts
on the last term on the last line in \eqref{dae:solv_1}, and using the
fact that $H_{1yy}=-A_0^2$ is even, we obtain that
\begin{equation}\label{dae:solv_2}
  x_j^{\prime} = -\frac{\mu^{\prime}(x_j)}{2} I_1 - \frac{1}{6H_{0j}^2} I_2
  \left(\lim_{y\to +\infty} H_{1y} + \lim_{y\to -\infty} H_{1y} \right)\,,
\end{equation}
in terms of the integral ratios $I_1$ and $I_2$ defined by
\begin{equation}\label{dae:int}
  I_1 \equiv \frac{\int A_0^2 \, dy}{\int A_{0y}^2 \, dy} \,, \qquad
  I_2 \equiv \frac{\int A_0^3 \, dy}{\int A_{0y}^2 \, dy } \,. \qquad
\end{equation}
By multiplying the ODE for $A_0$ in \eqref{dae:a0} first by $A_{0y}$
and then by $A_0$, we integrate the two resulting expressions to
obtain an algebraic system for $I_1$ and $I_2$, which yields
\begin{equation}\label{dae:int_res}
  I_1 = \frac{5}{\mu_j} \,, \qquad I_2 = 6 H_{0j} \,.
\end{equation}
Upon using \eqref{dae:int_res} in \eqref{dae:solv_2}, we conclude for
each $j=1,\ldots,N$ that
\begin{equation}\label{dae:solv_end}
  x_j^{\prime} = -\frac{5}{2} \frac{\mu^{\prime}(x_j)}{\mu(x_j)} -
 \frac{1}{H_{0j}}\left(\lim_{y\to +\infty} H_{1y} + \lim_{y\to -\infty} H_{1y} \right)\,.
\end{equation}
To determine $H_{0j}$ for $j=1,\ldots,N$ and the remaining term in
\eqref{dae:solv_end} we need to determine the outer solution.
 
Now in the outer region, defined away from ${\mathcal O}(\eps)$
regions near each $x_j$, $a$ is exponentially small. In the sense of
distributions we then use $A_0=H_{0j}\mu_j w(\sqrt{\mu_j}y)$ to calculate
across each $x=x_j$ that
\begin{equation}\label{dae:dist}
  \frac{1}{\eps} a^2 \rightarrow \left(\int A_0^2 \, dy\right) \delta(x-x_j)
  = \mu_j^{3/2} H_{0j}^2 \left(\int  w^2(z) \, dz\right) \delta(x-x_j)= 
  6 \mu_j^{3/2} H_{0j}^2 \delta(x-x_j) \,,
\end{equation}
owing to the fact that $\int w^2\, z=\int w \, dz=6$. In this way, the
outer problem for $h$ is
\begin{equation}\label{dae:hout}
  h_{xx} - h  = - 6 \sum_{j=1}^{N} H_{0j}^2 \mu_j^{3/2} \delta(x-x_j) \,, \quad
  |x|\leq L \,; \qquad h_x(\pm L,\sigma)=0 \,.
\end{equation}
The solution to \eqref{dae:hout} is
\begin{equation}\label{dae:hout_solv}
  h(x) = \sum_{i=1}^{N} H_{0i}^{2} \mu_i^{3/2} G(x;x_i) \,,
\end{equation}
 where $G(x;x_i)$ is the 1-D Green's function satisfying
\begin{equation}\label{dae:green}
  G_{xx} - G = -\delta(x-x_i) \,, \quad |x|\leq L\,; \qquad
   G_{x}(\pm L;x_i) = 0 \,.
\end{equation}
To match with the inner solutions near each $x_j$, we require for each
$j=1,..,N$ that
\begin{equation}\label{dae:h_match}
  h(x_j) = H_{0j} \,, \qquad \lim_{y\to \infty} H_{1y} + \lim_{y\to -\infty}
  H_{1y} = h_{x}(x_{j+}) + h_{x}(x_{j-}) \,. 
\end{equation}
In this way, by using \eqref{dae:h_match} in \eqref{dae:hout_solv}
and \eqref{dae:solv_end} we obtain the following DAE system for slow
spike motion:
\bsub \label{dae:full}
\begin{align}
  \frac{dx_j}{d\sigma} &= -\frac{5}{2}\frac{\mu^{\prime}(x_j)}{\mu_j}
   - \frac{12}{H_{j}} \left(\mu_j^{3/2} H_{j}^2 \langle G_x \rangle_j + 
 \sum_{\stackrel{i=1}{i\neq j}}^{N} \mu_i^{3/2} H_{i}^2 G_{x}(x_j;x_i)
  \right) \,, \label{dae:full_ode}\\
  H_{j} &= 6 \sum_{i=1}^{N} \mu_i^{3/2} H_{i}^2 G(x_j;x_i) \,,
        \label{dae:full_cons}
\end{align}
\esub
where $\mu_j\equiv \mu(x_j)$, $\langle G_{x}\rangle_j \equiv {\left[
    G_{x}(x_{j+};x_i)+G_{x}(x_{j-};x_i)\right]/2}$, and $G(x;x_j)$ is the
Green's function satisfying \eqref{dae:green}. In \eqref{dae:full}, we
have relabeled $H_{0j}$ by $H_j$.

A simple special case of \eqref{dae:full} is for the infinite-line
problem with $L\to \infty$, for which $G(x;x_i)=\frac{1}{2}e^{-|x-x_i|}$.
For this case, we calculate $\langle G_{x}\rangle_j=0$ and
$G_x(x_j;x_i)=-\frac{1}{2}\mbox{sign}(x_j-x_i) e^{-|x_j-x_i|}$. In this
way, we can rewrite \eqref{dae:full} as
\bsub \label{dae:inf}
\begin{align}
  \frac{dx_j}{d\sigma} &= -\frac{5}{2}\frac{\mu^{\prime}(x_j)}{\mu_j}
 + \frac{1}{H_{j}} \sum_{\stackrel{i=1}{i\neq j}}^{N} S_i \,\mbox{sign}(x_j-x_i)
            e^{-|x_j-x_i|}\,, \label{dae:inf_1}\\
  H_{j} &= \frac{1}{2} \sum_{i=1}^{N} S_i e^{-|x_j-x_i|} \,, \qquad
    H_{j} = \left( \frac{S_j}{6\mu_j^{3/2}}\right)^{1/2} \,. \label{dae:inf_2}
\end{align}
\esub

From \eqref{dae:full_ode}, we observe that the DAE dynamics for the
$j$-th spike is globally coupled to all of the other spikes through
full matrices. We now proceed as in \cite{iw} to derive an equivalent
representation of \eqref{dae:full_ode} that is based only on nearest
neighbor interactions.  To do so, we first write \eqref{dae:full}
compactly in matrix form as
\begin{equation}\label{dae:mat}
  \frac{ d\xb}{d\sigma} = - \frac{5}{2} \fb - 2 {\mathcal H}^{-1}
  {\mathcal P} {\mathcal G}^{-1} \hb \,, \qquad
   {\mathcal G}^{-1} \hb =  6 \, {\mathcal U} \hb^{2} \,, \qquad
\end{equation}
where ${\mathcal G}$ and ${\mathcal P}$ are defined in terms of the
Green's function by
\bsub
\begin{equation}\label{dae:gmat_1}
    {\cal G} \equiv \left ( 
\begin{array}{ccc}
   G(x_{1};x_1) & \cdots & G(x_{1};x_{N}) \\
     \vdots         & \ddots & \vdots               \\
  G(x_N;x_1)    & \cdots & G(x_N;x_N)
\end{array}
\right ) \,, \,\,\, {\mathcal P} \equiv
  \left ( 
\begin{array}{ccc}
  \langle G_x \rangle_1 & \cdots & G_{x}(x_{1};x_{N}) \\
     \vdots         & \ddots & \vdots               \\
 G_{x}(x_{N};x_1) & \cdots & \langle G_x \rangle_{N}
\end{array} \right) \,.
\end{equation}
In \eqref{dae:mat}, ${\mathcal U}$ and ${\mathcal H}$ are diagonal
matrices with diagonal entries $({\mathcal U})_{jj}=\mu(x_j)$ and
$({\mathcal H})_{jj}=H_{j}$ for $j=1,\ldots,N$, and we have defined
\begin{equation}\label{dae:gmat_2}
    \hb \equiv
\left (
\begin{array}{c}
 H_{1} \\ \vdots \\ H_{N}
\end{array}
\right ) \,, \qquad
 \hb^{2} \equiv
\left (
\begin{array}{c}
 H_{1}^{2}  \\ \vdots \\ H_{N}^{2 }
\end{array}
\right ) \,, \qquad
    \fb \equiv
\left (
\begin{array}{c}
  \frac{\mu^{\prime}(x_1)}{\mu(x_1)} \\ \vdots \\
  \frac{\mu^{\prime}(x_N)}{\mu(x_N)}
\end{array}
\right ) \,.
\end{equation}
\esub As shown in Appendix A of \cite{iw} (see also Appendix A of
\cite{iww}), the inverse ${\mathcal B}\equiv {\mathcal G}^{-1}$ of the
Green's matrix and the product ${\mathcal P}{\mathcal G}^{-1}$ are
each triangular matrices of the form \bsub\label{dae:triang}
\begin{equation}\label{dae:triang_mat}
 {\mathcal B} = \left( \begin{array}{cccc}
                 c_1 & d_1 & & 0\\
                 d_1 & \ddots & \ddots & \\
                   & \ddots & \ddots & d_{N-1} \\
          0 & & d_{N-1} & c_N \end{array} \right)\,, \quad
      2 {\mathcal P} {\mathcal B} \equiv {\mathcal A} =
      \left( \begin{array}{cccc}
                 e_1 & -d_1 & & 0\\
                 d_1 & \ddots & \ddots & \\
                   & \ddots & \ddots & -d_{N-1} \\
          0 & & d_{N-1} & e_N \end{array} \right)\,, 
\end{equation}
where the matrix entries are given by
\begin{equation} \label{dae:triang_ent}
\begin{split}
  c_{1} &= \coth(x_2-x_1) + \tanh(L+x_1) \,, \quad
   c_N = \coth(x_N-x_{N-1}) + \tanh(L-x_N) \,,\\
  c_j &= \coth(x_{j+1}-x_j) + \coth(x_j-x_{j-1}) \,, \quad j=2,\ldots N-1\,, \\
    e_1 &=\tanh(L+x_1)-\coth(x_2-x_1)\,,
        \quad e_N=\coth(x_N-x_{N-1}) - \tanh(L-x_N)\,, \\
  e_j &= \coth(x_{j}-x_{j-1}) - \coth(x_{j+1}-x_{j-1}) \,,\quad j=2,\ldots N-1\,, \\
    d_j &= -\csch(x_{j+1}-x_{j}) \,, \quad j=1,\ldots,N-1 \,.
\end{split}
\end{equation}
\esub
For the infinite-line problem, we calculate for the limit $L\to \infty$
that
\begin{equation}\label{dae:coeff_inf}
  \begin{split}
  c_1 &\to \frac{2}{1-e^{-2(x_2-x_1)}} \,, \quad
 c_N \to \frac{2}{1-e^{-2(x_N-x_{N-1})}} \,, \quad \mbox{as} \quad L\to \infty\,,\\
  e_1 &\to \frac{2}{1-e^{2(x_2-x_1)}} \,, \quad
  e_N \to -\frac{2}{1-e^{2(x_N-x_{N-1})}} \,,\quad \mbox{as} \quad L\to \infty\,.
 \end{split}
\end{equation}
Finally, upon substituting \eqref{dae:triang} into \eqref{dae:mat}, we
obtain the following more tractable, but equivalent, tridiagonal
representation of the DAE dynamics \eqref{dae:full}:
\begin{equation}\label{dae:reduce}
  \frac{ d\xb}{d\sigma} = - \frac{5}{2} \fb - {\mathcal H}^{-1} {\mathcal A}
  \hb \,, \qquad {\mathcal B} \hb = 6 \, {\mathcal U} \hb^2 \,.
\end{equation}

\section{Global Bifurcation Diagram of Spike Equilibria}
\label{pre:bif}

In this section we analyze bifurcation behavior for two-spike equilibria of
\eqref{dae:reduce} and study their stability properties in terms of
equilibrium points of the DAE system \eqref{dae:reduce}.
From \eqref{dae:reduce}, the equilibria satisfy the nonlinear
algebraic system $\Fb(x_1,x_2,H_1,H_2)={\bf 0}$ for
$\Fb\in \R^4$, given component-wise by
\begin{equation}\label{bif:F}
  \begin{split}
    {\mathcal F}_1 &\equiv -\frac{5}{2} \frac{\mu^{\prime}(x_1)}
      {\mu(x_1)} - e_1 + d_1 \frac{H_{2}}{H_{1}} \,, \qquad
    {\mathcal F}_2 \equiv -\frac{5}{2} \frac{\mu^{\prime}(x_2)}
    {\mu(x_2)} - e_2 - d_1 \frac{H_{1}}{H_{2}} \,,\\
    {\mathcal F}_3 &= 6 \left[\mu(x_1)\right]^{3/2} H_{1}^2 - c_1 H_{1} -
    d_1 H_{2}\,, \qquad
    {\mathcal F}_4 = 6 \left[\mu(x_2)\right]^{3/2} H_{2}^2 - d_1 H_{1} -
    c_2 H_{2}\,.
  \end{split}
\end{equation}
The linear stability properties of an equilibrium state
$(r_{+},r_{-},H_{+},H_{-})$ of the DAE dynamics \eqref{dae:reduce} is
based on the eigenvalues $\omega$ of the matrix eigenvalue problem
\begin{equation}\label{bif:jac}
  J \vb = \omega {\mathcal D} \vb \,,
\end{equation}
where $J\equiv D\Fb$ is the Jacobian of $\Fb$ and ${\mathcal D}$ is
the rank-defective diagonal matrix with matrix entries
$({\mathcal D})_{11}=1$, $({\mathcal D})_{22}=1$,
$({\mathcal D})_{33}=0$, and $({\mathcal D})_{44}=0$.  Since
$\mbox{rank}({\mathcal D})=2$, \eqref{bif:jac} has two infinite
eigenvalues. The signs of the real parts of the remaining two matrix
eigenvalues classify the linear stability of the equilibrium point for
\eqref{dae:reduce}. We will refer to these eigenvalues as the ``small
eigenvalues'' for spike stability in accordance with the term used in
\cite{iww} in the absence of a precursor field.

We now outline a simple approach for computing branches of solutions
to $\Fb={\bf 0}$ in terms of a parameter in the precursor
field $\mu(x)$. An alternative formulation is given in
\S \ref{pre:two_asymm} below. For the first approach, we introduce the
spike height ratio $s$ by
\begin{equation}
     s \equiv \frac{H_{2}}{H_{1}} \,, \label{bif:s}
\end{equation}
and reduce \eqref{bif:F} to the three-component system
$\Nb(x_1,x_2,s)=0$ with $\Nb\in \R^3$ defined by
\bsub \label{bif:3comp}
\begin{align}
    {\mathcal N}_1 &\equiv -\frac{5}{2} \frac{\mu^{\prime}(x_1)}
      {\mu(x_1)} - e_1 + d_1 s \,, \qquad
    {\mathcal N}_2 \equiv -\frac{5}{2} \frac{\mu^{\prime}(x_2)}
    {\mu(x_2)} - e_2 - \frac{d_1}{s} \,, \label{bif:N12}\\
    {\mathcal N}_3 &= s^2 \left[\mu(x_2)\right]^{3/2} (c_1+d_1 s) -
    \left[\mu(x_1)\right]^{3/2} (d_1+c_2 s) \,. \label{bif:N3}
 \end{align}
In terms of solutions to ${\mathcal N}_j=0$ for $j=1,\ldots,3$ the
spike heights are
\begin{equation}
  H_{1} = \frac{(c_1 + d_1 s)}{6\left[\mu(x_1)\right]^{3/2}} \,,
  \qquad H_{2}=s H_{1} \,. \label{bif:N_h}
\end{equation}
\esub In \eqref{bif:3comp} and \eqref{bif:F}, the constants $c_1$,
$c_2$, $d_1$, $e_1$, and $e_2$ are defined by (see
\eqref{dae:triang_ent}):
\begin{equation}\label{bif:coeff}
\begin{split}
  c_{1} &= \coth(x_2-x_1) + \tanh(L+x_1) \,, \quad
   c_2 = \coth(x_2-x_1) + \tanh(L-x_2) \,,\\
    e_1 &=\tanh(L+x_1)-\coth(x_2-x_1)\,,
          \quad e_2=\coth(x_2-x_1) - \tanh(L-x_2)\,, \\
    d_1 &= -\csch(x_2-x_1) \,.
\end{split}
\end{equation} 

For the special case where $\mu(x)$ is even, i.e. $\mu(x)=\mu(-x)$, we
label ``symmetric'' spike equilibria as those solutions of
\eqref{bif:3comp} for which $s=1$ and $x_2=-x_1$. For this case,
$c_1=c_2$, $e_2=-e_1$, and ${\mathcal N}_{3}(-x_2,x_2,1)=0$. Moreover,
we calculate that $e_2+d_1=\tanh(x_2) - \tanh(L-x_2)$, and so
\eqref{bif:3comp} reduces to finding a root $x_2$ on $0<x_2<L$ to
the scalar equation ${\mathcal S}(x_2)=0$ given by
\begin{equation}\label{bif:S}
 {\mathcal S}(x_2) \equiv \frac{\mu^{\prime}(x_2)}{\mu(x_2)} -
  \frac{2}{5} \left[\tanh(L-x_2) - \tanh(x_2) \right] \,.
\end{equation}
It readily follows that when $\mu(x)>0$ and $\mu^{\prime}(x)>0$, there
is always a root to ${\mathcal S}=0$ with $0<x_2<{L/2}$.  Our
bifurcation results shown below are for the quadratic precursor field
$\mu(x)=1+b x^2$ with $b\geq 0$, as given in \eqref{intro:prec}. For
this special choice of $\mu$, instead of computing $x_2=x_2(b)$
in \eqref{bif:S} using Newton iterations, we can solve ${\mathcal S}=0$
in \eqref{bif:S} in the explicit form $b=b(x_2)$, where
\begin{equation}\label{bif:symm_b}
  b = \frac{ \left[ \tanh(L-x_2) - \tanh(x_2) \right]}{
    x_2\left( 5 -x_2\left[\tanh(L-x_2) - \tanh(x_2)\right]\right)}
  \,.
\end{equation}
By varying $x_2$ on $0<x_2<{L/2}$ in \eqref{bif:symm_b}, and
keeping only points where $b>0$, we obtain a simple parametric
representation of the symmetric two-spike equilibrium solution branch
with $x_1=-x_2$. The common spike heights are given by
\begin{equation}\label{bif:symm_H}
    H_c\equiv H_{1,2} = \frac{1}{6\left[\mu(x_2)\right]^{3/2}}
     \left[ \tanh(x_2) + \tanh(L-x_2) \right] \,.
\end{equation}
The linear stability with respect to the DAE dynamics
\eqref{dae:reduce} at each value of $b$ on this symmetric solution
branch is obtained from a numerical computation of the matrix spectrum
of the generalized eigenvalue problem \eqref{bif:jac}.

To parameterize asymmetric two-spike equilibria for the special case
$\mu=1+bx^2$, we isolate $b$ from setting
${\mathcal N}_1={\mathcal N}_2=0$ in \eqref{bif:N12}. By equating the
resulting two expressions for $b$, we obtain an equation relating
$x_1$ and $x_2$, in which we treat $s$ as a parameter. The remaining
equation is ${\mathcal N}_3=0$ from \eqref{bif:N3}. In this way, for
$s\neq 1$, we calculate solutions $x_1=x_1(s)$, $x_2=x_2(s)$
to the two-component coupled system
\bsub \label{bif:M}
\begin{equation}\label{bif:M1}
 \begin{split}
  &\left(x_2^2-x_1^2\right)(e_1-d_1 s)\left( e_2
    + \frac{d_1}{s} \right) - 5
  \left[x_2 \left(e_1- d_1 s\right) - x_1\left(e_2 + \frac{d_1}{s}\right)
   \right] =0 \,, \\
  & \qquad s^2 \left[\mu(x_2)\right]^{3/2} (c_1+d_1 s) -
  \left[\mu(x_1)\right]^{3/2} (d_1+c_2 s) =0\,,
  \end{split}
\end{equation}
in which $\mu(x)=1+b x^2$, where $b$ is given by
\begin{equation}\label{bif:Mb}
   b = \frac{d_1 s - e_1}{5 x_1 + x_1^2 (e_1 - d_1 s)} \,.
\end{equation}
\esub The spike heights are then obtained from \eqref{bif:N_h} in
terms of the parameter $s$. This re-formulation of \eqref{bif:3comp}
gives a convenient approach for parameterizing solution branches of
asymmetric two-spike equilibria in terms of the spike height ratio
$s$. For the finite domain case $L<\infty$, the coefficients $c_1$,
$c_2$, $e_1$, $e_2$, and $d_1$, are given in \eqref{bif:coeff}, while
when $L=\infty$, we use $c_1=c_2=2/(1-e^{-2(x_2-x_1)})$ and
$e_1=-e_2=\frac{2}{1-e^{2(x_2-x_1)}}$.  Finally, at each point on
these solution branches the spectrum of the generalized eigenvalue
problem \eqref{bif:jac} is computed to determine the linear stability
of asymmetric spike equilibria to the small eigenvalues.

Although this approach works well for moderate values of $s$, for
either very large or small values of $s$ the nonlinear algebraic
system \eqref{bif:M} is rather poorly conditioned. As a result we need
an alternative approach to compute two-spike equilibria.

\subsection{Two-Spike Equilibria: An Alternative
  Parameterization}\label{pre:two_asymm}

An alternative approach to parameterize symmetric and asymmetric
two-spike equilibrium solution branches for the special case where
$\mu(x)$ is even is described in Appendix \ref{app:alt}. This approach
leads to a nonlinear algebraic system in terms of $r_{+}$, $r_{-}$,
and $\ell$, where $\ell$ is the symmetry point in the interval
$-r_{-}<\ell<r_{+}$ at which $h_x=0$. Here $x_2=r_{+}$ and
$x_{1}=-r_{-}$ are the two steady-state spike locations with spike
heights $H_{\pm}$. As shown in Appendix \ref{app:alt}, with this
formulation we must solve \bsub \label{bif:theo_all}
\begin{equation}\label{bif:theo}
  f(r_{+},\ell)=0 \,, \qquad    f(r_{-},-\ell)=0 \,, \qquad
  \xi(r_{+},\ell)-\xi(r_{-},-\ell)=0 \,,
\end{equation}
for $r_{\pm}$ and $\ell$, where $f(r,\ell)$ and $\xi(r,\ell)$ are 
defined by
\begin{equation}\label{alt:fxi}
    f(r,\ell) = \frac{\mu'(r)}{\mu(r)} + \frac{4}{5}
    \frac{\langle g_x(r,r;\ell) \rangle}{g(r,r;\ell)}\,, \qquad
\xi(r,\ell) = \frac{\mu^{-3/2}(r)}{6}\frac{g(\ell,r;\ell)}{g^2(r,r;\ell)}\,,
\end{equation}
\esub where $\langle g_x(r,r;\ell) \rangle$ indicates the average of
$g_x$ across $x=r$. Here $g(x,r;\ell)$ is the 1-D Green's function,
with Dirac point $r$ and left domain endpoint $\ell$, satisfying
\begin{equation}\label{bif:gfunc}
  g_{xx}-g = -\delta(x-r) \,, \quad \ell < x < L \,; \qquad
  g_x=0 \, \quad \mbox{at} \quad x=\ell\,,\, L \,.
\end{equation}
In the infinite domain case, where $L=\infty$, we calculate that
\begin{equation}\label{alt:g_inf}
  g(r,r;\ell) = \frac{1}{2}\left( 1 + e^{2(\ell-r)} \right)\,,
  \quad g(\ell,r;\ell) = e^{\ell-r}\,, \quad \langle g_x(r,r;\ell) \rangle
  = -\frac{1}{2} e^{2(\ell-r)}\,,
\end{equation}
so that \eqref{alt:fxi} becomes
\begin{equation}\label{alt:fxi_inf}
  f(r,\ell) = \frac{2br}{1+br^2} -\frac{4}{5\left(1 + e^{2(r-\ell)}\right)}\,,
  \quad \xi(r,\ell) =
  \frac{2(1+br^2)^{-3/2}}{3}\frac{e^{\ell-r}}{(1 + e^{2(\ell-r)})^2}
  \,.
\end{equation}
The spike heights for the inhibitor are defined in terms of $r_\pm$ by
\begin{equation}\label{alt:heights_inf}
  H_{\pm} = \frac{\mu^{-3/2}(r_\pm)}{6 g(r_\pm,r_\pm;\pm \ell)} =
  \frac{(1+br_\pm^2)^{-3/2}}{3(1+e^{2\left(\pm \ell-r_\pm \right)})}\,.
\end{equation}
Alternatively, for the finite domain case, we calculate from
\eqref{bif:gfunc} that
\begin{equation}\label{alt:g_fin}
  \begin{split}
    g(r,r;\ell) &= \frac{\cosh(r-\ell)\cosh(r-L)}{\sinh(L-\ell)}\,, \quad
    g(\ell,r;\ell)= \frac{\cosh(r-L)}{\sinh(L-\ell)}\,, \\
    &\qquad\qquad
  \langle g_x(r,r;\ell) \rangle = \frac{\sinh(2r-L-\ell)}{2\sinh(L-\ell)}\,,
   \end{split}
 \end{equation}
so that \eqref{alt:fxi} becomes
\begin{equation}\label{alt:fxi_fin}
  f(r,\ell) = \frac{2br}{1+br^2} + \frac{2\sinh(2r-L-\ell)}{5 \cosh(r-\ell)
    \cosh(r-L)}\,, \quad
\xi(r,\ell) = \frac{(1+br^2)^{-3/2}\sinh(L-\ell)}{6\cosh^2(r-\ell)\cosh(r-L)}\,.
\end{equation}
For this finite domain case, the spike heights are given by
\begin{equation}
  H_\pm = -\frac{(1+br_\pm^2)^{-3/2}\sinh(\pm \ell - L)}{6\cosh(\pm \ell - r_\pm)
    \cosh(r_\pm - L)}\,.
\end{equation}

\begin{figure}[htbp]
\begin{center}
  \includegraphics[width=0.49\linewidth,height=4.5cm]{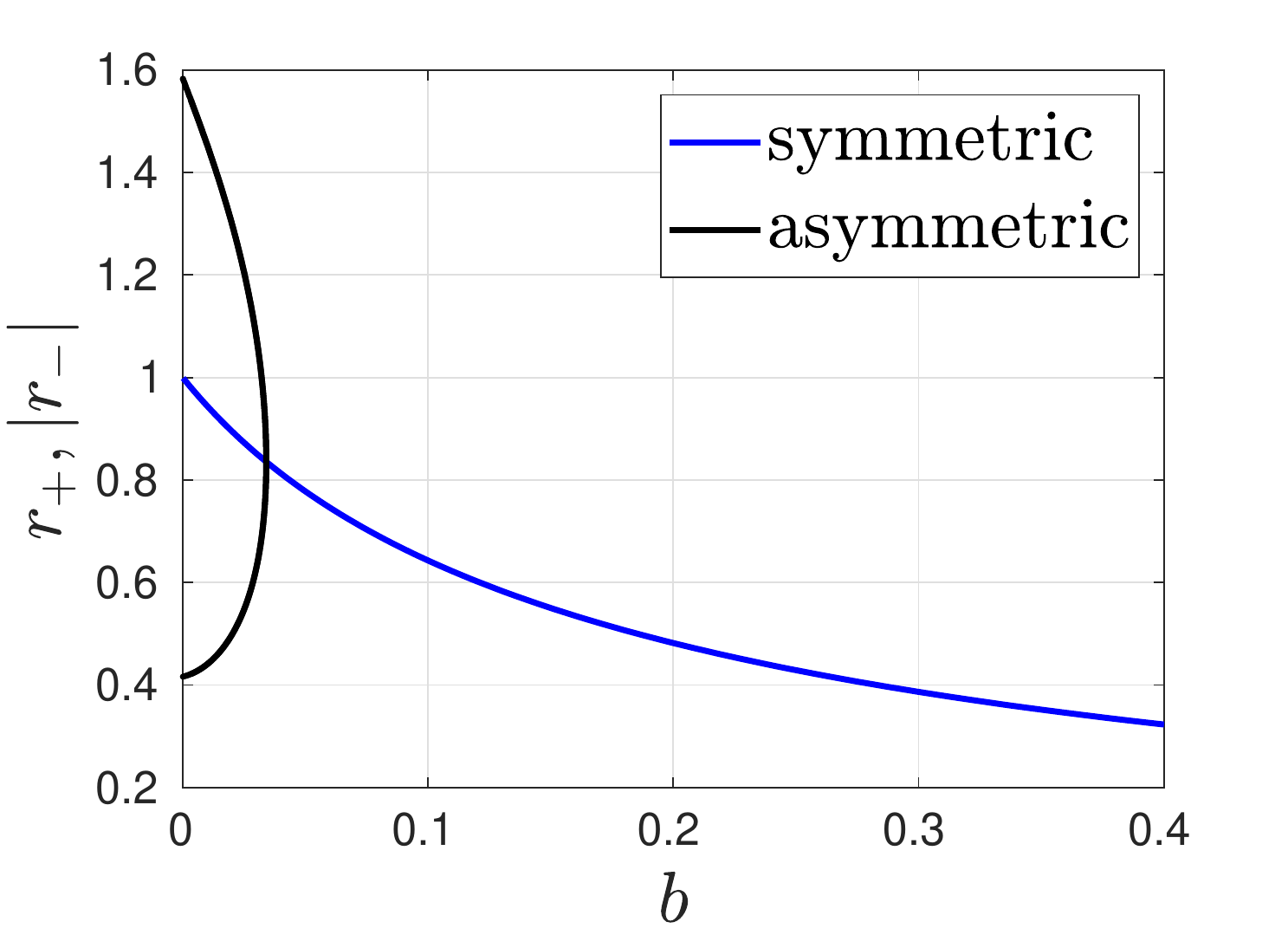}
  \includegraphics[width=0.49\linewidth,height=4.5cm]{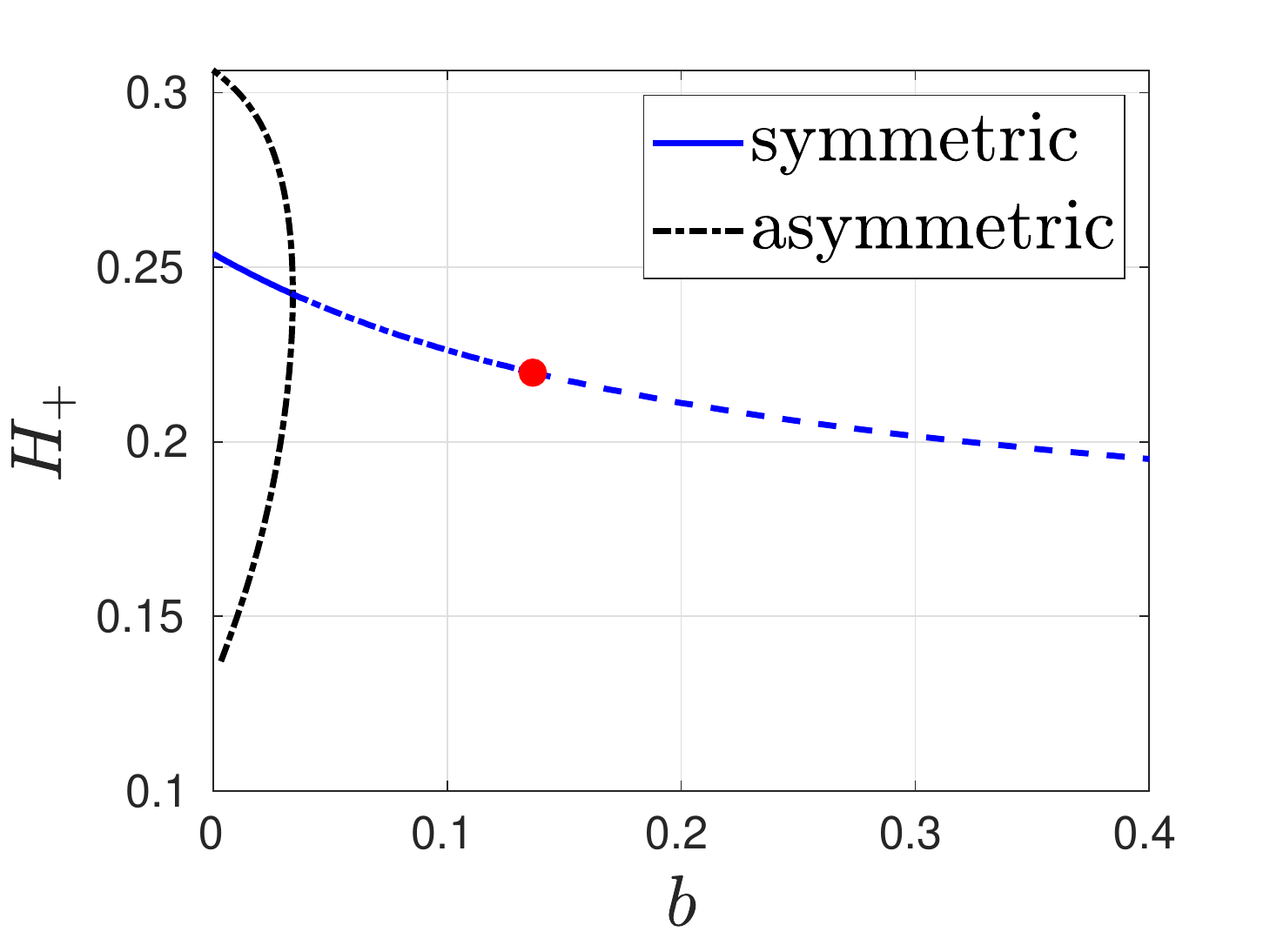}
  \caption{Left: steady-state spike locations $r_{+}$ and $-r_{-}$ for
    $L=2$ versus $b$ in \eqref{intro:prec}. Right: height $H_+$ of the
    rightmost spike versus $b$. Solid lines: linearly stable to both
    the small eigenvalues and the large (NLEP) eigenvalues when
    $\tau\ll 1$. Dash-dotted lines: unstable for the small eigenvalues
    but stable for the large eigenvalues when $\tau\ll 1$. Dashed
    line: stable to the small eigenvalues but unstable to the large
    eigenvalues when $\tau\ll 1$. Red dot: zero-eigenvalue crossing of
    the NLEP on the symmetric branch. Bifurcation from symmetric to
    asymmetric equilibria is subcritical.}\label{fig:L_2}
\end{center}
\end{figure}

\begin{figure}[htbp]
\begin{center}
  \includegraphics[width=0.49\linewidth,height=4.5cm]{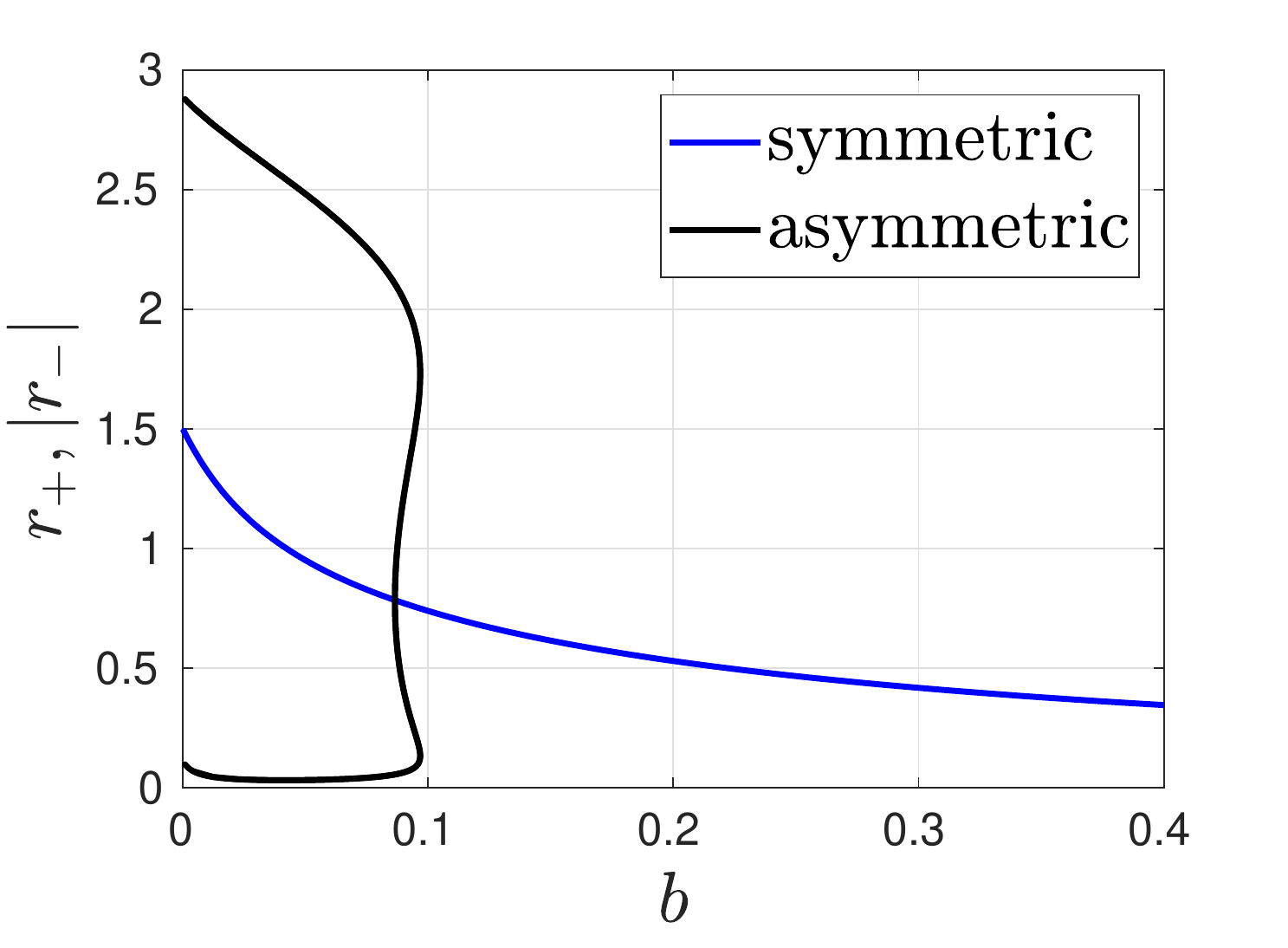}
  \includegraphics[width=0.49\linewidth,height=4.5cm]{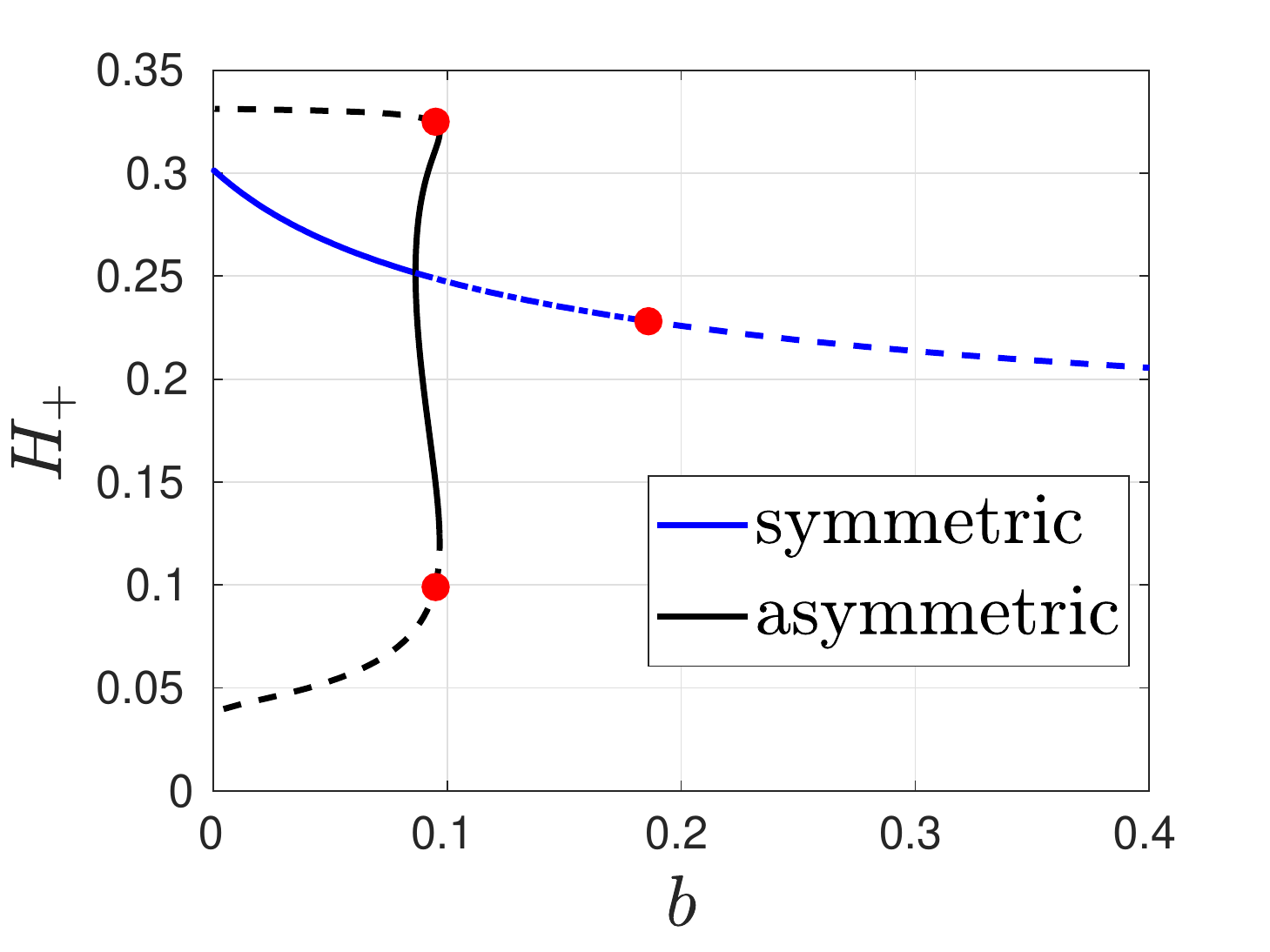}
  \caption{Similar caption as in Figs.~\ref{fig:L_5} and
    \ref{fig:L_2}. Left: steady-state spike locations $r_{+}$ and
    $-r_{-}$ for $L=3$ versus $b$. The pitchfork bifurcation is now
    supercritical. Right: height $H_+$ of the rightmost spike versus
    $b$. Solid lines: linearly stable to both the small eigenvalues
    and the large (NLEP) eigenvalues when $\tau\ll 1$. Dash-dotted
    lines: unstable for the small eigenvalues but stable for the large
    eigenvalues when $\tau\ll 1$.  Dashed line: stable to the small
    eigenvalues but unstable to the large eigenvalues when
    $\tau\ll 1$. There are only very small (nearly indistinguishable)
    zones along the asymmetric branches that are unstable to the small
    eigenvalues.  Red dots are where the NLEP has a zero-eigenvalue
    crossing.}\label{fig:L_3}
\end{center}
\end{figure}

\begin{figure}[htbp]
\begin{center}
  \includegraphics[width=0.49\linewidth,height=4.5cm]{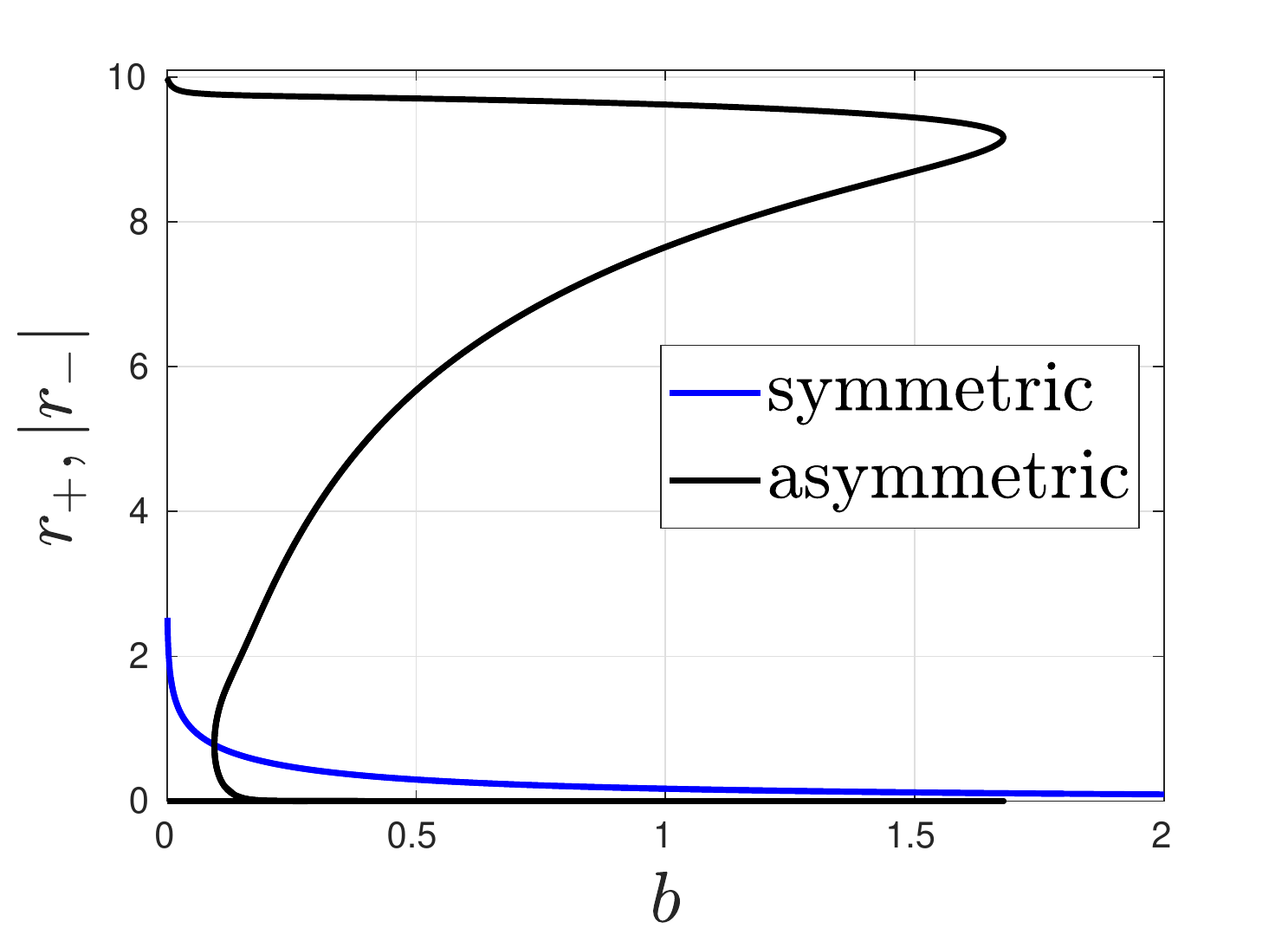}
  \includegraphics[width=0.49\linewidth,height=4.5cm]{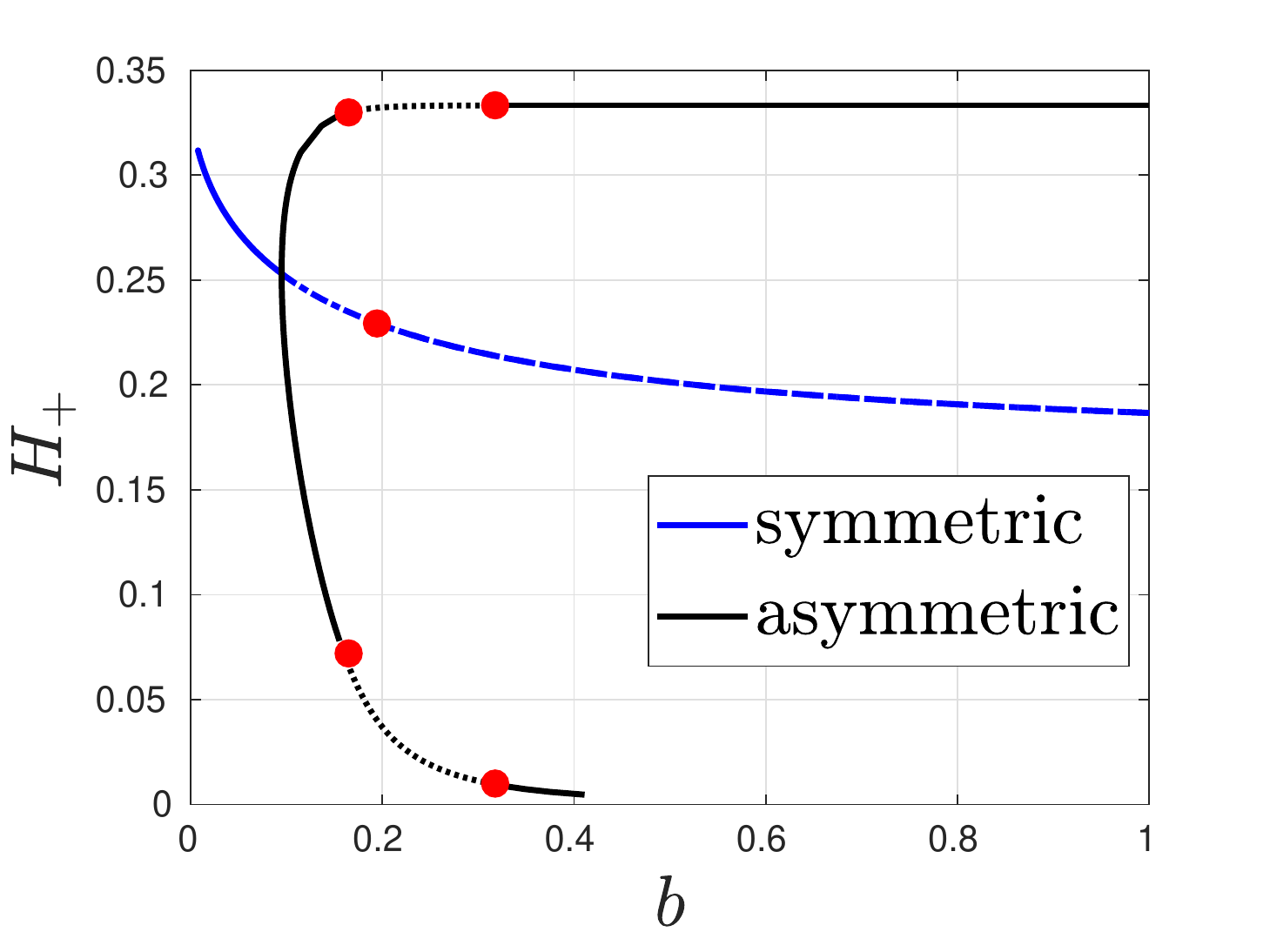}
  \caption{Left: steady-state spike locations $r_{+}$ and $-r_{-}$ for
    $L=10$ versus $b$. Right: height $H_+$ of the rightmost spike
    versus $b$. Solid lines: linearly stable to both the small
    eigenvalues and the large (NLEP) eigenvalues when $\tau\ll
    1$. Dash-dotted lines: unstable for the small eigenvalues but
    stable for the large eigenvalues when $\tau\ll 1$. Dashed line:
    stable to the small eigenvalues but unstable to the large
    eigenvalues when $\tau\ll 1$. Dotted line: unstable to both the
    small and large eigenvalues when $\tau\ll 1$. Red dots are where the
    NLEP has a zero-eigenvalue crossing. In the right panel we have
    not shown the hairpin turn that occurs when $b\approx 1.67$ that
    provides the connection between an interior spike and a boundary
    spike solution.}\label{fig:L_10}
\end{center}
\end{figure}

\begin{figure}[htbp]
\begin{center}
  \includegraphics[width=0.49\linewidth,height=4.5cm]{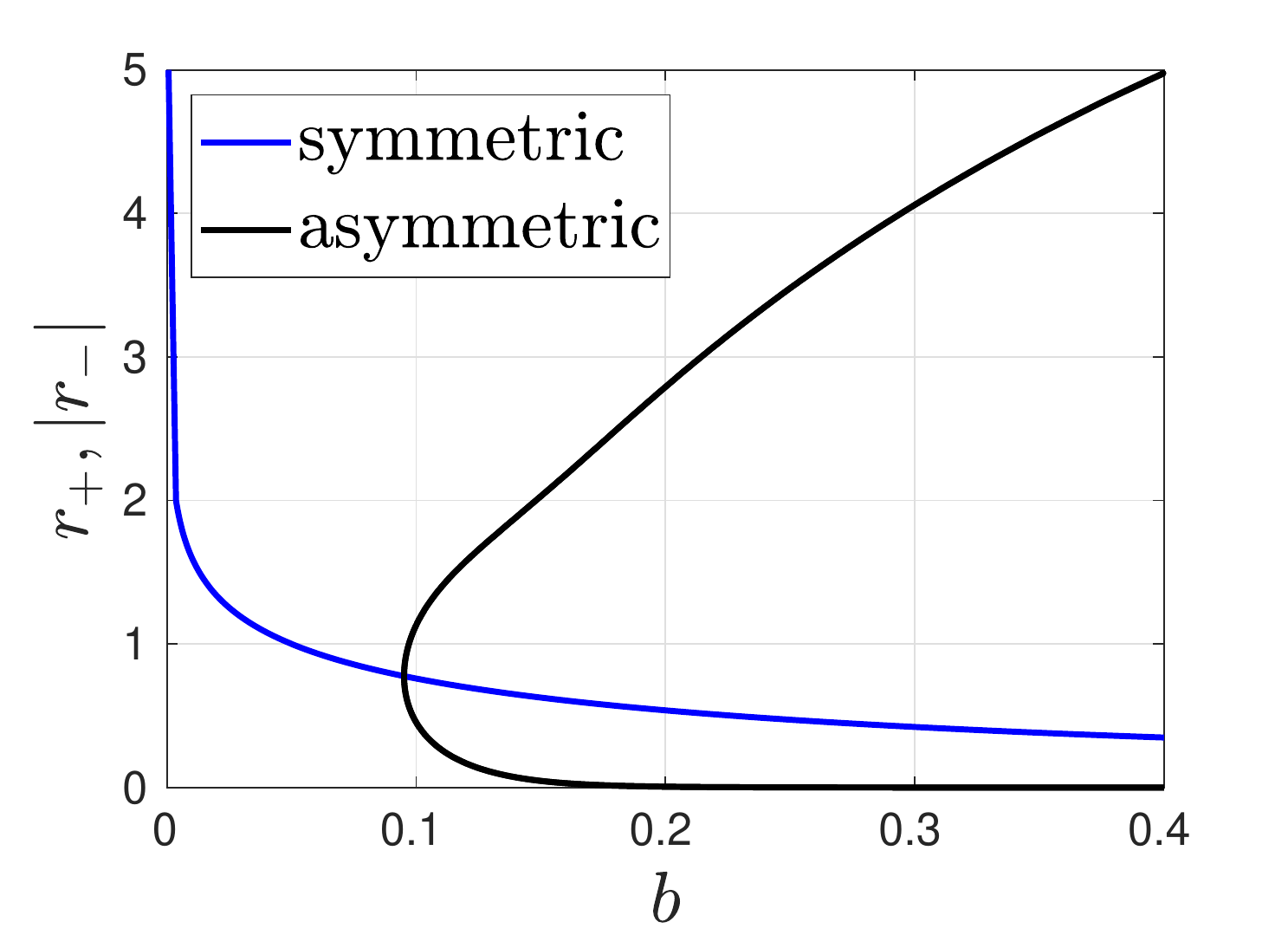}
  \includegraphics[width=0.49\linewidth,height=4.5cm]{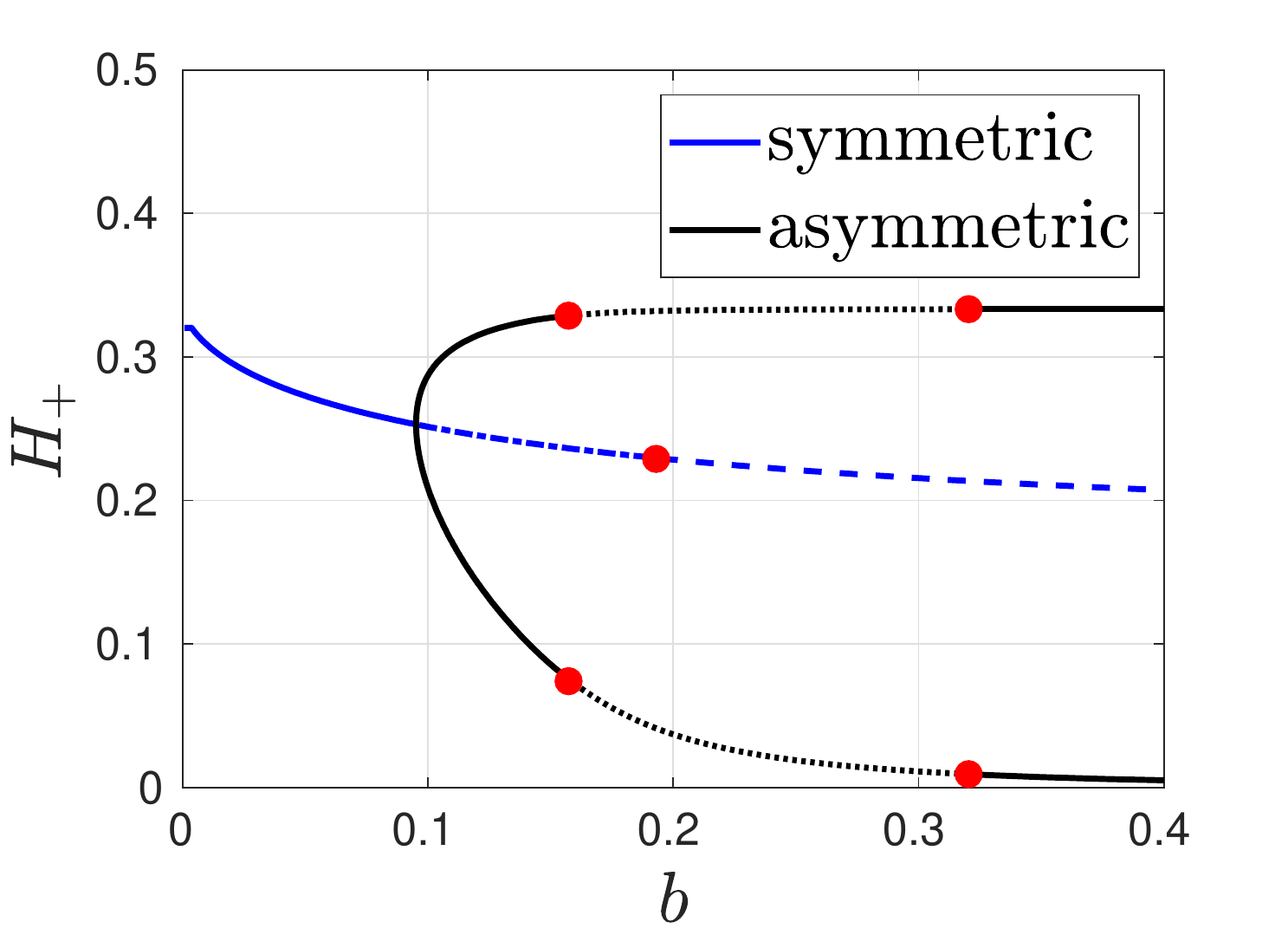}
  \caption{Left: steady-state spike locations $r_{+}$ and $-r_{-}$ for
    $L=\infty$ versus $b$. Right: height $H_+$ of the rightmost spike
    versus $b$. Solid lines: linearly stable to both the small
    eigenvalues and the large (NLEP) eigenvalues when $\tau\ll
    1$. Dash-dotted lines: unstable for the small eigenvalues but
    stable for the large eigenvalues when $\tau\ll 1$. Dashed line:
    stable to the small eigenvalues but unstable to the large
    eigenvalues when $\tau\ll 1$. Dotted line: unstable to both the
    small and large eigenvalues when $\tau\ll 1$. Red dots are where
    the NLEP has a zero-eigenvalue crossing. Observe that there is an
    intermediate range of $b$ along the asymmetric branches where the
    pattern is unstable to both the small and large eigenvalues. The
    asymmetric patterns re-stabilize for larger $b$ and results in a
    spike of large amplitude and another of negligible
    amplitude.}\label{fig:L_inf}
\end{center}
\end{figure}

\begin{figure}[htbp]
\begin{center}
\includegraphics[width=0.52\linewidth,height=4.8cm]{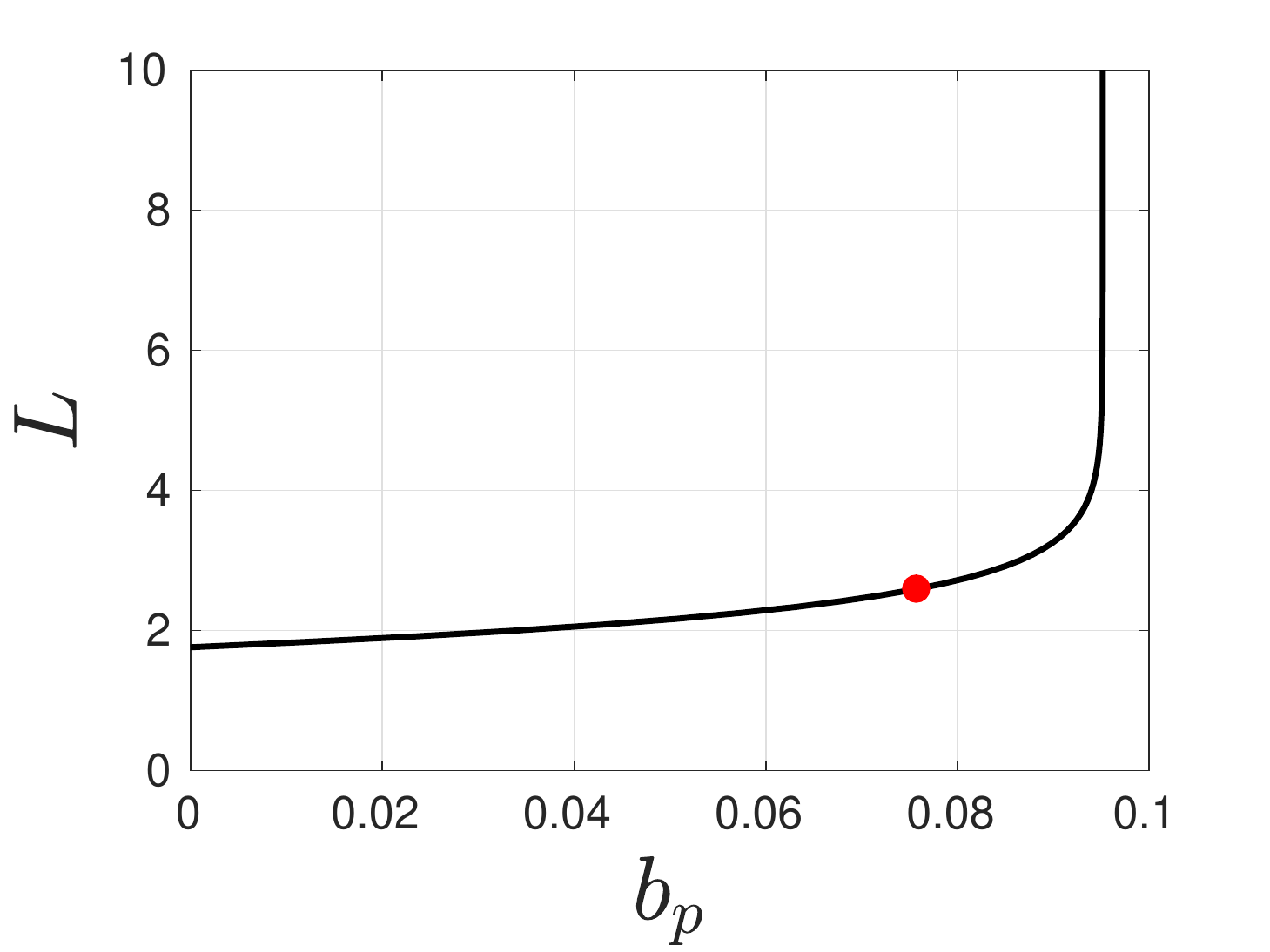}
\caption{Symmetry-breaking bifurcation point $b_p$ versus $L$ where the
  asymmetric branches of two-spike equilibria bifurcate from the
  symmetric branch. The red dot indicates the critical values
  $b_c \approx 0.0760,\, L_c \approx 2.597,\, r_{\pm,c} \approx 0.793$
  where this bifurcation switches between subcritical and
  supercritical. The bifurcation curve has a vertical asymptote
  $b \approx 0.095$ as $L \to \infty$.}\label{fig:b_L}
\end{center}
\end{figure}

To compute branches of two-spike equilibria as either $b$ or $L$ is varied,
we write \eqref{bif:theo_all} for $r_{\pm}$ and $\ell$ in the form
$\bm{F}({\bf u},\zeta)=0$, where
\begin{equation}\label{alt:sys}
  \bm{F}({\bf u},\zeta) \equiv \begin{pmatrix} f(r_+,l) \\ f(r_-,-l) \\
    \xi(r_+,l) - \xi(r_-,-l)  \end{pmatrix}\,, \quad
  \text{with} \quad {\bf u} \equiv (r_+, r_-, l)^T\,, \quad
  \zeta \equiv (b, L)^T\,.
\end{equation}

Families of solutions and branch points (corresponding to
symmetry-breaking pitchfork bifurcations) of this nonlinear system
were computed using the two software packages AUTO (cf.~\cite{auto})
and \textsc{coco} (cf.~\cite{coco}), thereby validating the diagrams
provided in Figs. \ref{fig:L_5}, \ref{fig:L_2}, \ref{fig:L_3},
\ref{fig:L_10}, \ref{fig:L_inf} and \ref{fig:b_L}. In Appendix
\ref{app:alt} we give explicit formulas for the Jacobian of $\bm{F}$
with respect to ${\bf u}$ and the parameter vector $\zeta$, since
providing analytical Jacobians significantly improves the performance
and accuracy of continuation routines as opposed to using numerical
Jacobians based on centered differences.

\subsection{Numerical Bifurcation Results for Two-Spike Equilibria}\label{pre:two_numer}

For $L=2$, in the left panel of Fig.~\ref{fig:L_2} we plot the
numerically computed steady-state spike locations versus the precursor
parameter $b$. In the right panel of Fig.~\ref{fig:L_2}, we plot the
corresponding height $H_{+}$ of the rightmost steady-state spike
versus $b$. In addition, in our plot of $H_+$ versus $b$ we indicate
by various line shadings the linear stability properties of the
steady-state solutions. We first observe that asymmetric two-spike
equilibria emerge from a subcritical symmetry-breaking bifurcation
from the branch of symmetric two-spike equilibria at the critical
value $b\approx 0.034$. However, the asymmetric solution branches are
all unstable with regards to the small eigenvalues, as indicated by
the dash-dotted black curves in the right panel of
Fig.~\ref{fig:L_2}. Below in the left panel of Fig.~\ref{fig:nlep_eig}
we show from a numerical computation of a vector-valued NLEP that
these asymmetric branches are all stable on an ${\mathcal O}(1)$
time-scale when $\tau$ is sufficiently small. These linear stability
properties are qualitatively similar to that for two-spike equilibria
of the GM model with no precursor field (cf.~\cite{ww_asy}).

In the left and right panels of Fig.~\ref{fig:L_3} and
Fig.~\ref{fig:L_5} we plot similar global bifurcation results for
two-spike equilibria when $L=3$ and $L=5$, respectively.  For these
values of $L$, we observe that the symmetry-breaking bifurcation is
now supercritical and that a large portion of the bifurcating
asymmetric two-spike branch of equilibria is linearly stable with
regards to the small eigenvalues. Moreover, as shown below in the
middle and right panels of Fig.~\ref{fig:nlep_eig}, these asymmetric
solution branches are all linearly stable for $\tau$ sufficiently
small with regards to the large eigenvalues for the range of values of
$H_+$ between the two red dots shown in the right panel of
Fig.~\ref{fig:L_3} for $L=3$ and of Fig.~\ref{fig:L_5} for
$L=5$. Overall, this establishes a parameter regime where linearly
stable asymmetric two-spike equilibria occur. For $L=3$, this
theoretical prediction of stable asymmetric two-spike equilibria is
confirmed below in Fig.~\ref{fig:pde_run3} of \S \ref{pre:numerics}
from full PDE simulations of \eqref{gm:full}. For $L=5$, a similar
validation of the linear stability theory through full PDE simulations
was given in Fig.~\ref{fig:pde_run1} of \S \ref{sec:intro}.

In Fig.~\ref{fig:L_10} we plot global bifurcation results for
two-spike equilibria when $L=10$. The right panel of
Fig.~\ref{fig:L_10} shows a parameter regime where stable asymmetric
two-spike equilibria can occur when $\tau\ll 1$. However, in contrast
to the global bifurcation diagrams when $L=2,3,5$, we observe that
when $L=10$ there are two zero-crossings for the NLEP on each
asymmetric solution branch, with the pattern being unstable to both
the small and large eigenvalues for some intermediate range of
$b$. This linear stability behavior with respect to the large
eigenvalues is confirmed below in the left panel of
Fig.~\ref{fig:nlep_eig_10inf} through numerical computations of the
spectrum of a vector-valued NLEP. Moreover, we observe from
Fig.~\ref{fig:L_10} that asymmetric patterns originating from a
symmetry-breaking bifurcation of symmetric two-spike equilibria are
path-connected through a saddle-node point of high curvature to an
unstable two-spike steady-state consisting of a boundary spike of
large amplitude and an interior spike of small amplitude.

Similar results are shown in Fig.~\ref{fig:L_inf} for the infinite
line problem where $L=\infty$. For this case, stable asymmetric
patterns occur near the symmetry-breaking bifurcation point. Moreover,
as for the case where $L=10$, along the asymmetric solution branch
there is an intermediate range of $b$ where the pattern is unstable to
both the small and large eigenvalues. This instability range of $b$
for the large eigenvalues is observed in Fig.~\ref{fig:nlep_eig_10inf}
below from our computations of the spectra of the vector-valued
NLEP. However, when $L=\infty$, there is no boundary spike solution
and, as observed in Fig.~\ref{fig:L_inf}, the asymmetric solution
branch no longer terminates at a finite value of $b$.

\subsection{Computation of a Degenerate Bifurcation Point}
\label{sec:fppp}

From the global bifurcation diagrams in Fig.~\ref{fig:L_2} and
Fig.~\ref{fig:L_3} we observe that the symmetry-breaking bifurcation
switches from subcritical to supercritical on the range $2<L<3$. We
now describe a procedure to accurately compute the critical precursor
parameter $b=b_c$ and critical domain half-length $L=L_c$ where this
switch occurs. The significance of these critical values is that for
$L>L_c$ the asymmetric solution branch is linearly stable
with regards to the small eigenvalues near the bifurcation point.

To formulate our procedure for computing these critical values we first
define
\begin{equation}
W(\ell) \equiv \xi (r_{+}(\ell),\ell)-\xi (r_{-}(\ell),-\ell) \,,  \label{458}
\end{equation}%
where $r_{\pm }=r_{\pm }(\ell)$ satisfy
\begin{equation*}
f(r_{\pm },\pm \ell)=0 \,.
\end{equation*}
Here $\xi(r,\ell)$ and $f(r,\ell)$ are defined in \eqref{alt:fxi}.
The asymmetric branch corresponds to a non-zero root of $W(\ell)$ and the
symmetry-breaking bifurcation occurs when $W^{\prime }(0)=0$. To
compute this point, denote $r=r_{\pm }(0)$, that is, the location of a
symmetric spike which satisfies $f(r,0)=0.$ Upon differentiating
\eqref{458} implicitly and evaluating at $\ell=0$ we obtain that
$r_{-}^{\prime }(0)=-r_{+}^{\prime }(0)=-r^{\prime}$, so that the
bifurcation occurs when the following system is satisfied:
\begin{equation}
  \ell=0\,,\quad f=0\,;\quad r^{\prime }=-\frac{f_{\ell}}{f_{r}}\,; \quad
  \xi_{r}r^{\prime }+\xi_{\ell}=0\,.  \label{1212}
\end{equation}%
In the left panel of Fig.~\ref{fig:maple} of Appendix \ref{app:alt} we
include the Maple code that computes this bifurcation point. For
example, when $L=2$ we obtain from solving \eqref{1212} that
$b=0.03406$ and $r=0.835585$.

Since $W(\ell)$ is an odd function we have for small $\ell$ that
\begin{equation*}
W(\ell)\sim \ell W^{\prime }(0)+\ell^{3}\frac{W^{\prime \prime \prime }(0)}{6}%
+O(\ell^{5})\,,
\end{equation*}%
with all even derivatives of $W$ being zero. The criticality of the
bifurcation depends on the sign of $W^{\prime \prime \prime }(0)$. A
positive sign corresponds to a supercritical bifurcation,\ whereas a
negative sign corresponds to a subcritical bifurcation. The change of
bifurcation occurs when
$W^{\prime \prime \prime }(0)=W^{\prime }(0)=0$. To compute
$W^{\prime \prime \prime }(0)$, we differentiate implicitly and set
$\ell=0$. We readily calculate that
\begin{gather*}
W^{\prime }(0) =\xi_{r}r^{\prime }+\xi_{\ell} \,, \qquad
  W^{\prime \prime }(0) =\xi_{rr}r^{\prime 2}+2\xi_{r\ell}r^{\prime }
      +\xi_{r}r^{\prime \prime }+\xi _{\ell\ell}\,, \\
W^{\prime \prime \prime }(0)=\xi _{rrr}r^{\prime 3}+3\xi _{rr\ell}r^{\prime
  2}+3\xi _{r\ell\ell}r^{\prime }+3\xi _{rr}r^{\prime }r^{\prime \prime } +
3\xi_{r\ell}r^{\prime \prime }+\xi _{r}r^{\prime \prime \prime } +
\xi_{\ell\ell\ell}\,.
\end{gather*}
The values of $r$, $r^{\prime }$ and $r^{\prime \prime }$ are obtained
by differentiating $f$ implicitly. This yields
\begin{gather*}
r^{\prime } =-\frac{f_{\ell}}{f_{r}} \,, \qquad
r^{\prime \prime } =-\frac{f_{rr}r^{\prime 2}+2f_{r\ell}r^{\prime }+
  f_{\ell\ell}}{f_r} \,, \\
r^{\prime \prime \prime } =-\frac{f_{rrr}r^{\prime 3}+3f_{rr\ell}r^{\prime
    2}+3f_{r\ell\ell}r^{\prime }+3f_{rr}r^{\prime }r^{\prime \prime }+
  3f_{r\ell}r^{\prime\prime }+f_{\ell\ell\ell}}{f_{r}}\,,
\end{gather*}
which are then evaluated at $\ell=0$. In this way, the set of
equations
\begin{equation}
l=0\,,\quad f=0\,;\quad W^{\prime }(0)=0\,,\quad W^{\prime \prime \prime }(0)=0\,,
\label{1214}
\end{equation}%
must be solved numerically to obtain the higher-order bifurcation
point.  The right panel of Fig.~\ref{fig:maple} of Appendix
\ref{app:alt} shows the Maple implementation. Although the system
\eqref{1214} is very large (its length is about 20,000 bytes in
Maple), its numerical solution is found instantaneously, yielding
\begin{equation}
L=L_c\equiv 2.5972\, \quad b=b_c\equiv 0.07596\,, \quad r=r_c\equiv .792655\,.
\end{equation}%
We conclude that the symmetry-breaking bifurcation is
supercritical when $L>2.5972$ and is subcritical when
$L<2.5972$.

\section{NLEP Stability Analysis}\label{pre:comp}

We now examine the stability on an ${\mathcal O}(1)$ time-scale of
steady-state spike equilibria of \eqref{gm:full}, labeled by $a_e$ and
$h_e$. We will derive a new vector-valued nonlocal eigenvalue problem
governing instabilities of the spike amplitudes on an
${\mathcal O}(1)$ time-scale. From this vector-NLEP, we will
analyze in detail the linear stability of the two-spike equilibria
constructed in \S \ref{pre:bif} to these ``large eigenvalues'' for the
choice $\mu=1+bx^2$.

To formulate the linear stability problem, we first introduce the
perturbation
\begin{equation}
   a(x,t) = a_{e} + e^{\lam t} \phi(x) \,, \qquad
   h(x,t) = h_{e} + e^{\lam t} \psi(x) \,, \label{stab:intro}
\end{equation}
into (\ref{gm:full}) and linearize. This leads to the singularly
perturbed eigenvalue problem
\bsub \label{stab:eig}
\begin{gather}
       \eps^{2} \phi_{xx} - \mu(x) \phi + \frac{2 a_e}{h_e} \phi
   - \frac{a_{e}^{2}}{h_{e}^{2}} \psi = \lam \phi \,, \quad
     |x|\leq L \,; \qquad \phi_x(\pm L)=0\,, \label{stab:eig_a} \\
 \psi_{xx} - (1+\tau \lam) \psi = -\frac{2}{\eps} a_e \phi\,,
  \qquad |x|\leq L \,; \qquad \psi_{x}(\pm L) = 0 \,. \label{stab:eig_b} 
\end{gather}
\esub 

In the inner region near a spike at $x=x_j$, we have from
\eqref{dae:a0_sol} that
\begin{equation*}
  a_e \sim \mu_j H_{j} w\left(\sqrt{\mu_j} y_j\right) \, \quad
  h_e\sim H_j\,, \quad \mbox{where} \quad y_j=\eps^{-1}(x-x_j) \,,
\end{equation*}
$\mu_j\equiv \mu(x_j)$, and $w(z)=\frac{3}{2}\sech^2({z/2})$. Here
$H_j$ is the spike height obtained from the steady-state of
\eqref{dae:reduce}. Next, we introduce the localized eigenfunction
\begin{equation}\label{stab:phi}
  \Phi_j(y_j) = \phi(x_j+\eps y_j) \,,
\end{equation}
and obtain from \eqref{stab:eig_a} that on $-\infty<y_j<\infty$, and
for each $j=1,\ldots,N$, 
\begin{equation}\label{stab:int}
  \frac{d^2 \Phi_j}{d y_j^2} - \mu_j \Phi_j + 2\mu_j
  w\left(\sqrt{\mu_j} y_j\right) \Phi_j - \mu_j^2 \left[
    w\left(\sqrt{\mu_j} y_j\right)\right]^2 \Psi_j = \lam \Phi_j \,,
\end{equation}
where $\Psi_j$ is a constant to be determined.  Then, we let
$z\equiv \sqrt{\mu_j} y$, and define
$\hat{\Phi}_j(z)\equiv \Phi_j\left({z/\sqrt{\mu_j}}\right)$, so that
\eqref{stab:int} becomes
\begin{equation}\label{stab:nlep_1}
  \frac{d^2 \hat{\Phi}_j}{d z^2} - \hat{\Phi}_j + 2 w(z)
  \hat{\Phi}_j - \mu_j \left[w(z)\right]^2 \Psi_j = \frac{\lam}{\mu_j}
  \hat{\Phi}_j \,, \quad -\infty<z<\infty \,.
\end{equation}
    
To determine $\Psi_j$, we must construct the outer solution for $\psi$
in \eqref{stab:eig_b}. In the sense of distributions we calculate for
$\eps\to 0$ that
\begin{equation}\label{stab:delta}
  \frac{2}{\eps} a_e \phi \rightarrow 2 H_j \sqrt{\mu_j}
  \left(\int_{-\infty}^{\infty} w(z) \hat{\Phi}_j(z) \, dz \right) \,
  \delta(x-x_j)\,.
\end{equation}
In this way, we obtain that the outer solution for $\psi$ in
\eqref{stab:eig_b} satisfies
\bsub \label{stab:out}
\begin{gather}
    \psi_{xx} - \theta_{\lambda}^2 \psi = -2 \sum_{j=1}^{n} H_{j} \sqrt{\mu_j}
    \left(\int_{-\infty}^{\infty} w(z) \hat{\Phi}_j(z) \, dz \right) \,
    \delta(x-x_j) \,, \quad |x|\leq L \,, \label{stab:out_1} \\
    \psi_x(\pm L) =0 \,, \qquad \theta_\lambda \equiv \sqrt{1+\tau \lambda}
     \,. \label{stab:out_2}
\end{gather}
\esub
In \eqref{stab:out_2} we must choose the principal branch of $\theta_\lambda$.
The constants $\Psi_j$ for $j=1,\ldots,N$ are obtained from the
matching condition that $\Psi_j=\psi(x_j)$ for $j=1,\ldots,N$.

By solving \eqref{stab:out} on each subinterval we readily derive a
linear algebraic system for
${\bf \Psi}\equiv \left(\Psi_1,\ldots,\Psi_N\right)^T$ in the form
\begin{equation}\label{stab:psi_vec}
  {\mathcal B}_{\lam} {\bf \Psi} = \frac{2}{\sqrt{1+\tau \lam}}\,
  {\mathcal U}^{1/2} {\mathcal H} \left( \int_{-\infty}^{\infty} w
   {\bf \Psi} \, dz \right) \,,
\end{equation}
where the diagonal matrices ${\mathcal U}$ and ${\mathcal H}$ have
diagonal entries $({\mathcal U})_{jj}=\mu(x_j)$ and
$({\mathcal H})_{jj}=H_{j}$ for $j=1,\ldots,N$. In \eqref{stab:psi_vec},
${\mathcal B}_{\lambda}$ is defined by
\bsub\label{stab:blam}
\begin{equation}\label{stab:bmatrix}
 {\mathcal B}_{\lam} = \left( \begin{array}{cccc}
                 c_{1\lam} & d_{1\lam} & & 0\\
                 d_{1\lam} & \ddots & \ddots & \\
                   & \ddots & \ddots & d_{N-1\lam} \\
          0 & & d_{N-1\lam} & c_{N\lam} \end{array} \right)\,, \quad
\end{equation}
where the matrix entries are given by
\begin{equation} \label{stab:bcoeff}
\begin{split}
  c_{1\lam} &= \coth(\theta_{\lam}(x_2-x_1)) + \tanh(\theta_{\lam}(L+x_1)) \,, \\
   c_{N\lam} &= \coth(\theta_{\lam}(x_N-x_{N-1})) + \tanh(\theta_\lam(L-x_N)) \,,\\
   c_{j\lam} &= \coth(\theta_{\lam}(x_{j+1}-x_j)) + \coth(\theta_{\lam}(x_j-x_{j-1}))
   \,, \quad j=2,\ldots N-1\,, \\
    d_{j\lam} &= -\csch(\theta_{\lam}(x_{j+1}-x_{j})) \,, \quad j=1,\ldots,N-1 \,.
\end{split}
\end{equation}
\esub

Next, upon substituting \eqref{stab:psi_vec} into \eqref{stab:nlep_1},
we obtain the following vector-valued NLEP for
$\hat{{\bf \Phi}}\equiv (\hat{\Phi}_1,\ldots,\hat{\Phi}_N)^T$ on
$-\infty<z<\infty$;
\bsub \label{stab:vnlep}
\begin{gather}
  {\mathcal L} \hat{{\bf \Phi}} - w^2 \frac{ \int_{-\infty}^{\infty} w {\mathcal
      E}_{\lam} \hat{ {\bf \Phi}} \, dz}{\int_{-\infty}^{\infty} w^2\,
    dz} = \lambda {\mathcal U}^{-1} \hat{ \bf \Phi} \,; \qquad
  \hat{{\bf \Phi}} \to {\bf 0} \quad \mbox{as} \,\,\, |z|\to \infty
  \,, \\
  {\mathcal E}_{\lam} \equiv \frac{12}{\sqrt{1+\tau\lambda}} \,
    {\mathcal U} {\mathcal B}_{\lambda}^{-1} {\mathcal U}^{-1} \left(
      {\mathcal U}^{3/2} {\mathcal H} \right) \,, \qquad {\mathcal L} \hat{{\bf
        \Phi}} \equiv \hat{{\bf \Phi}}^{\prime\prime} - \hat{{\bf
        \Phi}} + 2 w \hat{{\bf \Phi}} \,.\label{stab:vnlep_b}
\end{gather}
\esub
We then diagonalize ${\mathcal E}_{\lam}$ by finding the eigenvalues
${\mathcal E}_{\lam} \eb = \chi_{\lam} \eb$ and obtain that
\begin{equation}\label{stab:diag}
  {\mathcal E}_{\lam} = {\mathcal V} \Lambda {\mathcal V}^{-1}\,,
\end{equation}
where ${\mathcal V}$ is the matrix of eigenvectors of
${\mathcal E}_{\lam}$ and $\Lambda$ is the diagonal matrix of
eigenvalues with $(\Lambda)_{jj}=\chi_{\lam,j}$, for
$j=1,\ldots,N$. Then, by defining
$\tilde{{\bf \Phi}}={\mathcal V}^{-1}\hat{{\bf \Phi}}$, we obtain the
following vector-valued NLEP defined on $-\infty<z<\infty$ with
$\tilde{{\bf \Phi}}\to 0$ as $|z|\to \infty$:
\begin{equation}\label{stab:vfnlep}
     {\mathcal L} \tilde{{\bf \Phi}} - w^2 \Lambda \frac{
    \int_{-\infty}^{\infty} w \tilde{ {\bf \Phi}} \,
    dz}{\int_{-\infty}^{\infty} w^2\, dz} = \lambda {\mathcal C}
  \tilde{{\bf \Phi}} \,; \qquad
  {\mathcal C} \equiv {\mathcal V}^{-1} {\mathcal U}^{-1} {\mathcal V}
  \,.
\end{equation}
The key difference between this NLEP analysis and that for the
Gierer-Meinhardt model with no precursor field in \cite{iw} and \cite{iww}
is that the NLEP cannot be diagonalized into $N$ separate scalar NLEPs, one
for each eigenvalue of $\Lambda$. From \eqref{stab:vfnlep} we observe that
the NLEPs are coupled through the matrix ${\mathcal C}$.

We now study \eqref{stab:vfnlep} for our two-spike symmetric and asymmetric
equilibria constructed in \S \ref{pre:bif} for $\mu=1+bx^2$.

\subsection{NLEP Analysis: Symmetric 2-Spike Equilibria}
\label{stab-nlep:symm}

For the symmetric two-spike case with $x_2=-x_1$, we use
${\mathcal U}=\mu(x_2) I$ and ${\mathcal H}=H_c I$, to get from
\eqref{stab:vnlep_b} that
\begin{equation}\label{bif:symm_emat}
  {\mathcal E}_{\lambda} = \frac{12}{\sqrt{1+\tau\lambda}}
    \left[\mu(x_2)\right]^{3/2} H_{c}
  {\mathcal B}_{\lambda}^{-1} \,, \quad \mbox{where} \quad
   \left[\mu(x_2)\right]^{3/2} H_{c} = \tanh(x_2) + \tanh(L-x_2) \,,
\end{equation}
as obtained from \eqref{bif:symm_H}. We readily calculate the matrix
spectrum of ${\mathcal B}_{\lambda}$ as
\begin{equation} \label{bif:blambda}
  \begin{split}
    {\mathcal B}_{\lambda} \vb_1 &= \kappa_{1\lambda} \vb_1 \,; \quad
    \vb_1=(1,1)^T \,, \quad \kappa_{1\lambda} \equiv \tanh(\theta_\lambda x_2)
    + \tanh(\theta_\lambda(L-x_2)) \,,\\
    {\mathcal B}_{\lambda} \vb_2 &= \kappa_{2\lambda} \vb_2 \,; \quad
    \vb_2=(1,-1)^T \,, \quad \kappa_{2\lambda} \equiv \coth(\theta_\lambda x_2)
    + \tanh(\theta_\lambda(L-x_2)) \,.
  \end{split}
\end{equation}
In this way, for symmetric two-spike equilibria, we obtain that
\eqref{stab:vfnlep} is equivalent to the two scalar NLEPs, with
 NLEP {\em multipliers} $\chi_{1,\lambda}$ and $\chi_{2,\lambda}$, defined by
\bsub \label{stab:nlep_symm_all}
\begin{gather}
         {\mathcal L} \tilde{{\bf \Phi}} - w^2 \Lambda \frac{
    \int_{-\infty}^{\infty} w \tilde{ {\bf \Phi}} \,
    dz}{\int_{-\infty}^{\infty} w^2\, dz} = \frac{\lambda}
  {\left[\mu(x_2)\right]^{3/2} } \tilde{{\bf \Phi}} \,, \quad
  -\infty<z<\infty\,; \quad \tilde{{\bf \Phi}}\to 0 \quad \mbox{as}
  \quad |z|\to \infty \,; \label{stab:nlep_symm_1} \\
  (\Lambda)_{11}\equiv \chi_{1,\lambda} = \frac{2}{\sqrt{1+\tau\lambda}}
  \left( \frac{\tanh(x_2)+\tanh(L-x_2)}{\tanh(\theta_\lam x_2)+
  \tanh(\theta_\lam(L-x_2))}\right) \,, \label{stab:nlep_symm_chi1} \\
  (\Lambda)_{22}\equiv \chi_{2,\lambda} = \frac{2}{\sqrt{1+\tau\lambda}}
  \left( \frac{\tanh(x_2)+\tanh(L-x_2)}{\coth(\theta_\lam x_2)+
  \tanh(\theta_\lam(L-x_2))}\right) \,, \label{stab:nlep_symm_chi2}
\end{gather}
\esub
where $\theta_{\lam}=\sqrt{1+\tau\lam}$.

\begin{figure}[htbp]
\begin{center}
  \includegraphics[width=0.49\linewidth,height=4.5cm]{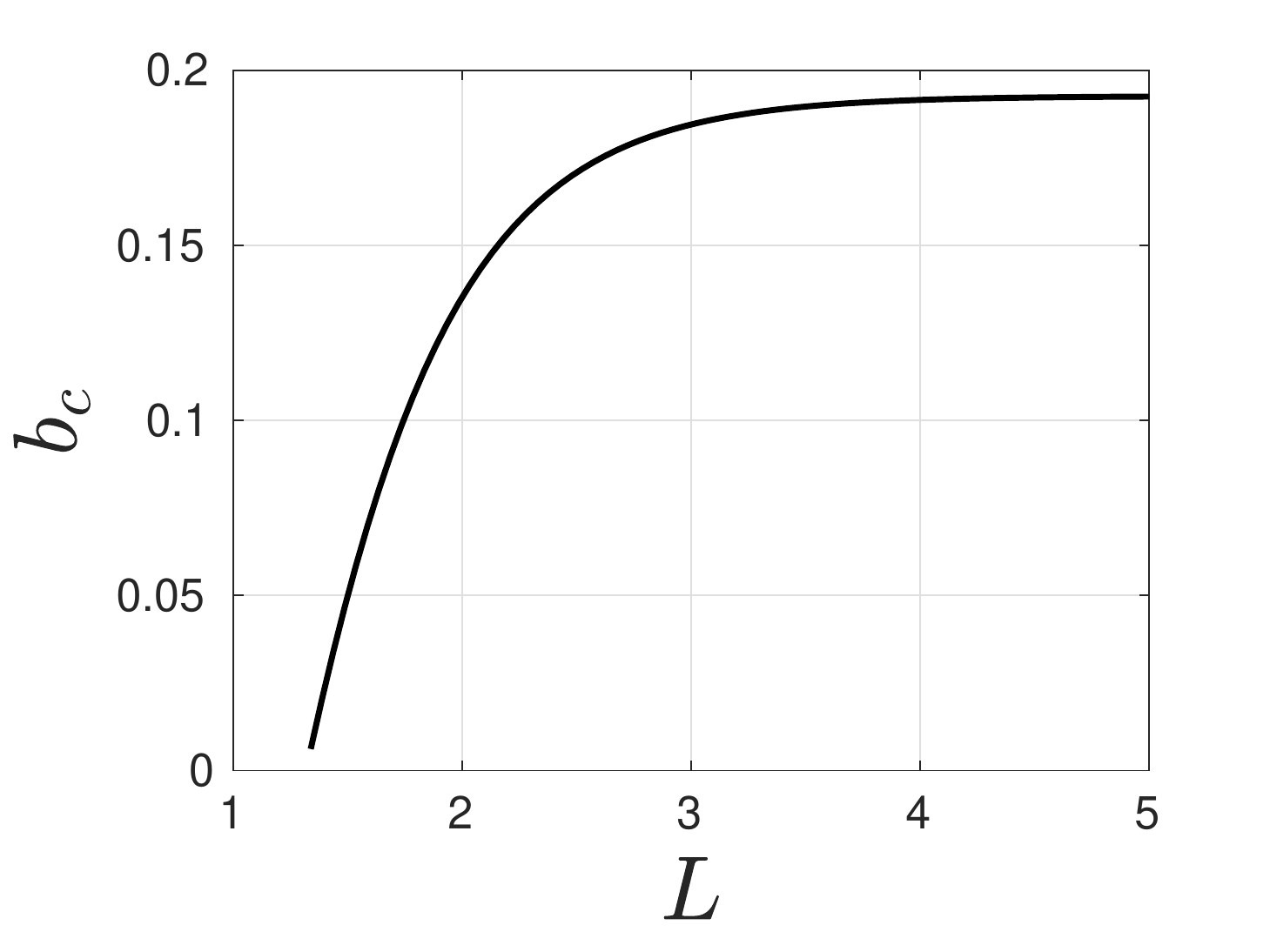}
  \includegraphics[width=0.49\linewidth,height=4.5cm]{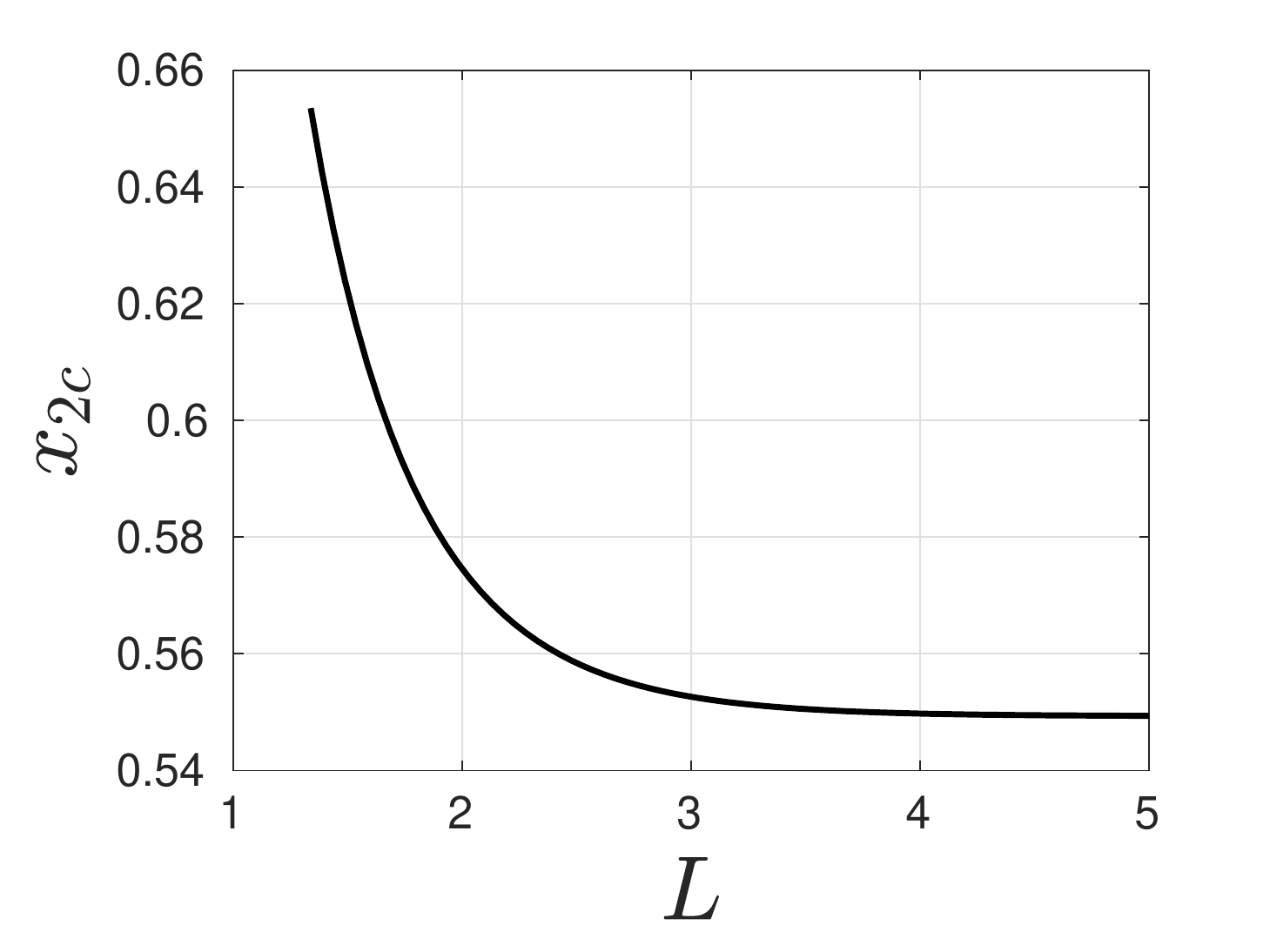}
  \caption{Critical values $b_c$ of the precursor parameter $b$ (left
    panel) and the spike location $x_{2c}$ (right panel) versus $L$
    where the NLEP \eqref{stab:nlep_symm_all} with multiplier
    $\chi_{2,\lambda}$ has a zero-eigenvalue crossing for the
    linearization of a symmetric two-spike steady-state. For
    $x_2<x_{2c}$, or equivalently for $b>b_{c}$, a competition instability
    on an ${\mathcal O}(1)$ time-scale occurs.}\label{fig:comp_tr}
\end{center}
\end{figure}

We first consider the {\em competition mode} corresponding to
$\vb_2=(1,-1)^T$ where the multiplier of the NLEP in
\eqref{stab:nlep_symm_1} is $\chi_{2,\lambda}$, which depends on
$\lambda$ through the product $\tau\lambda$, so that
$\chi_{2,\lambda}=\chi_{2,\lambda}(\tau\lambda)$. From Proposition 3.6
of \cite{swr}, we conclude for this competition mode that there is a
unique eigenvalue in $\mbox{Re}(\lambda)>0$ for any $\tau>0$ when
$\chi_{2,\lambda}(0)<2$. By using \eqref{stab:nlep_symm_chi2}, we calculate
that $\chi_{2,\lambda}(0)<2$ when
\begin{equation*}
    2\tanh(x_2) + 2\tanh(L-x_2)<\coth(x_2)+\tanh(L-x_2) \,,
\end{equation*}
which, after some algebra, reduces to
\begin{equation}\label{comp:x2_thres}
  \coth(x_2)\coth(L)>2 \qquad \Longrightarrow \qquad
  0<x_2< x_{2c}\equiv \frac{1}{2} \log\left( \frac{2+\coth{L}}{2-\coth{L}}
  \right) \,,
\end{equation}
provided that $L>L_c\equiv \log(2+\sqrt{3})\approx 1.3169$. We
conclude that a competition instability occurs whenever spikes become
too close. When $L<L_{c}$, a competition instability occurs for any
$x_{2}>0$.  Equivalently, from \eqref{bif:symm_b}, we conclude that on
the range $L>L_c$ a competition instability occurs along the symmetric
branch of equilibria whenever the precursor parameter $b$ satisfies
$b>b_c$, where
\begin{equation}\label{comp:bc}
  b_c = \frac{ \left[ \tanh(L-x_{2c}) - \tanh(x_{2c}) \right]}{
    x_{2c}\left( 5 -x_{2c}\left[\tanh(L-x_{2c}) - \tanh(x_{2c})\right]\right)}
  \,.
\end{equation}
In Fig.~\ref{fig:comp_tr} we plot $b_c$ and $x_{2c}$ versus $L$ on the
range $L>L_{c}\approx 1.3169$. Numerical values for $b_c$ for
different $L$ correspond to the red dots on the symmetric branches of
equilibria shown in Fig.~\ref{fig:L_5}, and in Figs.~\ref{fig:L_2},
\ref{fig:L_3}, \ref{fig:L_10}, \ref{fig:L_inf}. For $b<b_c$, or
equivalently for $x_2>x_{2c}$, Proposition 3.6 of \cite{swr} can be
used to prove that the two-spike symmetric steady-state is linearly
stable on ${\mathcal O}(1)$ time-scales whenever $\tau$ in
\eqref{gm:full} is below a Hopf bifurcation threshold $\tau_H$. We
refer the reader to \cite{swr} for the proof of this statement.

Next, we briefly consider the NLEP \eqref{stab:nlep_symm_all} for the
synchronous mode $\vb_1=(1,1)^T$, where the NLEP multiplier
$\chi_{1,\lambda}$ is given in \eqref{stab:nlep_symm_chi1}. We
calculate that $\chi_{1,\lambda}(0)=2$, for any $\tau>0$ and $b>0$. As
a result, from Theorem 2.4 of \cite{ww} (see also \cite{wei_surv}) we
conclude that the NLEP for the synchronous mode has no eigenvalues in
$\mbox{Re}(\lambda)>0$ when $\tau=0$, or when $\tau$ is sufficiently
small.  As similar to the analysis in \cite{ww} with no precursor, a
Hopf bifurcation can occur when $\tau$ exceeds a threshold, which now
depends on $b$ and $L$. We do not calculate this Hopf point
numerically here.

We summarize our NLEP stability result for the symmetric two-spike
steady-state branch as follows:

\begin{prop} Consider the two-spike symmetric steady-state solution
  for \eqref{gm:full} with precursor $\mu(x)=1+bx^2$, where the spike
  locations $x_1$ and $x_2$, with $x_2=-x_1$ are given in terms of $b$
  by \eqref{bif:symm_b}. Suppose that
  $L>L_c\equiv \log(2+\sqrt{3})\approx 1.3169$ and define the critical
  half-distance $x_{2c}$ between the spikes and the critical precursor
  parameter $b_c$ by \eqref{comp:x2_thres} and \eqref{comp:bc},
  respectively. Then, for any $b$ with $b>b_{c}$, or equivalently for
  any $x_{2}$ with $x_2<x_{2c}$, the NLEP \eqref{stab:nlep_symm_all}
  with multiplier $\chi_{2,\lambda}$ for the competition mode has a
  unique unstable eigenvalue in $\mbox{Re}(\lambda)>0$.
  Alternatively, if $b<b_{c}$, and for $0\leq \tau<\tau_H$, the
  two-spike symmetric steady-state is linearly stable on
  ${\mathcal O}(1)$ time-scales to the competition mode. Finally, the
  NLEP \eqref{stab:nlep_symm_all} for the synchronous mode, with
  multiplier $\chi_{1,\lambda}$, has no unstable eigenvalues when $\tau>0$
  is sufficiently small.
\end{prop}

\subsection{NLEP Analysis: Asymmetric 2-Spike Equilibria}
\label{stab-nlep:asymm}

We will analyze the NLEP \eqref{stab:vnlep} for two-spike asymmetric
equilibria for the special case where $\tau=0$. To do so, we set
${\mathcal F}_3={\mathcal F}_4=0$ in \eqref{bif:F} to calculate that
\begin{equation}\label{anlep:uhmat}
  {\mathcal U}^{3/2} {\mathcal H} = {\mathcal Z} \,, \qquad \mbox{where}
  \qquad  {\mathcal Z} \equiv \frac{1}{6}
     \left( \begin{array}{cc}
                 c_{1} + d_1 s &  0\\
                 0  & c_2 + {d_1/s}  \end{array} \right)\,, 
\end{equation}
with $s={H_2/H_1}$. As a result, since ${\mathcal U}$ and ${\mathcal Z}$
are diagonal matrices, we can write the NLEP in \eqref{stab:vnlep} when
$\tau=0$ as
\begin{equation}\label{anlep:nlep}
  {\mathcal L} \hat{{\bf \Phi}} - w^2 \frac{ \int_{-\infty}^{\infty} w {\mathcal
      E}_{\lam} \hat{ {\bf \Phi}} \, dz}{\int_{-\infty}^{\infty} w^2\,
    dz} = \lambda {\mathcal U}^{-1} \hat{ \bf \Phi} \,; \qquad
  {\mathcal E}_{\lam} \equiv 2 \, {\mathcal U}
  {\mathcal B}_{\lambda}^{-1} {\mathcal Z} {\mathcal U}^{-1}\,.
\end{equation}
Next, upon defining ${\mathcal A}$ by
${\mathcal A}={\mathcal Z}^{-1}{\mathcal B}_{\lambda}$, we calculate
its matrix spectrum ${\mathcal A}\vb = \kappa \vb$, which can be
written as ${\mathcal B}_{\lambda} \vb = \kappa {\mathcal Z} \vb$. By
using \eqref{stab:blam} for ${\mathcal B}_{\lambda}$ with $\tau=0$,
and \eqref{anlep:uhmat} for ${\mathcal Z}$, we conclude that $\kappa$
must satisfy \bsub \label{anlep:det}
\begin{equation}
   \mbox{det} \left( \begin{array}{cc}
                 c_{1} -\kappa (c_1 + d_1 s) &  d_1\\
                       d_1  & c_2 -\kappa\left(c_2+ \frac{d_1}{s}\right)
                     \end{array} \right)=0 \,,
\end{equation}
which yields that $\kappa$ satisfies the quadratic equation
\begin{equation}\label{anlep:quad}
  \kappa^2\left( c_1 c_2 + c_2 d_1 s + d_1^2 + \frac{c_1 d_1}{s} \right)
  -\kappa\left(2c_1 c_2 + d_1 s c_2 + \frac{d_1 c_1}{s}\right) + c_1c_2-d_1^2=0
  \,.
\end{equation}
\esub
Observe that $\kappa_1=1$ is always an eigenvalue, and so $\kappa_2$ can
readily be found. A simple calculation yields that the matrix spectrum of
${\mathcal Z}^{-1}{\mathcal B}_{\lambda}$ is
\begin{equation}\label{anlep:amat_spec}
  \begin{split}
    \kappa_1 &=1 \,, \quad \vb_1= \left(\begin{array}{c}
                     1 \\
                     s \end{array} \right) \,, \\
  \kappa_2&=\frac{c_1 c_2 - d_1^2}{c_1c_2 + d_1^2 + d_1\left(c_2 s + {c_1/s}
    \right)} \,, \quad \vb_2 =\left(\begin{array}{c}
                    -d_1 \\
  c_1 - \kappa_2(c_1+d_1s) \end{array} \right) \,.
\end{split}                                  
\end{equation}
Next, we define the eigenvector matrix ${\mathcal V}$, the
diagonal matrix $\Lambda$, and the matrix ${\mathcal C}$ by
\begin{equation}\label{anlep:vmat_lambda}
  {\mathcal V} \equiv \left( \begin{array}{cc}
     1 &  -d_1\\
     s  & c_1 -\kappa_2(c_1+d_1 s) \end{array} \right)\,, \qquad
    \Lambda \equiv \left( \begin{array}{cc}
      2 &  0\\
       0  & {2/\kappa_2} \end{array} \right)\,, \qquad
   {\mathcal C}\equiv {\mathcal V}^{-1} {\mathcal U}^{-1}{\mathcal V} \,,
\end{equation}   
so that ${\mathcal E}_{\lambda}=2 {\mathcal U} {\mathcal A}^{-1}{\mathcal
  U}^{-1} =\left( {\mathcal U}{\mathcal V}\right) \Lambda \left(
  {\mathcal U}{\mathcal V}\right)^{-1}$. Finally, by setting
${\bf \tilde{\Phi}} =({\mathcal U}{\mathcal V})^{-1}{\bf \hat{\Phi}}$,
we obtain the vector-valued NLEP \eqref{stab:vfnlep}, where $\Lambda$
and ${\mathcal C}$ are defined explicitly in
\eqref{anlep:vmat_lambda}.

In the context of spike stability, the vector-valued NLEP
\eqref{stab:vfnlep} is a new linear stability problem, for which the
NLEP stability results for the scalar case in \cite{wei_surv}, \cite{ww},
and \cite{dgk3} are not directly applicable. Analytically, it is
challenging to provide necessary and sufficent conditions to guarantee that
the NLEP \eqref{stab:vfnlep} has no eigenvalues in $\mbox{Re}(\lambda)>0$.
However, one can analyze any zero-eigenvalue crossings, by using the
well-known identity $L_0 w=w^2$. By setting ${\bf \tilde{\Phi}}=
(0,w)^T$, we observe from \eqref{stab:vfnlep} that a zero-eigenvalue
crossing will occur when $\kappa_2=2$. By using
\eqref{anlep:amat_spec} for $\kappa_2$, a zero-eigenvalue crossing occurs
when
\begin{equation}\label{anlep:thresh}
  c_1 c_2 + 3 d_1^2 = 2 |d_1| \left( c_2 s + \frac{c_1}{s}\right) \,.
\end{equation}
Here $c_1$, $c_2$ and $d_1$ are determined in terms of the steady-state
spike locations $x_1$ and $x_2$ by \eqref{bif:coeff}, while $s={H_2/H_1}$
parameterizes the branch of asymmetric two-spike equilibria
in either \eqref{bif:3comp}, or equivalently \eqref{bif:M}. An interpretation
of the zero-eigenvalue crossing is given in the following remark.

\begin{remark}
  Equilibria of the DAE system \eqref{dae:reduce} are solutions to the
  nonlinear algebraic system $\Fb(x_1,x_2,H_1,H_2)={\bf 0}$ for
  $\Fb\in \R^4$, as given in \eqref{bif:F}. For a fixed $x_1$ and
  $x_2$, we claim that the linearization of the subsystem
  ${\mathcal F}_3={\mathcal F}_4=0$ in \eqref{bif:F} for the spike
  amplitudes is not invertible when the NLEP has a zero-eigenvalue
  crossing. To see this, we calculate along solutions to \eqref{bif:F}
  that
\begin{equation*}
\begin{split}
 J_3 &\equiv \left( \begin{array}{cc}
                {\mathcal F}_{3H_1} & {\mathcal F}_{3H_2} \\
                  {\mathcal F}_{4H_1} & {\mathcal F}_{4H_2} 
              \end{array} \right)
     = \left( \begin{array}{cc}
                12\mu_1^{3/2}H_1 -c_1 & -d_1 \\
                  -d_1 & 12 \mu_2^{3/2} H_2 - c_2 
              \end{array} \right) =
       \left( \begin{array}{cc}
                c_1+2d_1 s & -d_1 \\
                  -d_1 & c_2 + 2{d_1/s} 
              \end{array} \right) \,.
\end{split}
\end{equation*}
A simple calculation shows that $\mbox{det}(J_3)=0$ if and only if
\begin{equation}\label{j3:det}
     c_1 c_2 + 3 d_1^2 = -2 d_1 \left(c_2 s + \frac{c_1}{s}\right)\,,
\end{equation}
which is the condition derived in \eqref{anlep:thresh} for the
zero-eigenvalue crossing of the NLEP.
\end{remark}

The condition \eqref{anlep:thresh} for a zero-eigenvalue crossing
is indicated by the red dots on the asymmetric branches of equilibria shown
in Fig.~\ref{fig:L_5}, and in Figs.~\ref{fig:L_2}, \ref{fig:L_3},
\ref{fig:L_10}, \ref{fig:L_inf}. For the corresponding scalar NLEP case,
where ${\mathcal C}$ is a multiple of the identity, the rigorous results of
\cite{wei_surv} prove that $\mbox{Re}(\lambda)\leq 0$ if and only if
$\kappa_2<2$, and that an unstable real eigenvalue exists if $\kappa_2>2$.
We now investigate numerically whether these optimal linear stability
results persist for the vector-valued NLEP.

\subsubsection{Numerical Computation of the Vector-Valued NLEP}
\label{anlep:num_nlep}

We compute the discrete eigenvalues of the vector-valued NLEP
\eqref{stab:vfnlep} for
${\bf \tilde{\Phi}}\equiv \left(\tilde{\Phi}_1,
  \tilde{\Phi}_2\right)^T$, where $\Lambda$ and ${\mathcal C}$ are
defined in \eqref{anlep:vmat_lambda}.  To do so, we use a second-order
centered finite difference discretization of the NLEP, where the
nonlocal term is discretized using the trapezoidal rule. We discretize
\eqref{stab:vfnlep} on $0\leq z\leq z_M$ using the nodal values
\begin{align*}
  z_j = h (j-1) \,, \quad h & \equiv \frac{z_M}{n-1}
  \,, \quad w_j = w(z_j)=\frac{3}{2}\mbox{sech}^{2}\left(\frac{z_j}{2}\right)\,,
  \quad j=1,\ldots,n \,, \\
  {\bf \Psi} &\equiv \left(\Psi_{1,1},\ldots,\Psi_{1,n},\Psi_{2,1},\ldots,
  \Psi_{2,n}\right)^T \,,
\end{align*}
where $\Psi_{1,j}\approx \tilde{\Phi}_1(z_j)$ and
$\Psi_{2,j}\approx \tilde{\Phi}_2(z_j)$ for $j=1,\ldots,n$. We impose
that ${\bf \tilde{\Phi}^{\prime}}=0$ at $z=0,z_M$, which is
discretized by centered differences. The resulting block-structured
matrix eigenvalue problem for the pair ${\bf \Psi}\in \R^{2n}$ and $\lambda$
is given by
\bsub \label{nlep:block}
\begin{equation}
  \left( {\mathcal K}_n + {\mathcal M}_n\right) {\bf \Psi} = \lambda
  {\mathcal P}_n {\bf \Psi} \,, \label{nlep:block_1}
\end{equation}
where the matrices ${\mathcal K}_n\in \R^{2n,2n}$,
${\mathcal M}_n\in \R^{2n,2n}$ and ${\mathcal P}_n\in \R^{2n,2n}$, are
defined by
\begin{equation}\label{nlep:block_2}
  {\mathcal K}_n \equiv \left( \begin{array}{cc}
      {\mathcal K} &  0\\
     0  & {\mathcal K} \end{array} \right)\,,  \quad
  {\mathcal M}_n \equiv \left( \begin{array}{cc}
      {\mathcal M} &  0\\
     0  & \kappa_2^{-1} {\mathcal M} \end{array} \right)\,,  \quad
  {\mathcal P}_n \equiv \left( \begin{array}{cc}
      c_{11} I &  c_{12} I\\
     c_{21} I & c_{22} I \end{array} \right)\,.
\end{equation}
Here $I\in \R^{n,n}$ is the identity, and $c_{ij}$ for
$1\leq i,j\leq 2$ are the matrix entries of the $2\times 2$ matrix
${\mathcal C}$ defined in \eqref{anlep:vmat_lambda}. In
\eqref{nlep:block_2}, the $n\times n$ tridiagonal matrix
${\mathcal K}$ and the full $n\times n$ matrix ${\mathcal M}$ are
defined, respectively, by
\begin{equation}\label{nlep:block_3}
  \begin{split}
    {\mathcal K}_{1,2}= {\mathcal K}_{n,n-1} &= \frac{2}{h^2} \,, \quad
    {\mathcal K}_{ii} = -\frac{2}{h^2} -1 + 2 w_i \,, \quad \mbox{for}
    \quad i=1,\ldots n\,, \\
  {\mathcal K}_{i,i+1}&={\mathcal K}_{i,i-1}=\frac{1}{h^2} \,, \quad
  \mbox{for} \quad i=2,\ldots,n-1 \,,
  \end{split}
\end{equation}
and
\begin{equation}\label{nlep:block_4}
  {\mathcal M} \equiv -\frac{2h}{3} \left( \begin{array}{ccccc}
   w_1^2\left(\frac{w_1}{2}\right) &  w_1^2 w_2 & \hdots & w_1^2 w_{n-1} &
    w_1^2 \left(\frac{w_n}{2}\right)  \\
   \vdots & \vdots & \vdots & \vdots & \vdots \\
   \vdots & \vdots & \vdots & \vdots & \vdots \\
   w_n^2\left(\frac{w_1}{2}\right)  &  w_n^2 w_2  & \hdots & w_n^2 w_{n-1} &
    w_n^2 \left(\frac{w_n}{2}\right)  \end{array} \right)\,.
\end{equation}
\esub

\begin{figure}[htbp]
\begin{center}
  \includegraphics[width=0.32\linewidth,height=4.5cm]{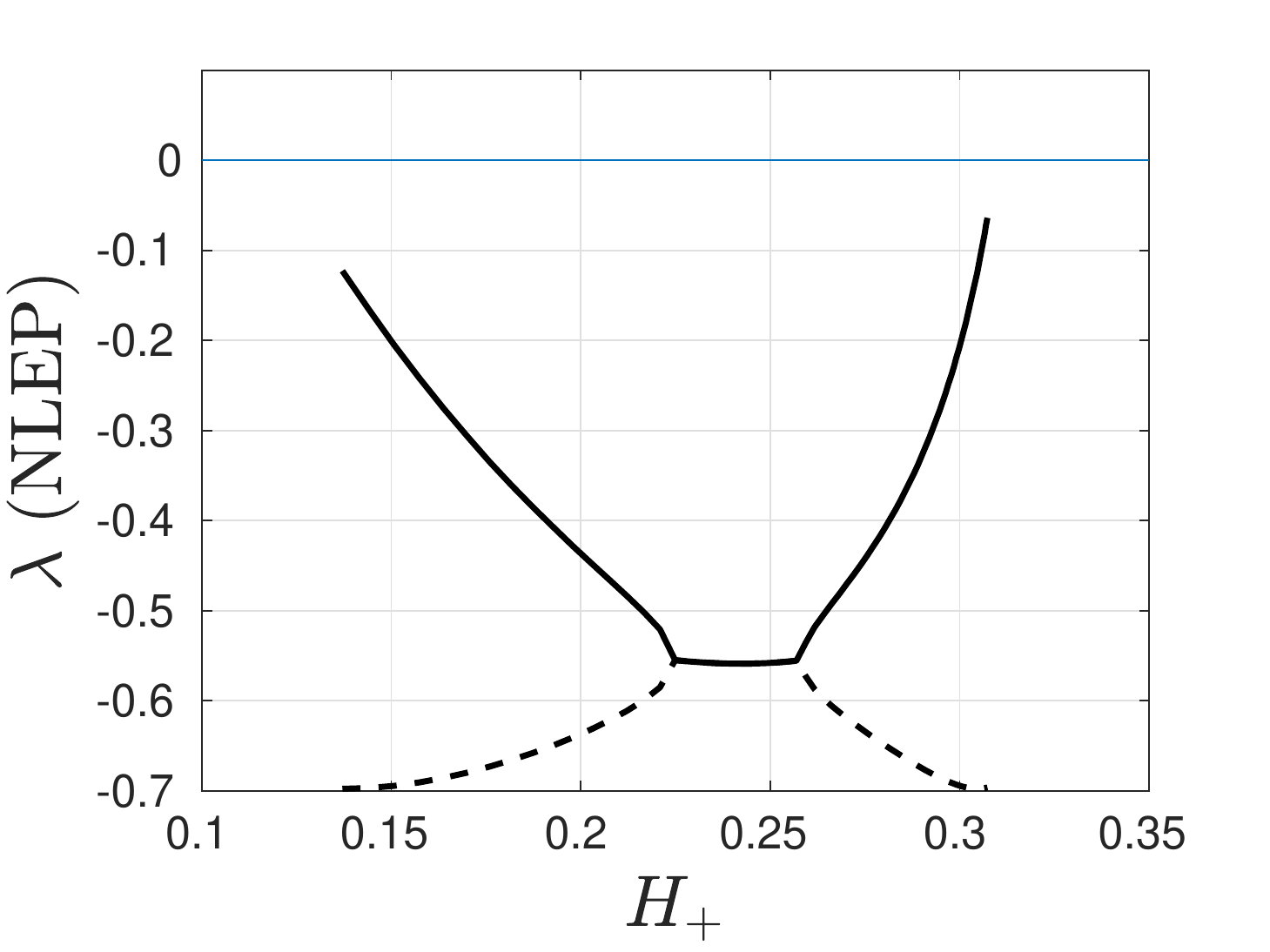}
  \includegraphics[width=0.32\linewidth,height=4.5cm]{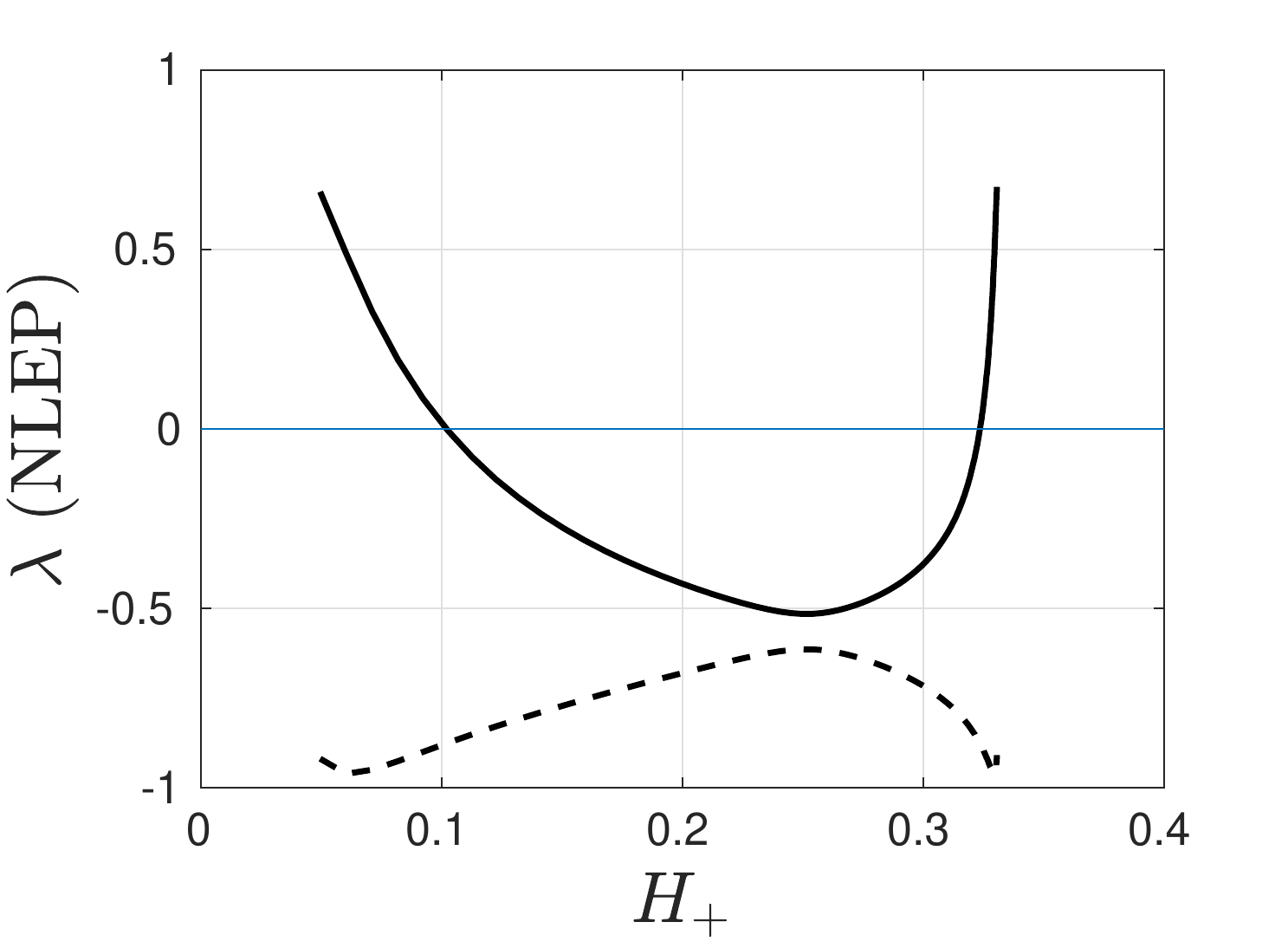}
  \includegraphics[width=0.32\linewidth,height=4.5cm]{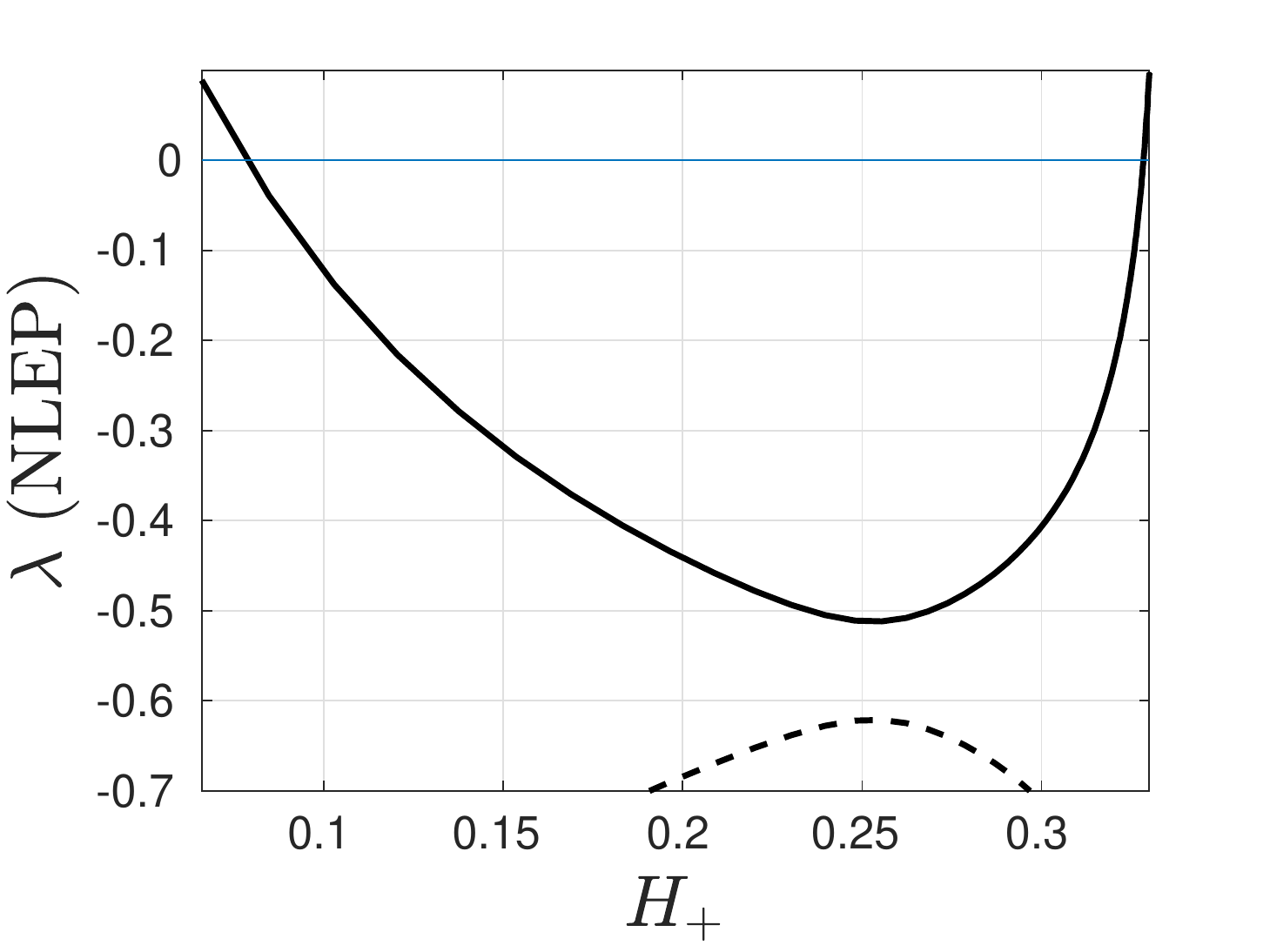}
  \caption{Plot of the first (heavy solid) and second (dashed)
    eigenvalues (ordered by the largest real parts), as computed from
    the discretization of the vector-valued NLEP \eqref{stab:vfnlep}
    versus the height $H_+$ of the rightmost spike along the
    asymmetric solution branches shown in Figs.~\ref{fig:L_2},
    \ref{fig:L_3} and \ref{fig:L_5} for domain half-lengths $L=2$
    (left), $L=3$ (middle) and $L=5$ (right), respectively. Numerical
    evidence shows that when $\kappa_2<2$, the vector NLEP has no
    unstable eigenvalues, and that a unique positive eigenvalue occurs
    when $\kappa_2>2$.  Here $\kappa_2$ is defined in
    \eqref{anlep:amat_spec} and the zero-eigenvalue crossing occurs
    when $\kappa_2=2$, leading to \eqref{anlep:thresh}. The thin
    horizontal blue line is the zero-eigenvalue
    crossing.} \label{fig:nlep_eig}
\end{center}
\end{figure}

For $n=250$ and $z_M=15$, the matrix spectrum of \eqref{nlep:block} is
computed numerically using a generalized matrix eigenvalue solver from
EISPACK at each point along the asymmetric solution branches of
two-spike equilibria. In Fig.~\ref{fig:nlep_eig} we plot the first two
eigenvalues of \eqref{nlep:block}, defined as those with the largest
real parts, versus the height $H_+$ of the rightmost spike for
$L=2,3,5$. In terms of $H_+$, we recall that the asymmetric branches
of equilibria for these values of $L$ were shown in the right panels
of Figs.~\ref{fig:L_2}, \ref{fig:L_3} and \ref{fig:L_5},
respectively. From Fig.~\ref{fig:nlep_eig} we observe that the first
two eigenvalues are real-valued except for a small range of $H_+$ when
$L=2$, where they form a complex conjugate pair. These numerical
results confirm the zero-eigenvalue crossing condition
\eqref{anlep:thresh}, obtained by setting $\kappa_2=2$, as evidenced
by the intersection of the heavy-solid curves and the horizontal blue
lines in Fig.~\ref{fig:nlep_eig}. However, most importantly, the
results in Fig.~\ref{fig:nlep_eig} establish numerically that the
vector-valued NLEP \eqref{stab:vfnlep}, which is valid for $\tau=0$,
has no unstable discrete eigenvalues whenever $\kappa_2<2$, and that
there is a unique unstable discrete eigenvalue when
$\kappa_2>2$. Increasing the number of gridpoints $n$ or the cutoff
$z_M$ did not alter the results to two decimal places of accuracy.

\begin{figure}[htbp]
\begin{center}
  \includegraphics[width=0.45\linewidth,height=4.5cm]{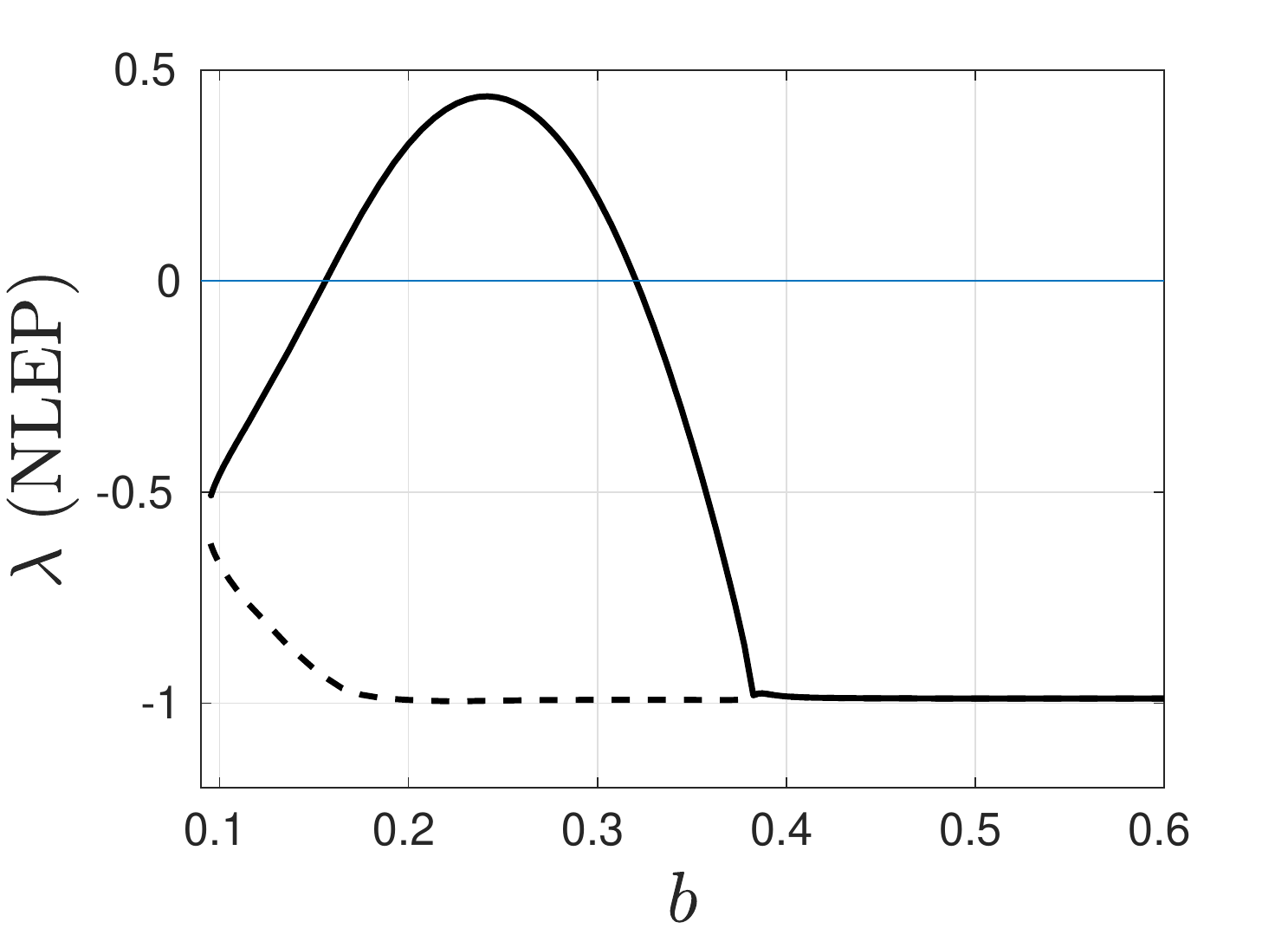}
  \includegraphics[width=0.45\linewidth,height=4.5cm]{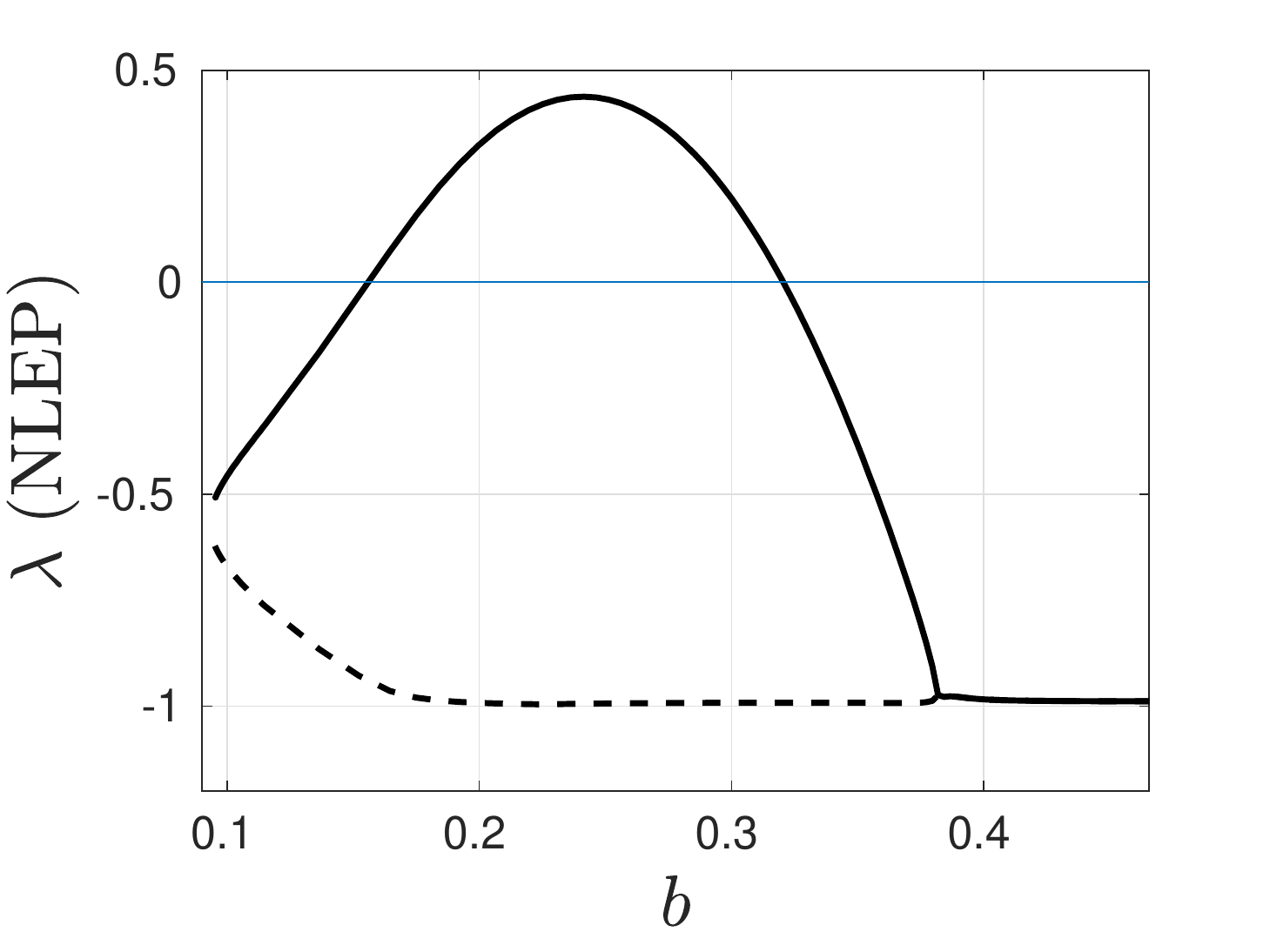}
  \caption{Plot of the first (heavy solid) and second (dashed)
    eigenvalues (ordered by the largest real parts), as computed from
    the discretization of the vector-valued NLEP \eqref{stab:vfnlep}
    versus the precursor parameter $b$ along the asymmetric solution
    branches shown in Figs.~\ref{fig:L_10} and \ref{fig:L_inf} for a
    domain half-length $L=10$ (left panel) and an infinite domain
    $L=\infty$ (right panel), respectively. The NLEP has two
    zero-eigenvalue crossings (intersection with the horizontal blue
    line) on each portion of the asymmetric branch at parameter values
    where $\kappa_2=2$ (see Fig.~\ref{fig:L_10} and
    Fig.~\ref{fig:L_inf}). Between the zero-eigenvalue crossings the
    vector NLEP has a unique unstable real
    eigenvalue.} \label{fig:nlep_eig_10inf}
\end{center}
\end{figure}

For $L=10$ and for the infinite domain problem with $L=\infty$, in
Fig.~\ref{fig:nlep_eig_10inf} we plot the first two eigenvalues of
\eqref{nlep:block} versus the precursor parameter $b$ along the
asymmetric solution branches of Fig.~\ref{fig:L_10} and
Fig.~\ref{fig:L_inf}.  From Fig.~\ref{fig:nlep_eig_10inf} we observe
that along these solution branches the NLEP has two zero-eigenvalue
crossings, corresponding to where $\kappa_2=2$, and that the vector
NLEP has a unique unstable eigenvalue between these crossings. This
linear stability behavior is encoded in the global bifurcation
diagrams for $L=10$ and $L=\infty$ shown in the right panels of
Fig.~\ref{fig:L_10} and Fig.~\ref{fig:L_inf}, respectively.

\section{Validation from PDE Simulations}\label{pre:numerics}

In this section, we validate our global bifurcation and linear
stability results for the precursor field $\mu(x)=1+bx^2$ from
time-dependent PDE simulations of \eqref{gm:full}.  In our
simulations, we give initial conditions for \eqref{gm:full} that
correspond to a two-spike quasi-equilibrium solution, where the spike
heights satisfy the constraint in \eqref{dae:reduce} for given spike
locations $x_1$ and $x_2$ at $t=0$.

\begin{figure}[htbp]
\begin{center}
  \includegraphics[width=0.24\linewidth,height=4.5cm]{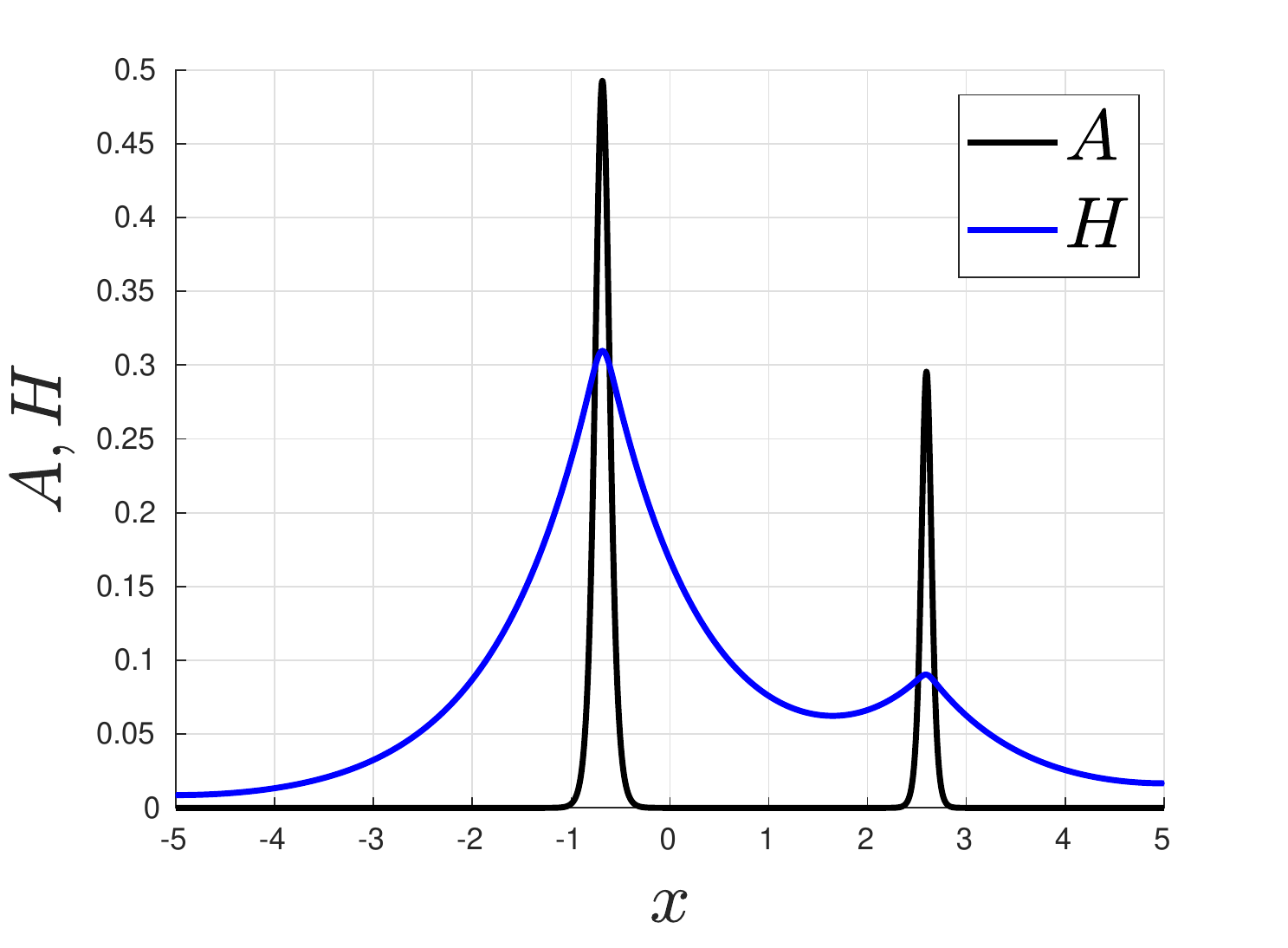}
  \includegraphics[width=0.24\linewidth,height=4.5cm]{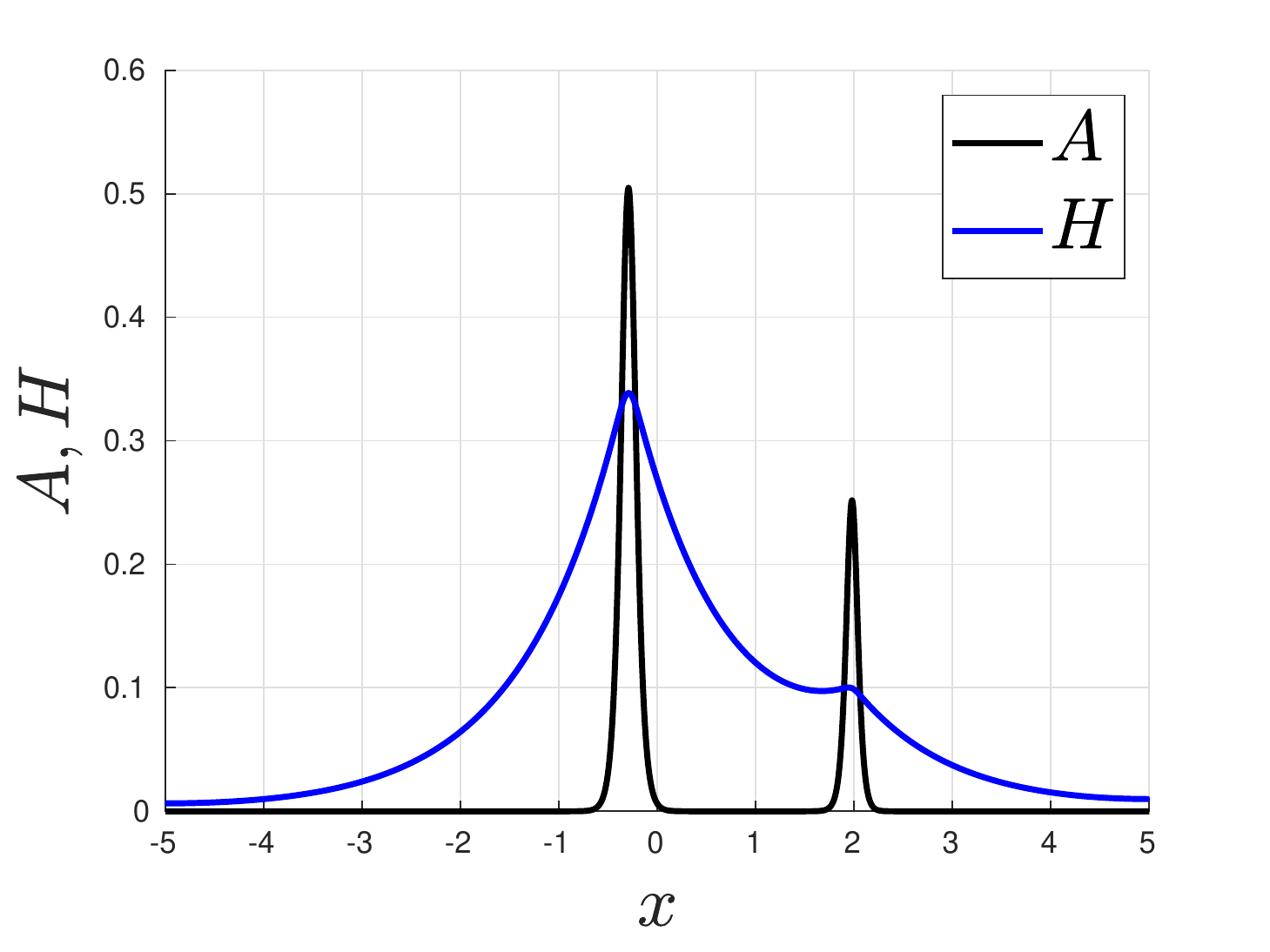}
  \includegraphics[width=0.24\linewidth,height=4.5cm]{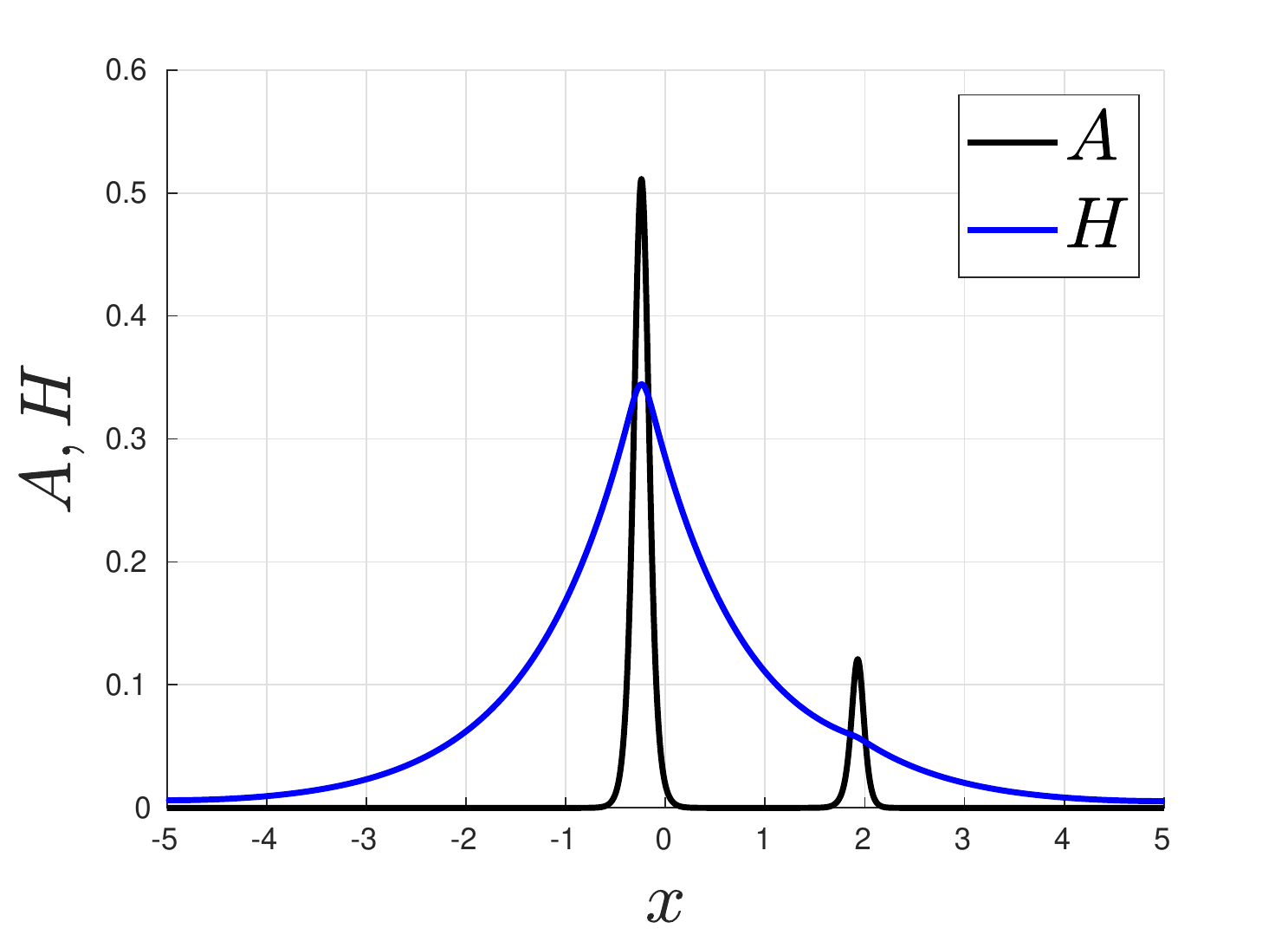}
  \includegraphics[width=0.24\linewidth,height=4.5cm]{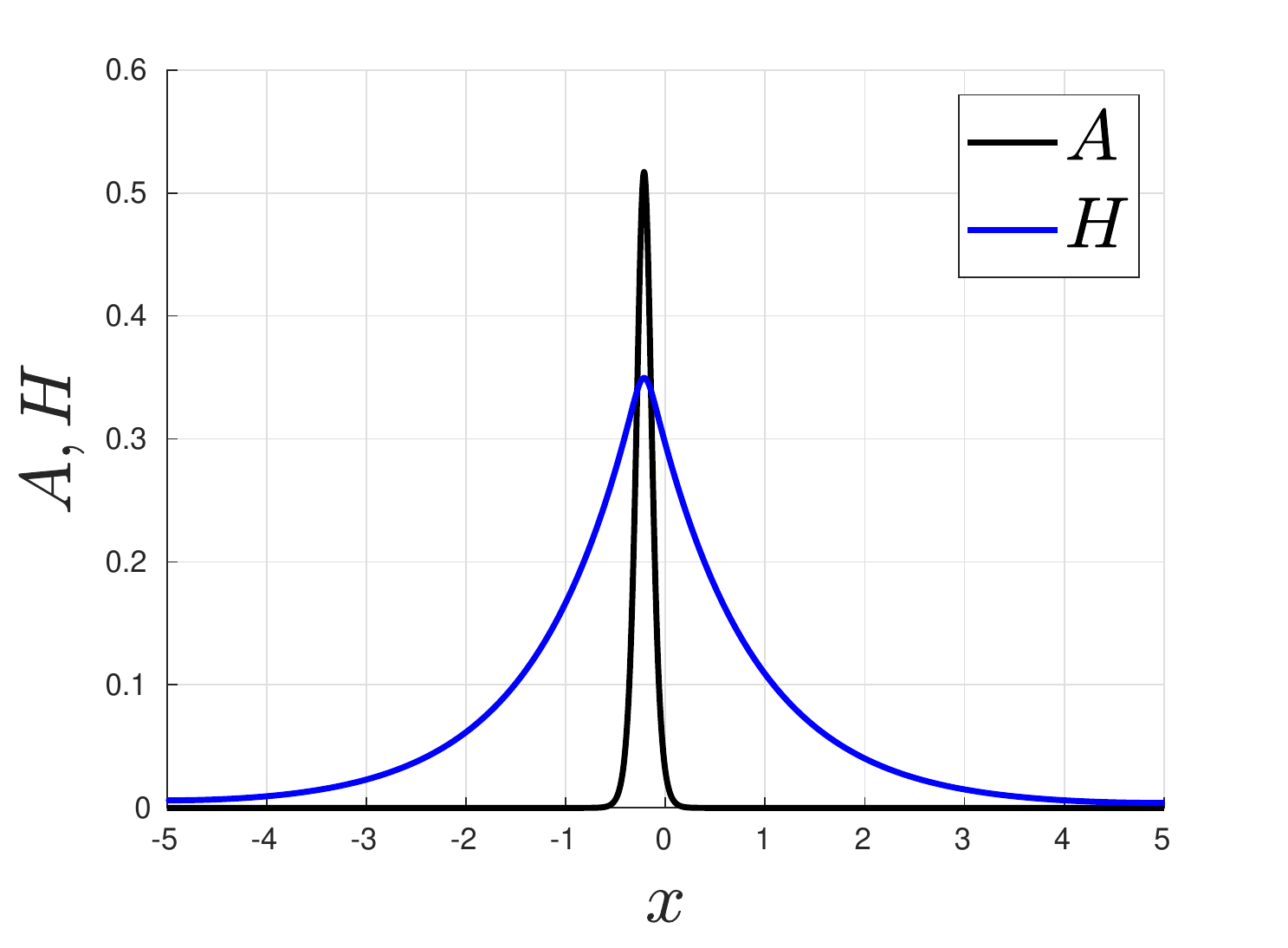}
  \caption{Time-dependent PDE simulations of \eqref{gm:full} with
    $L=5$, $\eps=0.05$, and $\tau=0.25$ for a precursor
    $\mu(x)=1+bx^2$ with $b=0.18$. Initial condition is a
    quasi-equilibrium two-spike solution with spike locations
    $x_1(0)=-1$ and $x_2(0)=3$. Plots of $A$ and $H$ versus $x$ at
    four different times showing that one spike is annihilated as time
    increases.  For $b=0.18$, the right panel in Fig.~\ref{fig:L_5}
    shows that there is no stable asymmetric two-spike pattern. Left:
    $t=180$. Left Middle: $t=335$. Right Middle: $t=650$. Right: $t=800$.
  } \label{fig:pde_run2}
\end{center}
\end{figure}

For $L=5$ and $b=0.12$, the results from the PDE simulations shown in
Fig.~\ref{fig:pde_run1} confirm that a quasi-equilibrium two-spike
pattern tends to a stable asymmetric two-spike equilibrium on a long
time scale, as predicted by the bifurcation diagram shown in the right
panel of Fig.~\ref{fig:L_5}. The other parameter values are shown in
caption of Fig.~\ref{fig:pde_run1}.  In contrast, if $b=0.18$, from the
PDE simulation results shown in Fig.~\ref{fig:pde_run2} we observe that
a two-spike quasi-equilibrium solution undergoes a competition instability
leading to the destruction of a spike. For this parameter set, there is
no stable  asymmetric two-spike steady-state pattern as observed from the
right panel of Fig.~\ref{fig:L_5}.

\begin{figure}[htbp]
\begin{center}
  \includegraphics[width=0.32\linewidth,height=4.5cm]{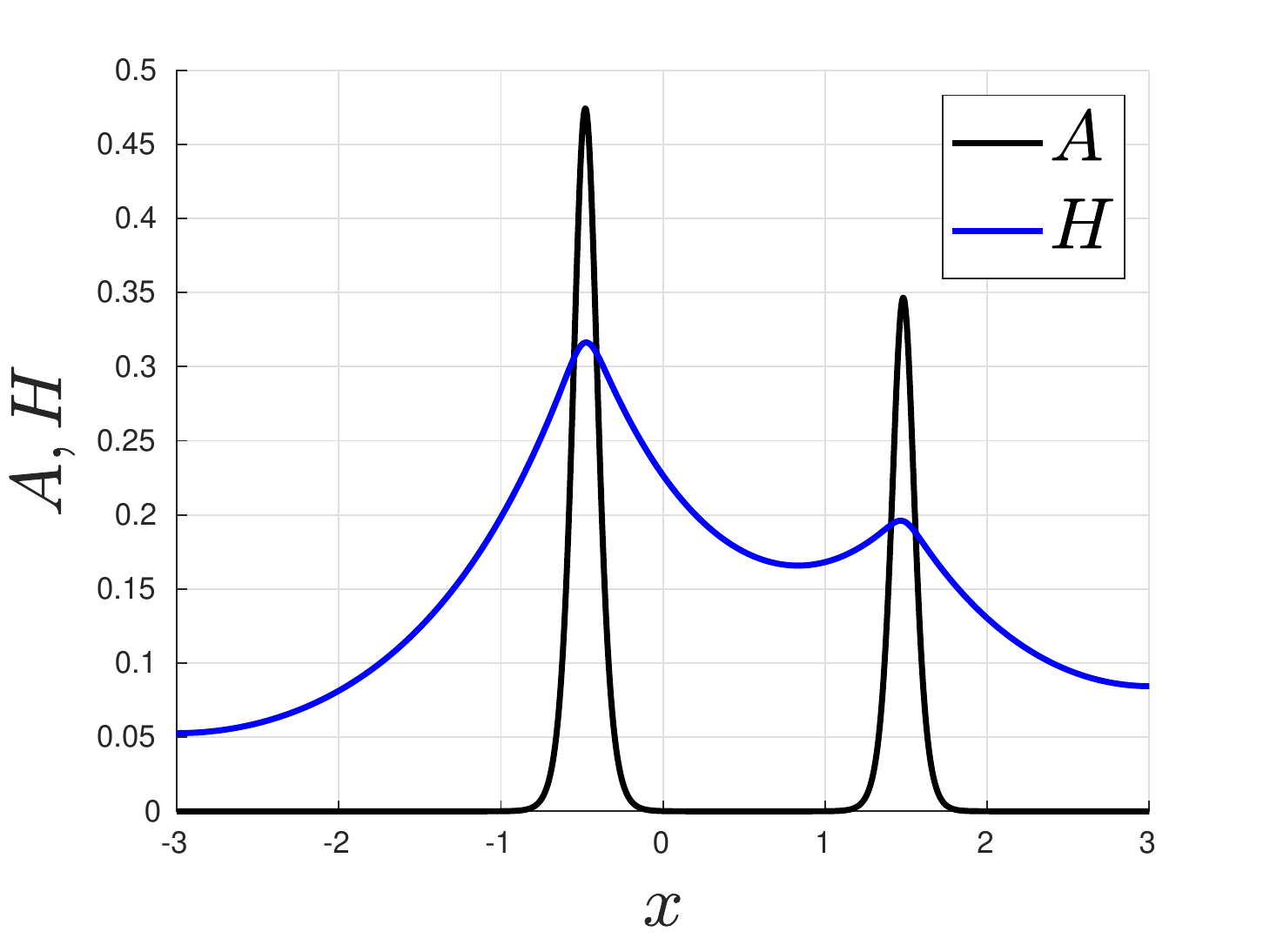}
  \includegraphics[width=0.32\linewidth,height=4.5cm]{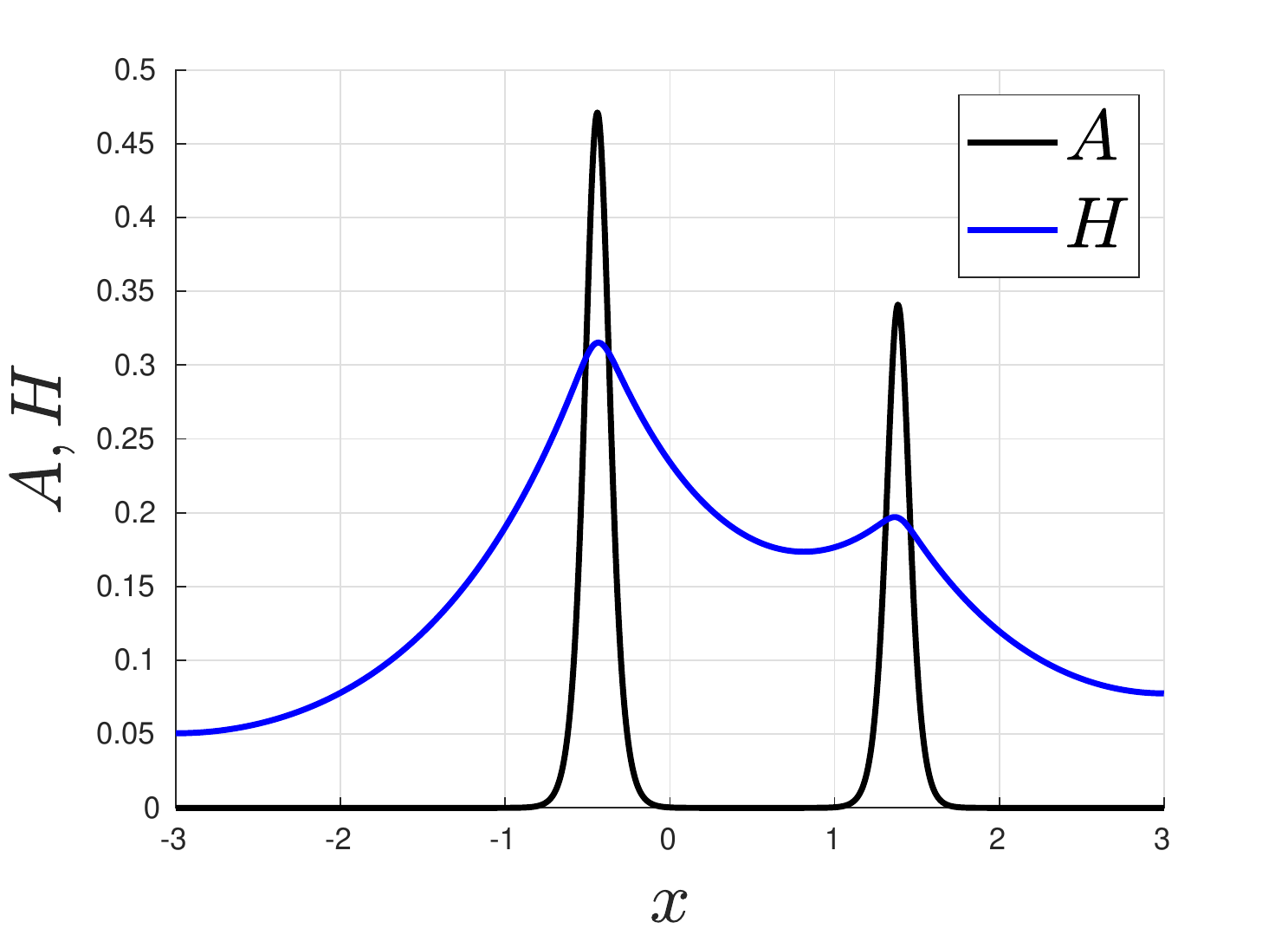}
  \includegraphics[width=0.32\linewidth,height=4.5cm]{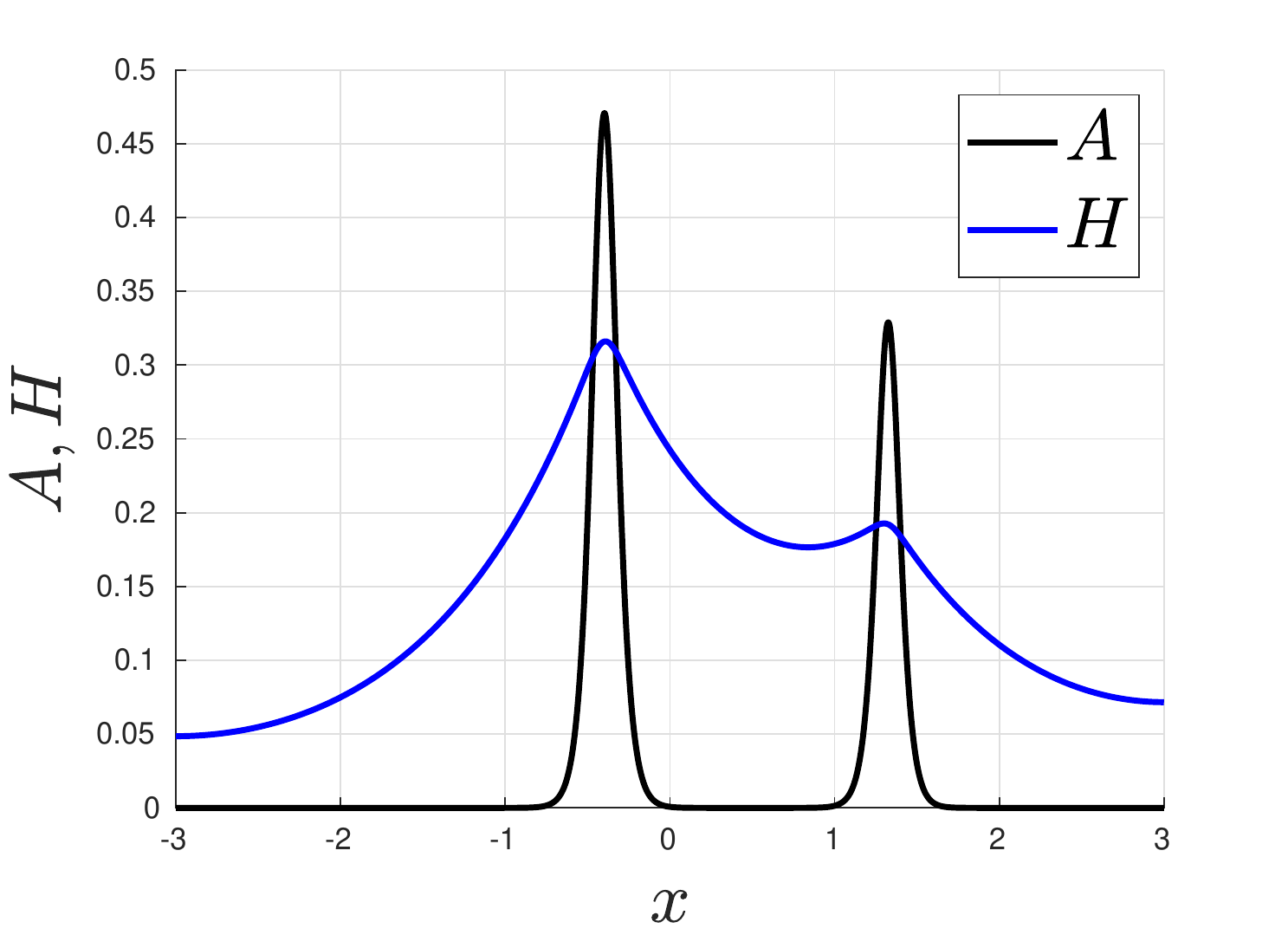}
  \caption{Time-dependent PDE simulations of \eqref{gm:full} with
    $L=3$, $\eps=0.05$, and $\tau=0.15$ for a precursor
    $\mu(x)=1+bx^2$ with $b=0.09$. Initial condition is a
    quasi-equilibrium two-spike solution with spike locations
    $x_1(0)=-0.5$ and $x_2(0)=1.5$. Plots of $A$ and $H$ versus $x$ at
    three different times showing the convergence towards a stable
    asymmetric two-spike pattern as predicted from the right panel of
    Fig.~\ref{fig:L_3}.  Left: $t=31$. Middle: $t=301$. Right:
    $t=900$. As $t$ increases there is only a slight adjustment of the
    pattern.}
   \label{fig:pde_run3}
\end{center}
\end{figure}

Similarly, for $L=3$ and $b=0.09$, we observe from the full numerical
results shown in Fig.~\ref{fig:pde_run3} that the quasi-equilibrium
two-spike pattern converges as $t$ increases to a stable asymmetric
steady-state pattern. As shown in the bifurcation diagram given in the
right panel of Fig.~\ref{fig:L_3} there is a stable asymmetric
two-spike steady-state for these parameter values.

\begin{figure}[htbp]
\begin{center}
  \includegraphics[width=0.24\linewidth,height=4.5cm]{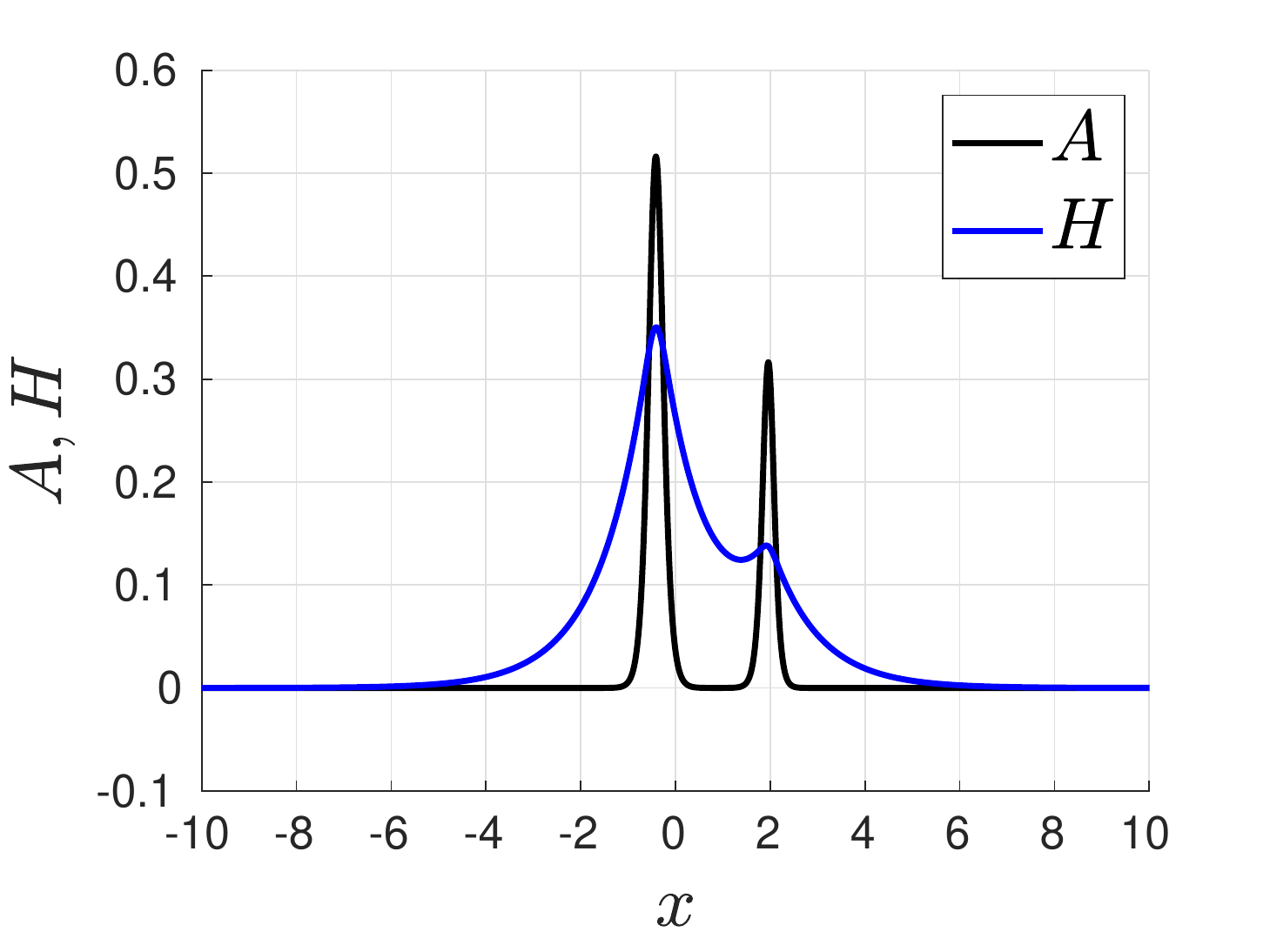}
  \includegraphics[width=0.24\linewidth,height=4.5cm]{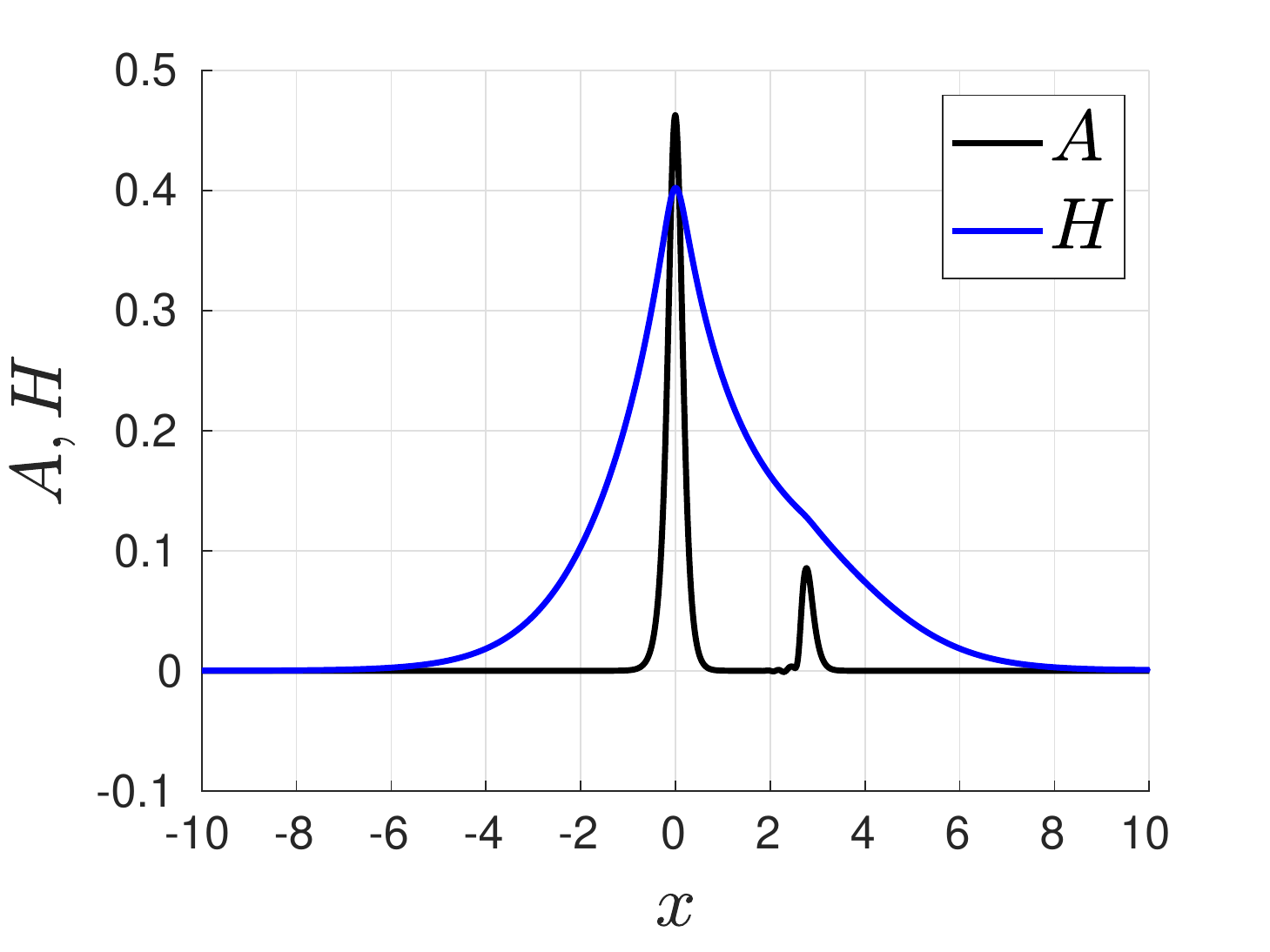}
  \includegraphics[width=0.24\linewidth,height=4.5cm]{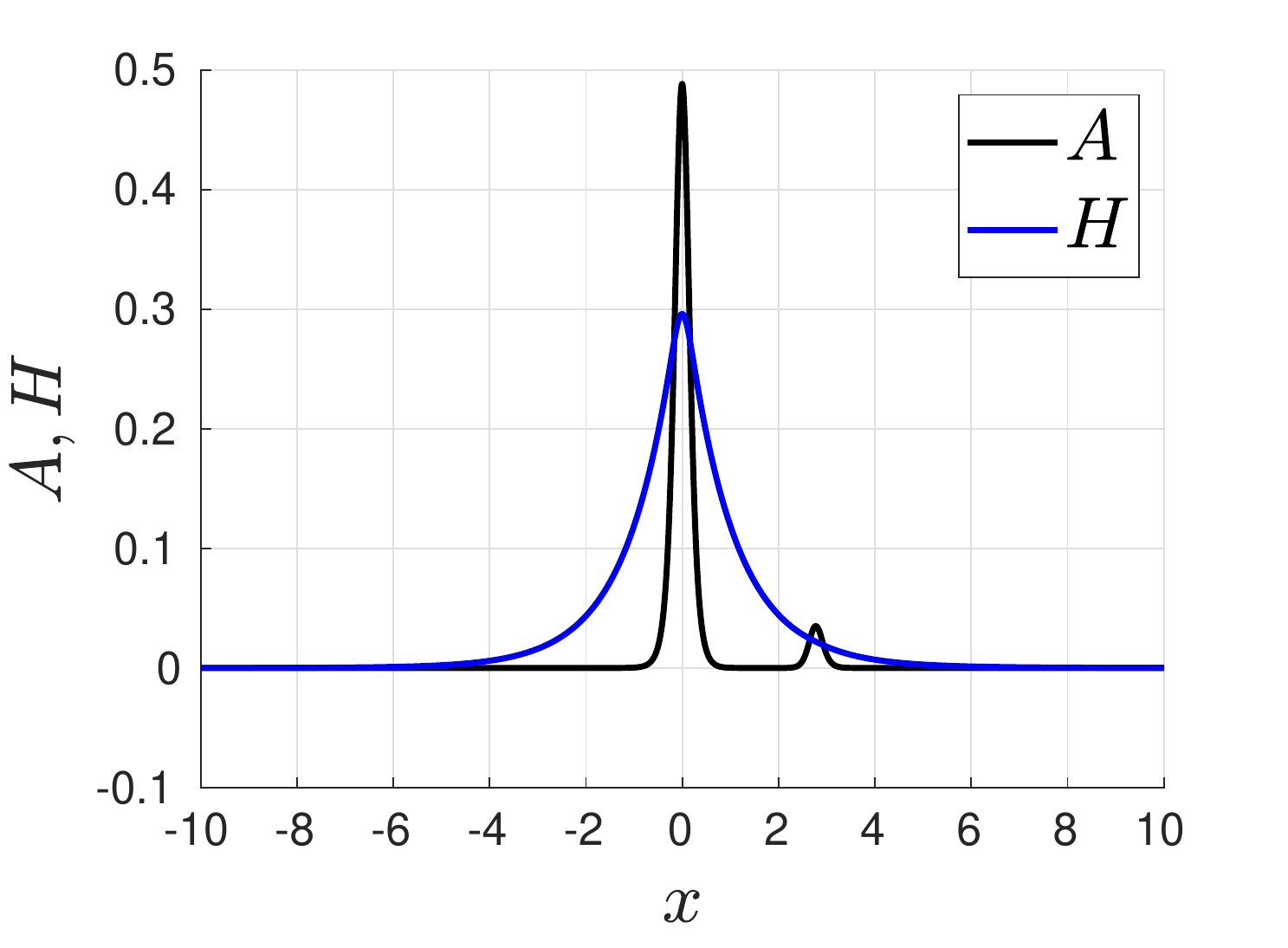}
  \includegraphics[width=0.24\linewidth,height=4.5cm]{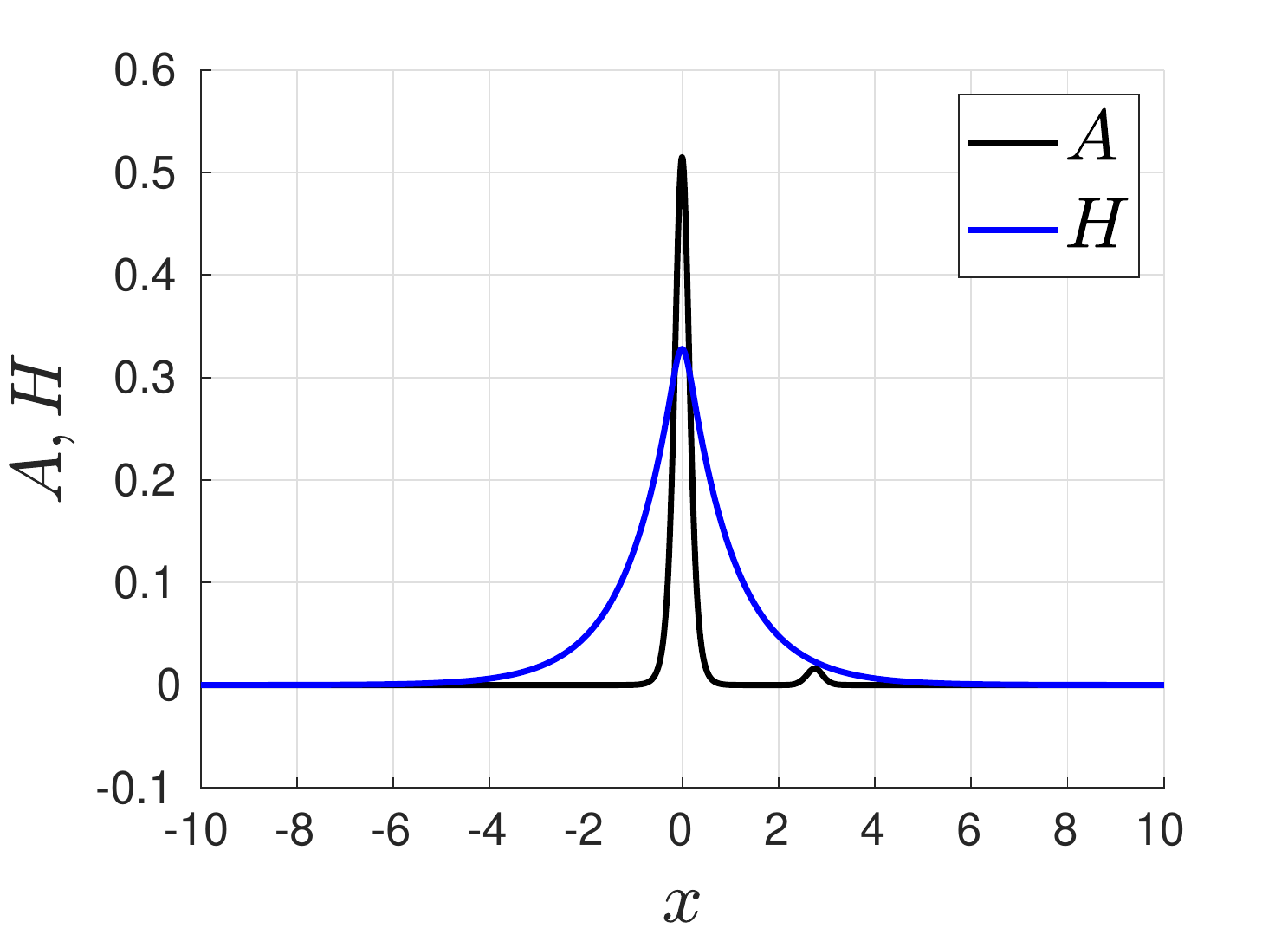}
  \caption{Left panel: steady-state of time-dependent PDE simulations
    of \eqref{gm:full} with $L=10$, $\eps=0.10$, and $\tau=0.15$ for
    $\mu(x)=1+bx^2$ with $b=0.15$. Other panels: PDE simulations of
    \eqref{gm:full} when $b$ is increased to $b=0.20$ (other
    parameters the same). For $b=0.20$, the NLEP stability theory in
    Fig.~\ref{fig:nlep_eig_10inf} predicts no stable asymmetric
    two-spike steady-state. The PDE numerical results show a collapse
    of the small spike. Left middle: $t=0$. Right middle: $t= 0.61$.
    Right: $t=1.2$. For the PDE simulations with $b=0.15$ and
    $b=0.20$, the initial condition was a $2\%$ perturbation of the
    asymmetric steady state shown in the global bifurcation diagram
    Fig.~\ref{fig:L_10}.}
   \label{fig:pde_L10}
\end{center}
\end{figure}

Finally, for $L=10$, in Fig.~\ref{fig:pde_L10} we show results for
two-spike solutions computed from PDE simulations of \eqref{gm:full}
for $b=0.15$ and for $b=0.20$. In the left panel of
Fig.~\ref{fig:pde_L10} we show a stable asymmetric two-spike
steady-state for $b=0.15$ as computed numerically from
\eqref{gm:full}, starting from an initial condition chosen to be close
to the stable asymmetric pattern predicted from the global bifurcation
diagram in Fig.~\ref{fig:L_10}. For $b=0.20$, where no such stable
asymmetric pattern exists from Fig.~\ref{fig:L_10}, the PDE
simulations shown in the other three panels in Fig.~\ref{fig:pde_L10}
confirm the instability and show the annihilation of the small spike
as time increases.

\section{Discussion}\label{sec:discussion}

For the GM model \eqref{gm:full} with a precursor field
$\mu(x)=1+b x^2$, we have shown that a linearly stable asymmetric
two-spike steady-state pattern can emerge from a supercritical
pitchfork bifurcation at some critical value of $b$ along a symmetric
branch of two-spike equilibria. For this symmetry-breaking
bifurcation, the critical value of $b$ depends on the domain
half-length $L$. From a linearization around the steady-state of a DAE
system of ODEs for the spike locations and spike heights, we have
shown numerically that some portions of the asymmetric branches of
equilibria are linearly stable to the small eigenvalues. Moreover,
from a combined analytical and numerical investigation of the spectrum
of a novel class of vector-valued NLEP, we have shown that portions of
the branches of asymmetric two-spike equilibria are linearly stable to
${\mathcal O}(1)$ time-scale spike amplitude instabilities. Overall,
our combined analytical and numerical study establishes the
qualitatively novel result that linearly stable asymmetric two-spike
equilibria can occur for the GM model with a precursor
field. Asymmetric two-spike equilibria in 1-D for the GM model are all
unstable in the absence of a precursor field \cite{ww_asy}.

Although we have only exhibited stable asymmetric patterns for the GM
model with a specific precursor field with two spikes, the analytical
framework we have employed applies to multiple spikes, to other precursor
fields, and to other singularly perturbed RD systems. In particular, 
the equilibria of the DAE system \eqref{dae:mat} could be used to
compute the bifurcation diagram of symmetric and asymmetric spike
equilibria for more than two spikes.

There are two open directions that warrant further investigation.  One
specific focus would be to extend NLEP stability theory for scalar
NLEPs to establish analytically necessary and sufficient conditions
for the vector-valued NLEP \eqref{stab:vfnlep} to admit no eigenvalues
in $\mbox{Re}(\lambda)>0$. In this NLEP we would allow ${\mathcal C}$
in \eqref{stab:vfnlep} to be an arbitrary matrix with positive
eigenvalues. A second open direction would be to extend the 1-D theory
for the GM model with a precursor field to a 2-D setting in order to
construct stable asymmetric spot patterns in a 2-D domain.
  
\section*{Acknowledgements}\label{sec:ak}
Theodore Kolokolnikov and Michael Ward were supported by NSERC
Discovery grants. Fr\'ed\'eric Paquin-Lefebvre was supported by a UBC
Four-Year Graduate Fellowship.

\begin{appendix}
\renewcommand{\theequation}{\Alph{section}.\arabic{equation}}
\setcounter{equation}{0}

\section{Alternative Formulation of Two-Spike Equilibria}\label{app:alt}

In this appendix we briefly outline the derivation of the coupled
system \eqref{bif:theo_all} characterizing two-spike equilibria for
the special case where $\mu(x)$ is even in $x$. We center the spikes
at $x_2=r_{+}$ and $x_1=-r_{-}$, and we let $\ell$ be the unknown
location, with $x_1<\ell<x_2$, where $h_x(\ell)=a_{x}(\ell)=0$. We
label the spike heights as $H_{\pm}=h(\pm r_{\pm})$.

To proceed, we first construct a steady-state spike at $x=r_{+}$ on
the interval $(\ell,L)$ with $h_x=0$ and $a_x=0$ at $x=\ell,L$. A
similar construction is made for the interval $(-L,\ell)$ with a spike
at $x=-r_{-}$. Then, since $\mu(x)$ is even, we can write the two
steady-state conditions in a compact unified form, with the remaining
equation resulting from adjusting $h(\ell)$ so that $h(x)$ is
continuous across $x=\ell$.

For the right interval $\ell<x<L$ with a spike at $x=r_{+}$, we
proceed as in the derivation of \eqref{dae:full} to obtain that
$r_{+}$ satisfies
\begin{equation}\label{alt:right}
   -\frac{\mu^{\prime}(r_{+})}{\mu(r_{+})} 
-\frac{4}{5} \frac{ \langle g_{1x}\rangle\vert_{x=r_{+}}}{g_1\vert_{x=r_{+}}}=0\,,
\end{equation}
where $\langle g_{1x}\rangle$ is the average of $g_{1x}$ across
$x=r_{+}$. Here $g_1(x,r_{+})$ is the 1-D Green's function satisfying
\begin{equation}\label{alt:g1}
  g_{1xx}-g_1 = -\delta(x-r_{+}) \,, \quad \ell < x < L \,; \qquad
  g_{1x}=0 \, \quad \mbox{at} \quad x=\ell\,,\, L \,.
\end{equation}
The inhibitor field $h(x)$ and the spike height $H_{+}=h(r_{+})$ are given by
\begin{equation}\label{alt:hright}
  h(x)=6 H_{+}^2 \mu_{+}^{3/2} g_1(x,r_{+}) \,, \qquad
  H_{+}= \frac{\mu_{+}^{-3/2}}{6 g_1\vert_{x=r_{+}}} \,,
\end{equation}
where $\mu_{+}\equiv \mu(r_{+})$.  Similarly, for the left interval
$-L<x<\ell$ with a spike at $x=-r_{-}$, we obtain that $r_{-}$
satisfies
\begin{equation}\label{alt:left}
   -\frac{\mu^{\prime}(-r_{-})}{\mu(-r_{-})} 
-\frac{4}{5} \frac{ \langle g_{2x}\rangle\vert_{x=-r_{-}}}{g_2\vert_{x=-r_{-}}}=0\,,
\end{equation}
where $g_2(x,r_{-})$ satisfies
\begin{equation}\label{alt:g2}
  g_{2xx}-g_2 = -\delta(x+ r_{-}) \,, \quad -L < x < \ell \,; \qquad
  g_{2x}=0 \, \quad \mbox{at} \quad x=\ell\,,\, -L \,.
\end{equation}
The inhibitor field $h(x)$ and the spike height $H_{-}=h(-r_{-})$ are given by
\begin{equation}\label{alt:hleft}
  h(x)=6 H_{-}^2 \mu_{-}^{3/2} g_2(x,r_{-}) \,, \qquad
  H_{-}= \frac{\mu_{-}^{-3/2}}{6 g_2\vert_{x=-r_{-}}} \,,
\end{equation}
where $\mu_{-}=\mu(-r_{-})$.

Since $\mu(x)$ is even, we have $\mu(-r_{-})=\mu(r_{-})$ and
$\mu^{\prime}(-r_{-})=-\mu^{\prime}(r_{-})$. Next, we set
$\tilde{x}=-x$ in \eqref{alt:g2} and label
$\tilde{g}_2(\tilde{x}, r_{-})\equiv g_2(-\tilde{x},r_{-})$, so that
\eqref{alt:left} becomes
\begin{equation}\label{alt:t_left}
   -\frac{\mu^{\prime}(r_{-})}{\mu(r_{-})} 
   -\frac{4}{5} \frac{ \langle \tilde{g}_{2\tilde{x}}
     \rangle\vert_{\tilde{x}=r_{-}}}{g_2\vert_{\tilde{x}=r_{-}}} = 0\,,
\end{equation}
where $\tilde{g}_2(\tilde{x},r_{-})$ satisfies
\begin{equation}\label{alt:t_g2}
  \tilde{g}_{2\tilde{x}\tilde{x}}-\tilde{g}_2 =
  -\delta(\tilde{x}-r_{-}) \,, \quad -\ell
  < \tilde{x} < L \,; \qquad
  g_{2\tilde{x}}=0 \, \quad \mbox{at} \quad \tilde{x}=-\ell\,,\, L \,.
\end{equation}

To combine \eqref{alt:right} and \eqref{alt:t_left} into a unified
expression it is convenient to define $g(x,r;\ell)$ as in
\eqref{bif:gfunc}, so that $g_1(x,r_{+})=g(x,r_{+};\ell)$ and
$\tilde{g}_2(x,r_{-})=g(x,r_{-};-\ell)$. In this way,
\eqref{alt:right} and \eqref{alt:t_left} reduce to $f(r_{+},\ell)=0$
and $f(r_{-},-\ell)=0$, where $f(r,\ell)$ is defined in
\eqref{alt:fxi}. The condition that the inhibitor field is continuous
across $x=\ell$, as obtained by equating the two expressions for
$h(\ell)$ in \eqref{alt:hright} and \eqref{alt:hleft}, yields the
continuity condition $\xi(r_{+},\ell)=\xi(r_{-},-\ell)$ as written in
\eqref{alt:fxi}.

The computation of two-spike equilibria reduces to finding roots of
$\bm{F}({\bf u},\zeta)=0$, as defined in \eqref{alt:sys} as the
parameter vector $\zeta \equiv (b, L)^T$ is varied.  To compute paths
of solutions we employ the software packages AUTO
(cf.~\cite{auto}) and \textsc{coco} (cf.~\cite{coco}) and provide
the Jacobian matrices
\begin{align}\label{app:jac_1}
  D_u\bm{F} &= \begin{pmatrix} \frac{\partial f}{\partial r}(r_+,\ell) & 0 &
    \frac{\partial f}{\partial l}(r_+,\ell) \\ 
    0 & \frac{\partial f}{\partial r}(r_-,-\ell) & -\frac{\partial f}{\partial
      \ell}(r_-,\ell) \\
    \frac{\partial \xi}{\partial r}(r_+,\ell) &
    -\frac{\partial \xi}{\partial r}(r_-,-\ell) &
    \frac{\partial \xi}{\partial l}(r_+,\ell) +
    \frac{\partial \xi}{\partial l}(r_-,-\ell) \end{pmatrix},\,\\
  D_\zeta\bm{F} &= \begin{pmatrix} \frac{\partial f}{\partial b}(r_+,\ell)
    & \frac{\partial f}{\partial L}(r_+,\ell) \\
    \frac{\partial f}{\partial b}(r_-,-\ell) & \frac{\partial f}{\partial L}
    (r_-,-\ell) \\
    \frac{\partial \xi}{\partial b}(r_+,\ell) -
    \frac{\partial \xi}{\partial b}(r_-,-\ell) &
    \frac{\partial \xi}{\partial b}(r_+,\ell) -
    \frac{\partial \xi}{\partial L}(r_-,-\ell)
\end{pmatrix}\,.
\end{align}
By using \eqref{alt:fxi_fin} for $f$ and $\xi$, we can calculate the
entries in the Jacobians analytically as
\begin{equation}\label{app:jac_ent}
\begin{split}
  \frac{\partial f}{\partial r} &= 
  \left[ \frac{4\cosh(2r-\ell-L) - 2(\tanh(r-\ell) +
      \tanh(r-L))\sinh(2r-\ell-L)}{5 \cosh(r-L)\cosh(r-\ell)} \right] + \frac{2b(1-br^2)}{(1+br^2)^2}\,, \\
 \frac{\partial f}{\partial \ell} &= \frac{2}{5}\left[
   \frac{\sinh(2r-\ell-L)\tanh(r-\ell) - \cosh(2r-\ell-L)}{\cosh(r-L)
     \cosh(r-\ell)} \right]\,, \\
\frac{\partial f}{\partial b} &= \frac{2r}{(1+br^2)^2}\,, \\
\frac{\partial f}{\partial L} &= \frac{2}{5}\left[
  \frac{\sinh(2r-\ell-L)\tanh(r-L) - \cosh(2r-\ell-L)}{\cosh(r-L)\cosh(r-\ell)}
\right]\,, \\
\frac{\partial \xi}{\partial r} &= \frac{\sinh(\ell-L)}{6(1+br^2)^{5/2}}
\left[\frac{3br + (1+br^2)(2\tanh(r-\ell) + \tanh(r-L))}
  {\cosh^2(r-\ell)\cosh(r-L)}\right]\,, \\
\frac{\partial \xi}{\partial \ell} &=
\frac{(1+br^2)^{-3/2}}{6}\left[\frac{2\tanh(r-\ell)\sinh(L-\ell) -
    \cosh(L-l)}{\cosh^2(r-\ell)\cosh(r-L)}\right]\,, \\
\frac{\partial \xi}{\partial b} &= -\frac{r^2(1+br^2)^{-5/2}}{4}
\left[\frac{\sinh(L-\ell)}{\cosh^2(r-\ell)\cosh(r-L)}\right]\,, \\
\frac{\partial \xi}{\partial L} &= \frac{(1+br^2)^{-3/2}}{6}
\left[\frac{\cosh(L-\ell) + \sinh(L-\ell)\tanh(r-L)}
  {\cosh^2(r-\ell)\cosh(r-L)}\right]\,. 
\end{split}
\end{equation}

\begin{figure}[tbh]
\begin{center}
  \includegraphics[width=0.50\textwidth]{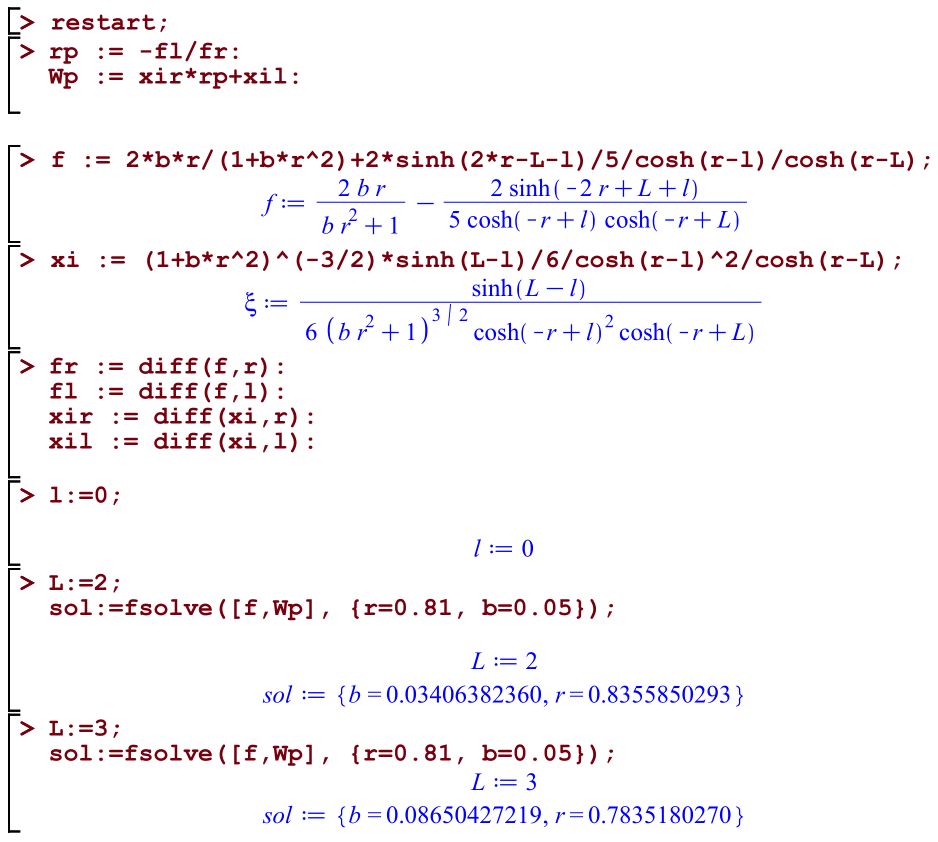}
  \includegraphics[width=0.49\textwidth]{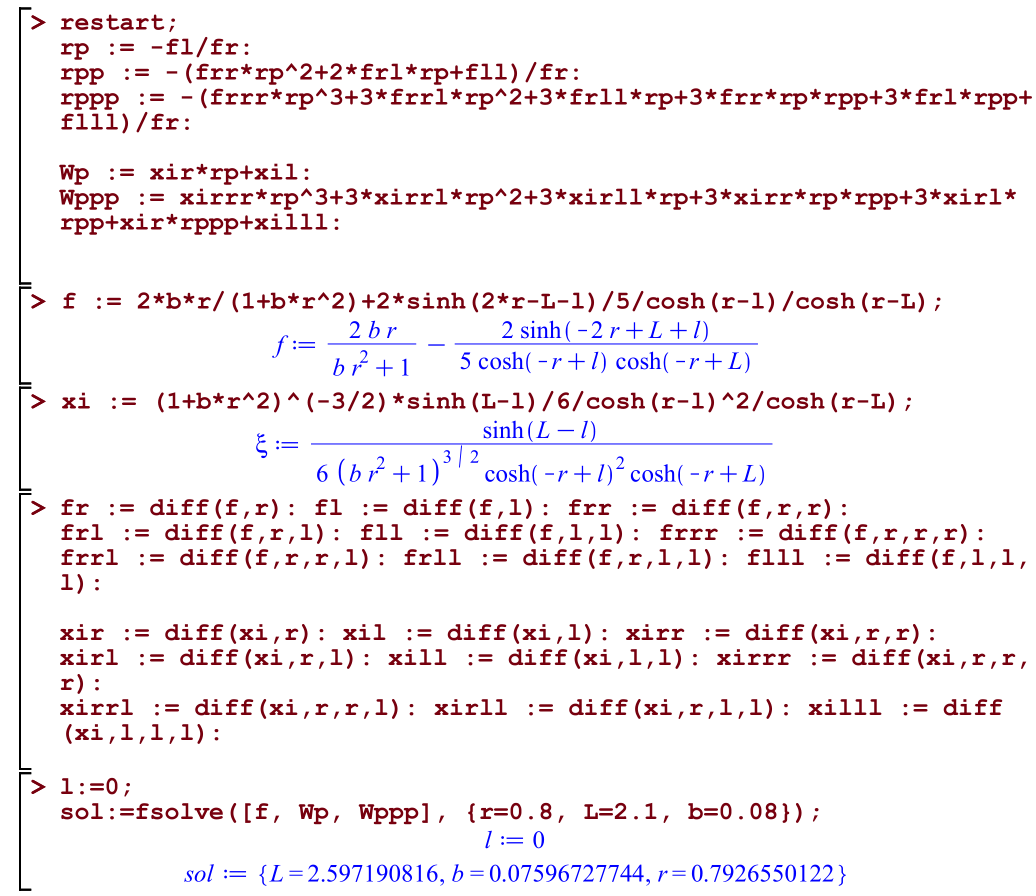}
\end{center}
\caption{Maple code to compute the bifurcation point (left panel) from
  \eqref{1212} and the second-order bifurcation point (right panel)
  from \eqref{1214}, which corresponds to the switch between a
  subcritical and a supercritical symmetry-breaking bifurcation.}
\label{fig:maple}
\end{figure}

Finally, in Fig.~\ref{fig:maple} we include the Maple code used to
compute the symmetry-breaking bifurcation point as well as parameter
set where this bifurcation switches from subcritical to supercritical.
This was described in \eqref{1212} and \eqref{1214} of \S
\ref{sec:fppp}.

\end{appendix}

\end{document}